\documentclass[preprint,12pt]{elsarticle}
\usepackage{graphicx}
\usepackage{epsfig}
\usepackage{amssymb}
\usepackage{adjustbox}
\usepackage{color}
\usepackage{array}
\usepackage[linesnumbered,boxed]{algorithm2e}

\usepackage{nfontdef}

\newcommand{\ignore}[1]{}

\newcommand{\vek}[1]{\mathchoice{\displaystyle\boldsymbol#1}
{\textstyle\boldsymbol#1}{\scriptstyle\boldsymbol#1}
{\scriptscriptstyle\boldsymbol#1}}

\journal{}

\begin{document}

\begin{frontmatter}

\title{Compression and Reconstruction of Random Microstructures using
  Accelerated Lineal Path Function}

\author[ctu]{Jan Havelka}
\ead{jan.havelka.1@fsv.cvut.cz}

\author[ctu]{Anna Ku\v{c}erov\'a\corref{auth}}
\ead{anicka@cml.fsv.cvut.cz}

\author[ctu]{Jan S\'{y}kora}
\ead{jan.sykora.1@fsv.cvut.cz}

\cortext[auth]{Corresponding author. Tel.:~+420-2-2435-5326;
fax~+420-2-2431-0775}

\address[ctu]{Department of Mechanics, Faculty of Civil Engineering,
  Czech Technical University in Prague, Th\'{a}kurova 7, 166 29 Prague
  6, Czech Republic}

\begin{abstract}

Microstructure reconstruction and compression techniques are designed
to find a microstructure with desired properties. While the
microstructure reconstruction searches for a microstructure with
prescribed statistical properties, the microstructure compression
focuses on efficient representation of material morphology for a
purpose of multiscale modelling. Successful application of those
techniques, nevertheless, requires proper understanding of underlying
statistical descriptors quantifying material morphology. In this paper
we focus on the lineal path function designed to capture namely
short-range effects and phase connectedness, which can be hardly
handled by the commonly used two-point probability function. The usage
of the lineal path function is, however, significantly limited by huge
computational requirements. So as to examine the properties of the
lineal path function within the computationally exhaustive compression
and reconstruction processes, we start with the acceleration of the
lineal path evaluation, namely by porting part of its code to the
graphics processing unit using the CUDA (Compute Unified Device
Architecture) programming environment. This allows us to present a
unique comparison of the entire lineal path function with the commonly
used rough approximation based on the Monte Carlo and/or sampling
template. Moreover, the accelerated version of the lineal path
function is then compared with the two-point probability function
within the compression and reconstruction of two-phase
morphologies. Their significant features are thoroughly discussed and
illustrated on a set of artificial periodic as well as real-world
random microstructures.
\end{abstract}


\end{frontmatter}
\section{Introduction}
\label{sec:intro}

Computational modelling of random heterogeneous materials is a
nontrivial multi-disciplinary problem with a wide range of
relevant engineering applications. The FE$^{2}$-methods have been
developed as promising techniques for material modelling and used
to derive effective models at the scale of interest. The unifying
theoretical framework is provided by homogenization theories
aiming at the replacement of the heterogeneous microstructure by
an equivalent homogeneous material, see~\cite{Torquato:2006}.
Currently, two main approaches are available: (i) computational
homogenization and (ii) effective media theories.

The latter approach aims at estimating the material response
analytically on the basis of limited geometrical information (e.g. the
volume fractions of constituents) of the analysed medium.  Structural
imperfections are introduced in a cumulative sense using one of the
averaging schemes, e.g. the Mori-Tanaka method~\cite{Vorel:2009:SEM}.
The computational requirements are very low, however, such an
analytical solution is available only for a limited spectrum of
microstructural geometries such as media with a specific shape of
inclusions.

Methods based on computational homogenization are more general in
application. They study the distribution of local fields within a
typical heterogeneity pattern using a numerical method. It is
generally accepted that detailed discretisation techniques, and
the finite element method in particular, remain the most powerful
and flexible tools available.  Despite the tedious computational
time, it provides us the details of local fields, see
e.g.~\cite{Sykora:2012:JCAM,Sykora:2013:AMC}. However, the
principal requirement is to find a representative volume element
(RVE), which can be intriguing in case of real-world random
microstructures. Recent
studies~\cite{Zeman:2007:MSMSE,Schroder:AAM:2011} suggest that
structure preserving spatial geometrical statistics such as a
statistically equivalent periodic unit cell (SEPUC) -- also known
as a statistically similar representative volume element (SSRVE) --
is computationally very efficient comparing to the classical
concept of the RVE.

A relatively new concept of microstructure modelling is based on the
production of a set of structures morphologically similar to the
original media, so called Wang tiles,
see~\cite{Novak:PR:2012,Novak:MSMSE:2013}.  It is an approach that
allows us to obtain aperiodic local fields in heterogenous media with
a small set of statistically representative tiles. The main advantage
of the stochastic Wang tillings is the computational efficiency and
long range spatial correlations, which are neglected in classical
homogenization techniques, see~\cite{Novak:PR:2012}. The tiles can be
in some cases produced by a computational efficient image quilting
algorithm \cite{Doskar:2014:PRE} or generally also by optimising a
chosen statistical descriptor.

The present paper is devoted to statistical descriptors defining
statistically/morphologically similar material structures (cells or
tiles). Such structures are generally obtained by a process of
microstructure reconstruction~\cite{Yeong:1998:PRE} or 
compression~\cite{Povirk:AMM:2005} so as to represent the
microstructure as accurately as possible in terms of the selected
statistical descriptor. In particular, we focus on two commonly
used descriptors, the two-point probability function and the
lineal path function,
see~\cite{Li:2012:CMS,Singh:MSE:2008,Schroder:AAM:2011}. The goal
of this paper is to investigate in more detail the properties and
differences of these two descriptors within the compression and
reconstruction process. So as to achieve this goal, we concentrate
on calculation of the entire lineal path function instead of its
often used rough discretisation by a sampling template evaluated
approximately using a Monte Carlo-based procedure,
see~\cite{Zeman:2003}.  Since the evaluation of the entire lineal
path function can be computationally extremely exhaustive, we
present certain acceleration steps on the algorithmic as well as
on the implementation side, where the significant speed up is
achieved namely by porting the algorithm to the graphics
processing unit (GPU) using the CUDA environment.

This article has been organised in the following way. The next
section describes a theoretical formulation of the both
descriptors. Section~\ref{sec:imple} is devoted to acceleration of the
lineal path function and presents the resulting speed-up obtained
at GPU in comparison with the sequential CPU formulation.
Section~\ref{sec:optim} briefly introduce the optimisation
algorithm employed for microstructure compression and
reconstruction discussed in Sections~\ref{sec:recon}
and~\ref{sec:compress}, respectively. Final summary of the
essential findings are provided in Section~\ref{sec:concl}.

\section{Statistical description of random media}
\label{sec:statdes}

The morphology quantification for random heterogeneous materials
starts from the introduction of the concept of an ensemble
established by Kr\"{o}ner~\cite{Kroner:JMPS:1977} and
Beran~\cite{Beran:1968}. Proposed mathematical formulations are
considered as one of the milestones in statistical physics and the
basic idea is that macroscopic observables can be calculated by
performing averages over the systems in the ensemble. In other
words, the ensemble represents the collection of geometrical
systems having different microstructures but being completely
identical from a macroscopic point of view~\cite{Zeman:2003}.

A variety of statistical descriptors were developed to describe the
morphology of a multi-phase random heterogenous
material~\cite{Zeman:2003,Zeman:2007:MSMSE} based on the concept of
an ensemble. In the present work, the two-point probability function and
the lineal path function are investigated as frequently used
descriptors. Therefore, this section provides their brief analytical
description and classical numerical implementation.

As a preamble, throughout this paper we consider an ensemble of a
two-phase medium consisting of a black and white phase labelled by
superscripts $i,\,j\in\{\mathrm{b},\mathrm{w}\}$. We also model
the medium only as a two-dimensional system, where the position of
an arbitrary point $\vek{x_a}$ is defined by the Cartesian
coordinates $\vek{x_a} = ( x_a, y_a)$. Nevertheless, the extension
into the three-dimensional systems is very straightforward.

\subsection{Two-point probability function}
\label{subsec:TPPF}

More formally, the two-point probability function
$S_{2}^{ij}({\vek{x}_1}, {\vek{x}_2})$\footnote{with $S_{2}^{i}$
  abbreviating $S_{2}^{ii}$} quantifies the probability of finding
simultaneously the phase $i$ and the phase $j$ at two arbitrarily
chosen points ${\vek{x}_1}$ and ${\vek{x}_2}$, respectively, and
can be written in the form,
see~\cite{Zeman:2003,Lombardo:2009:IJMCE},
\begin{equation}
S_{2}^{ij}({\vek{x}_1}, {\vek{x}_2}) =
\langle\chi^{i}(\vek{x}_1,\alpha) \chi^{j}(\vek{x}_2,
\alpha)\rangle,
\label{eq:s2def}
\end{equation}
where the symbol $\langle\cdot\rangle$ denotes the ensemble average of
the product of characteristic functions
$\chi^{i}(\vek{x}_a,\alpha)$, which are equal to one when the point
$\vek{x}_a$ lies in the phase $i$ in the sample $\alpha$ and equal
to zero otherwise:
\begin{equation}
\chi^{i}(\vek{x}_a,\alpha) = \left\{
\begin{array}{l}
1, \quad \mbox{if } \vek{x_a} \in D^{i}(\alpha) \\
0, \quad \mbox{otherwise}
\end{array}.
\right. \label{eq_charf}
\end{equation}
\begin{figure} [t!]
\centering
\begin{tabular}{@{}c@{}|@{}c@{}@{}c@{}}
\begin{adjustbox}{valign=m}
    \begin{tabular}{c}
    \includegraphics*[width=25mm,keepaspectratio]{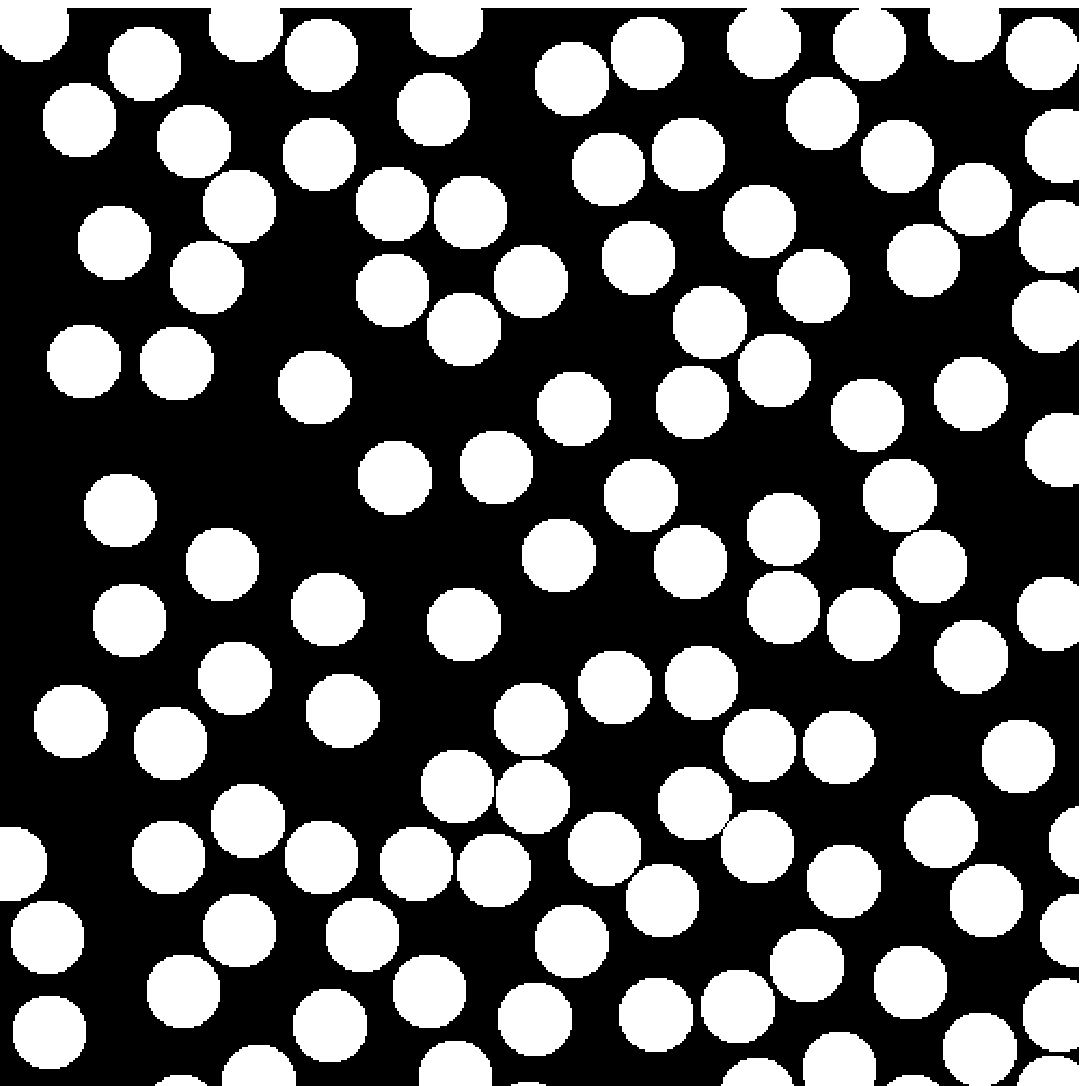} \\
    (a)\\
    \end{tabular}
\end{adjustbox}
&
\begin{adjustbox}{valign=m}
    \begin{tabular}{c}
    \includegraphics*[width=50mm,keepaspectratio]{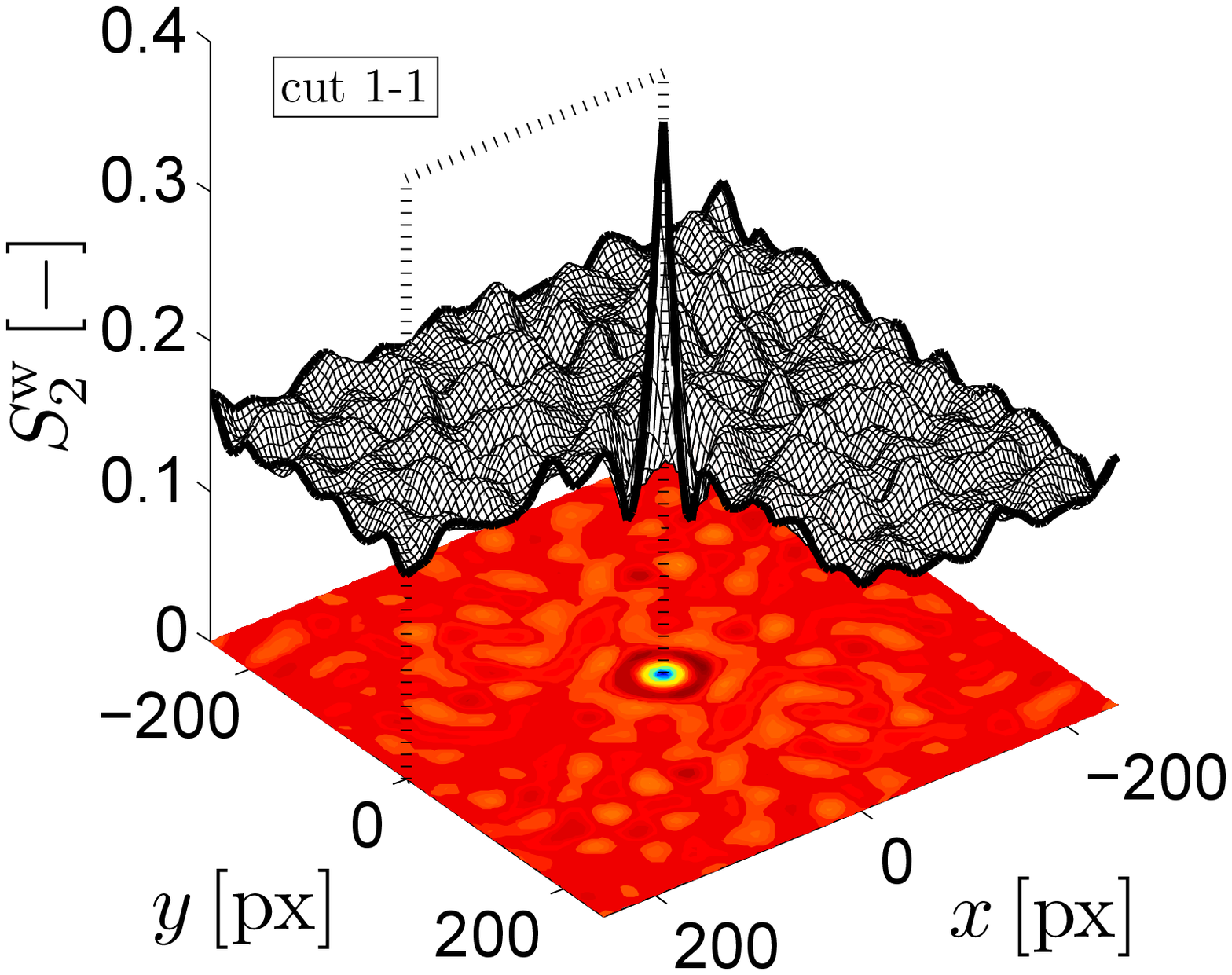}\\
    (b)\\
    \includegraphics*[width=50mm,keepaspectratio]{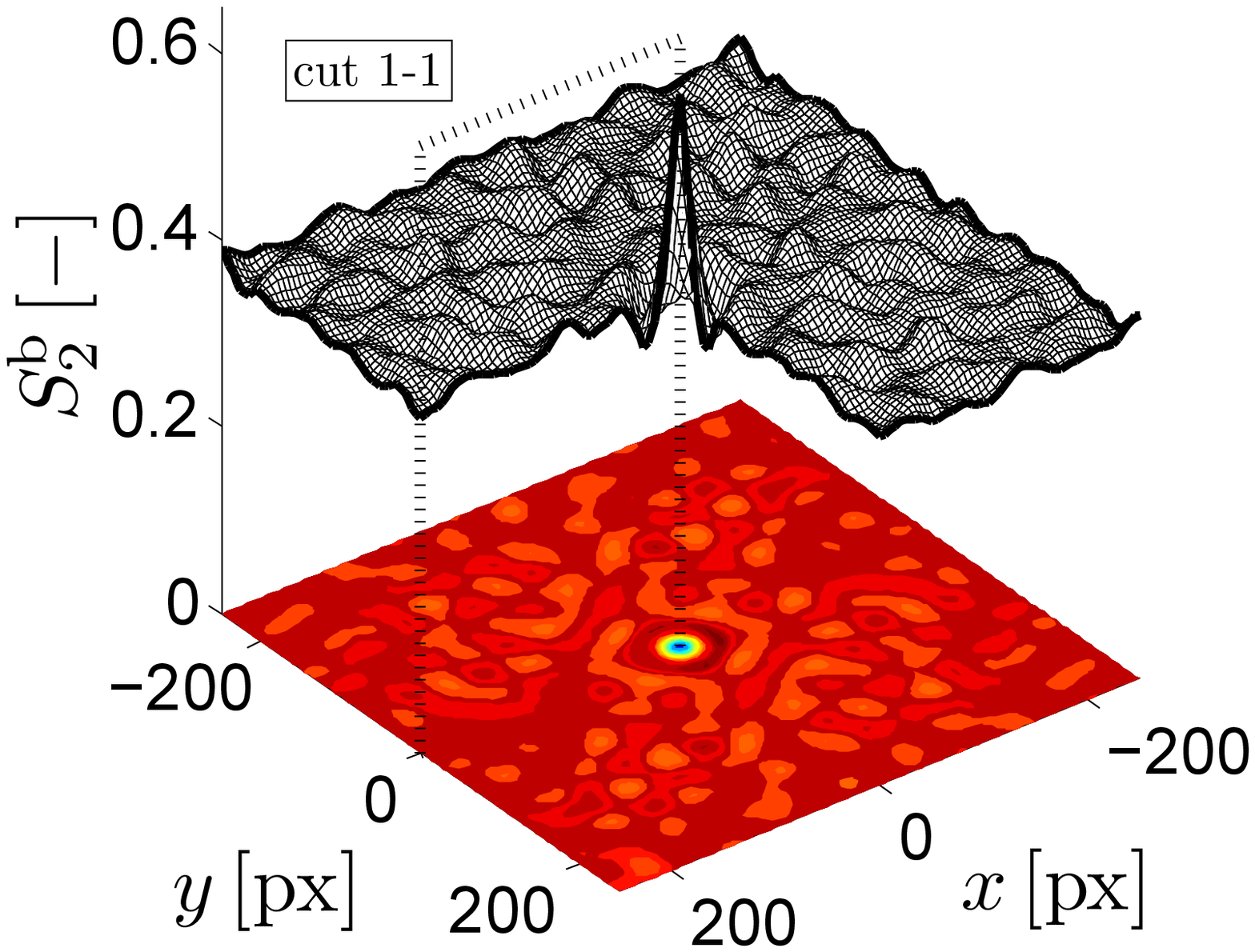}\\
    (c)\\
    \end{tabular}
\end{adjustbox}
&
\begin{adjustbox}{valign=m}
    \begin{tabular}{c}
    \includegraphics*[width=55mm,keepaspectratio]{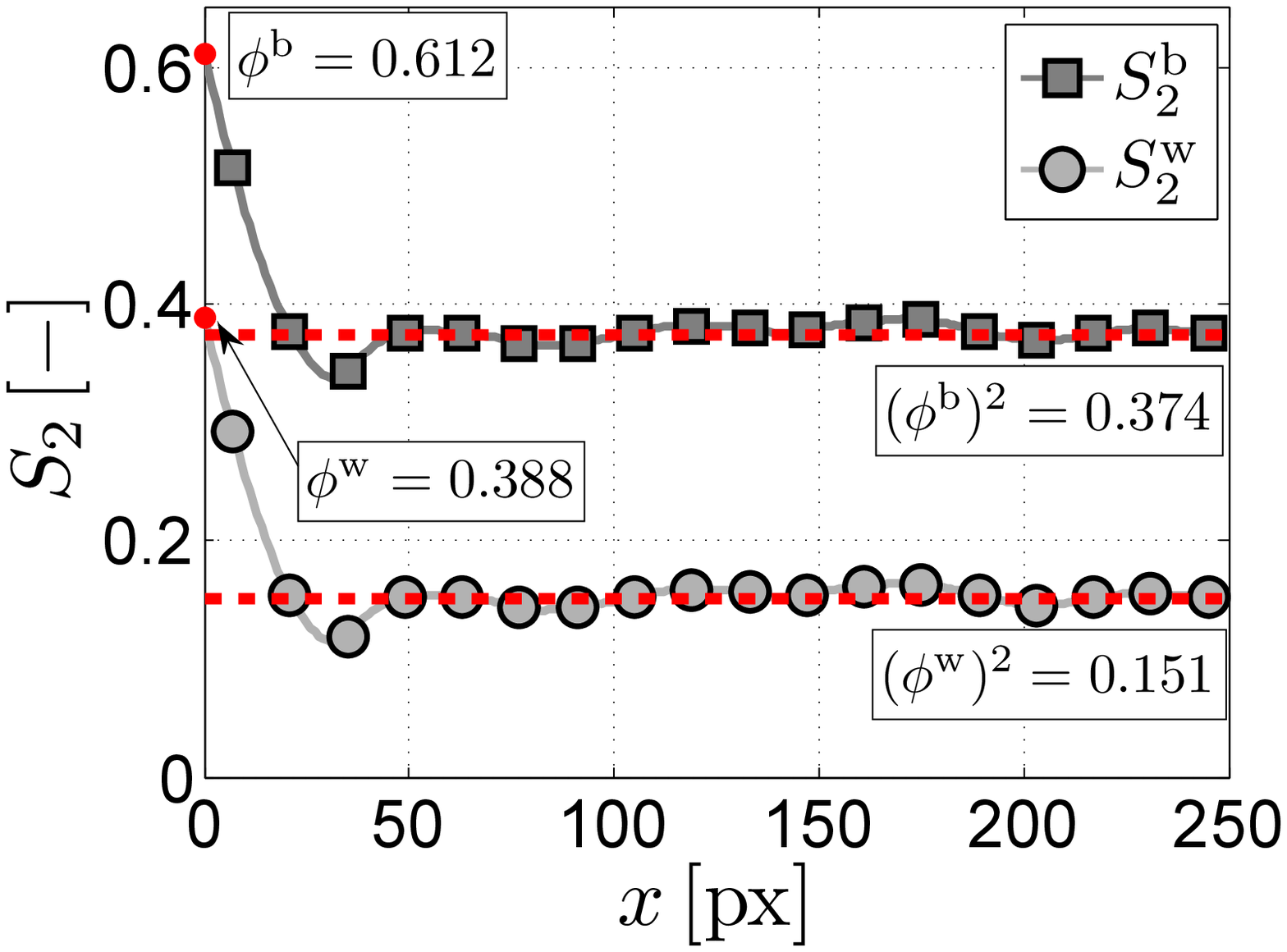} \\
    (d)\\
    \end{tabular}
\end{adjustbox}
\end{tabular}
\caption{Illustration of the two-point probability function: (a)
  Example of a homogeneous system, size $500\times500$ [px]; (b) $S_{2}^{\mathrm{w}}$-function; (c) $S_{2}^{\mathrm{b}}$-function;
  (d) Comparison of $S_2$-functions in cut 1-1.}
\label{fig:s2ilustr}
\end{figure}
In Eq.~(\ref{eq_charf}), $D^{i}(\alpha)$ denotes the domain
occupied by the $i$-th phase. In general, the evaluation of these
characteristics may prove to be prohibitively difficult.
Fortunately for homogeneous systems, the $S_{2}^{i}$ depends only
on the relative position of the two points $\vek{x} = \vek{x}_2 -
\vek{x}_1$ and has following asymptotic properties,
see~\cite{Yeong:1998:PRE},
\begin{eqnarray}
  S_{2}^{i}(|\vek{x}| = 0) & = & \phi^{i}, \label{eq:s2zero}\\
  \lim_{|\vek{x}| \rightarrow\infty}S_{2}^{i}(\vek{x}) & = & (\phi^{i})^2, \label{eq:s2lim}
\end{eqnarray}
where $\phi^{i}$ is the volume fraction of the $i$-th phase.
Eq.~\eqref{eq:s2zero} follows from definition~\eqref{eq:s2def} and
means that the probability of a randomly thrown point (i.e. vector of
zero length) falling into the phase~$i$ is equal to the volume
fraction of the phase~$i$. On the other hand, Eq.~\eqref{eq:s2lim}
assumes that the system has no long-range correlations and thus,
falling of the two distant points $\vek{x_1}$ and $\vek{x_2}$ into the
phase $i$ are independent events, each having the probability equal to
$\phi^{i}$, see Fig.~\ref{fig:s2ilustr} as an illustrative example of such
a system.

Even though we aim at characterization of generally non-periodic
media by a SEPUC, whose boundaries are constructed as periodic, it
has been demonstrated in~\cite{Gajdosik:2006:PEM} that assumption
of periodic boundaries does not introduce a systematic bias in the
values of statistical descriptors. On the other hand, the
assumption of the periodicity simplifies the computation of the
two-point probability function, because we do not need to consider
all the possible orientations of the vector $\vek{x}$.
\begin{figure} [t!]
\centering
\begin{tabular}{ccc}
  \includegraphics*[width=60mm,keepaspectratio]{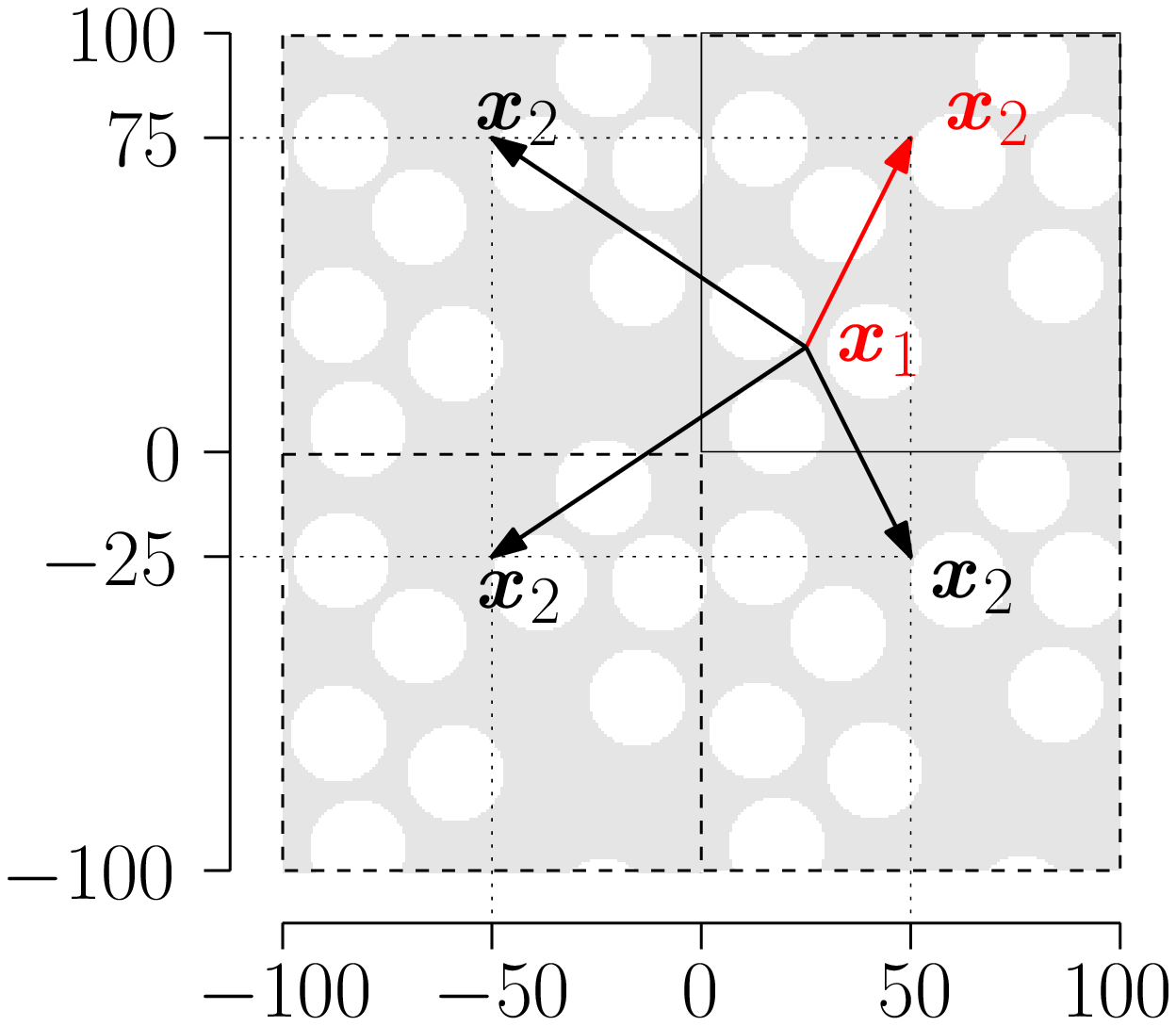} &
  \includegraphics*[width=60mm,keepaspectratio]{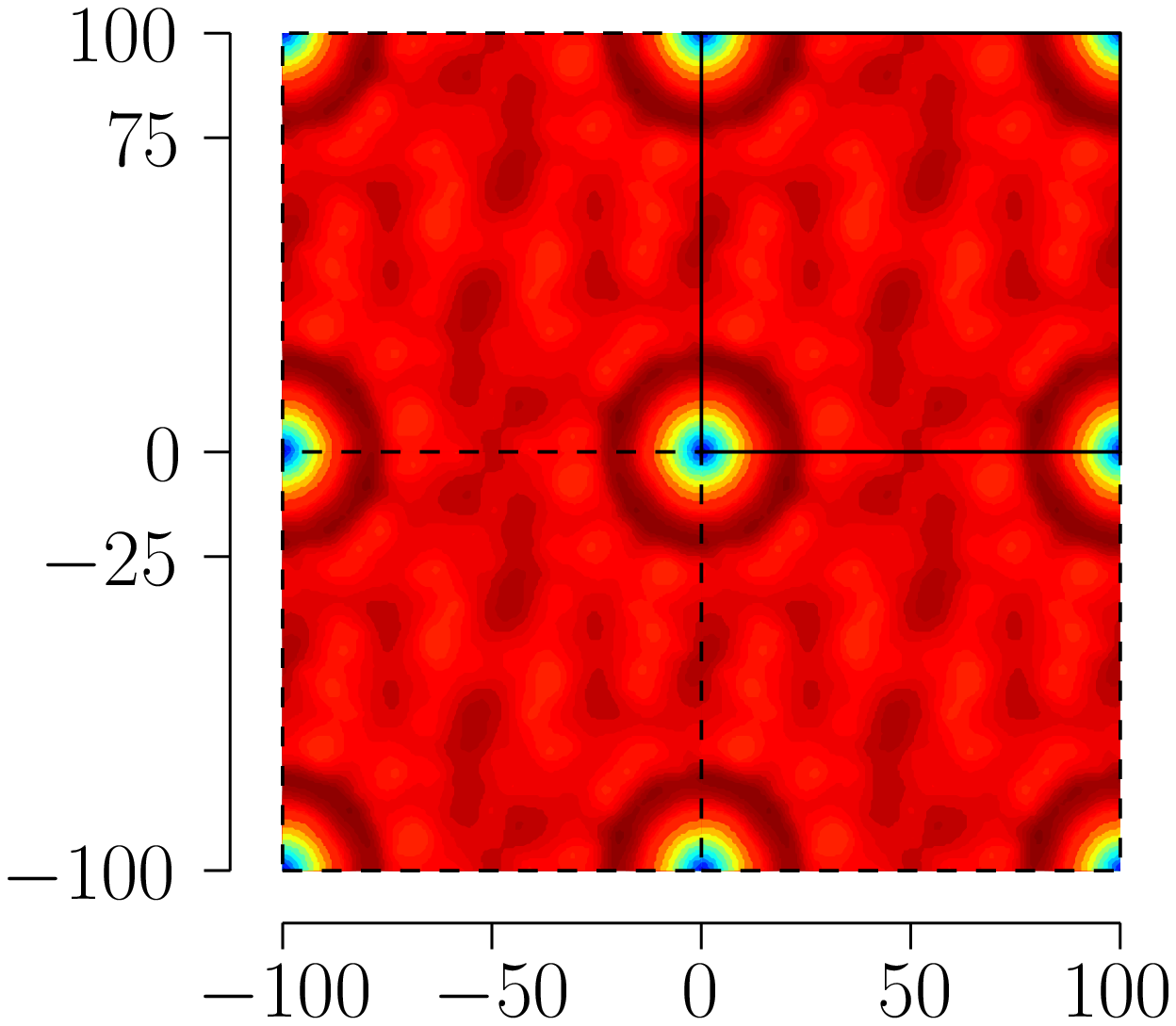} \\
  (a) & (b) \\
\end{tabular}
\caption{Vectors connecting identical points in periodical setting (a)
  and the corresponding identical values of the periodic two-point
  probability function (b)} \label{fig:periodicity}
\end{figure}
As shown in Fig.~\ref{fig:periodicity}a, four differently oriented
vectors are actually connecting the identical points and obviously
have the same value of the two-point probability function. As a
consequence, the evaluation of the two-point probability function for
vectors oriented into the first quadrant includes the information about
all the other vector orientations, see Fig.~\ref{fig:periodicity}b.

The last note concerns particularly the two-phase medium, where the two-point
probability functions of particular phases are related according to
the following equation
\begin{equation}
  S_{2}^{i}(\vek{x}) = (\phi^{i})^2 - (\phi^{j})^2 + S_{2}^{j}(\vek{x}),
\label{eq:S2add}
\end{equation}
i.e. they differ only by a constant as also visible in
Fig.~\ref{fig:s2ilustr}d.  Since the constant is given by known volume
fractions of particular phases, only one
two-point probability function needs to be determined to describe the
two-phase medium.  For this reason, we may drop the superscript of
$S_{2}^{i}(\vek{x})$ and write only the two-point
probability function as $S_{2}(\vek{x})$.

\begin{figure} [t!]
\centering
\includegraphics*[width=50mm,keepaspectratio]{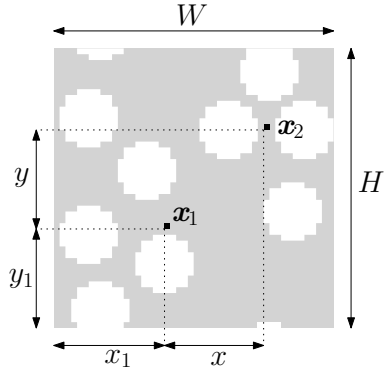}
\caption{Illustration of a digitised image} \label{fig:s2px}
\end{figure}

Implementation of the two-point probability function is based on the
assumption of a discrete description of a studied system, mainly binary
images in our case. The general and simple Monte Carlo-based
evaluation strategy throws randomly two points into the investigated
medium and counts successful ``hits'' of both points into the phase
$i$.  This approach is, however, not only approximate, but also very
computationally demanding. Therefore, another practical method was
introduced on the basis of rewriting the two-point probability
function as an autocorrelation of the characteristic function $\chi^{i}$
for a periodic medium as, see~\cite{Zeman:2003},
\begin{equation}
  S_{2}^{i}(x,y) =
  \frac{1}{WH}\sum_{x_1=0}^{W-1}\sum_{y_1=0}^{H-1}\chi^{i}(x_1,y_1)\chi^{i}((x_1+x)\%W,(y_1+y)\%H),
\label{eq:S2dis}
\end{equation}
where the symbol $\%$ is the modulo, $\chi^{i}(x_1,y_1)$ denotes the value
of $\chi^{i}$ for the pixel located in the $y_1$-th row and the $x_1$-th
column of the digitised image with the dimensions $W \times H$, $x$
and $y$ are the vertical and horizontal distances between two pixels,
see Fig.~\ref{fig:s2px}.
According to~\cite{Lombardo:2009:IJMCE}, the Eq.~(\ref{eq:S2dis}) can
be computed in an efficient way using the Fast Fourier Transform.
Applying this, the reformulation of the two-point probability function
$S_{2}^{i}$ for a periodic medium can be written as
\begin{equation}
S_{2}^{i}(x,y) =
\frac{1}{WH}\mbox{IDFT}\left\{\mbox{DFT}\left\{\chi^{i}(x,y)\right\}
\overline{ \mbox{DFT} \left\{\chi^{i}(x,y)\right\} } \right\},
\end{equation}
where IDFT is the inverse Discrete Fourier Transform (DFT), the symbol
$\bar{\cdot}$ stands for the complex conjugate. This method is very
efficient and its accuracy depends only on the selected resolution of the
digitised medium, see~\cite{Zeman:2003,Zeman:2007:MSMSE}. The Fast
Fourier Transform, which needs only $\mathcal{O}(WH\log(WH)+WH)$
operations, is used to perform the numerical computations presented
below.

\subsection{Lineal path function}
\label{subsec:LPF}

\begin{figure} [t!]
\centering
\begin{tabular}{@{}c@{}|@{}c@{}@{}c@{}}
\begin{adjustbox}{valign=m}
    \begin{tabular}{c}
    \includegraphics*[width=25mm,keepaspectratio]{CCcomposite_5e2x5e2.eps} \\
    (a)\\
    \end{tabular}
\end{adjustbox}
&
\begin{adjustbox}{valign=m}
    \begin{tabular}{c}
    \includegraphics*[width=50mm,keepaspectratio]{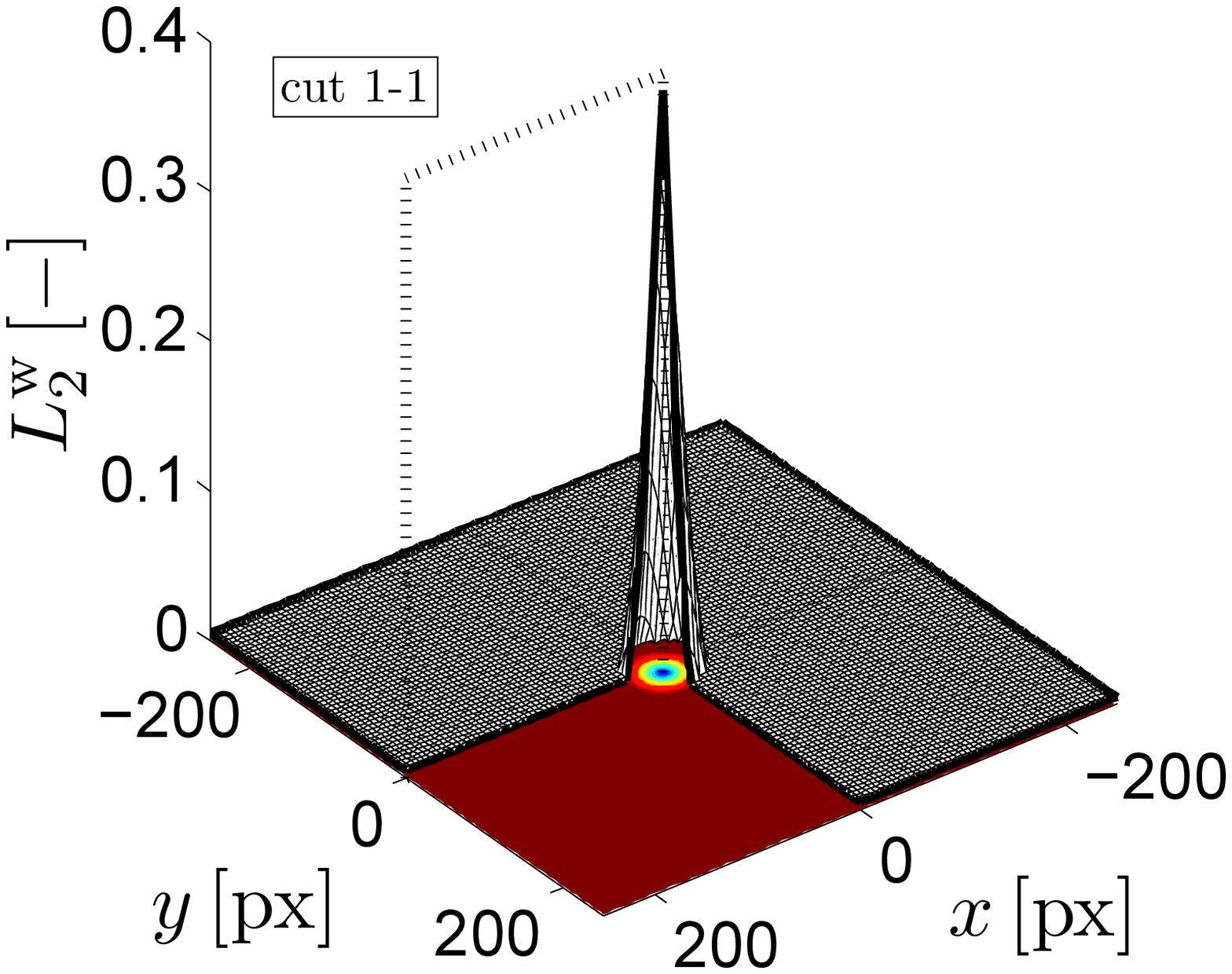}\\
    (b)\\
    \includegraphics*[width=50mm,keepaspectratio]{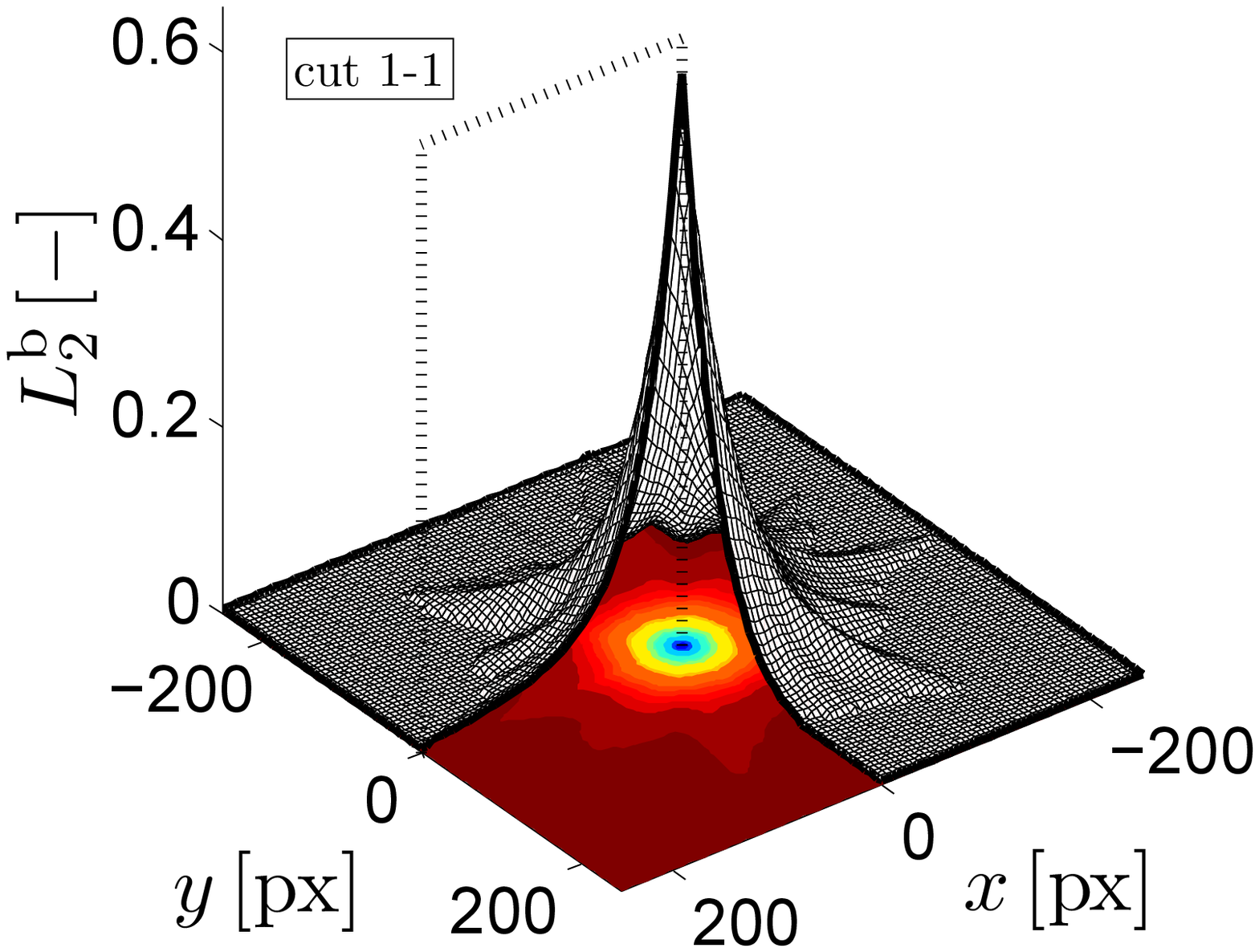}\\
    (c)\\
    \end{tabular}
\end{adjustbox}
&
\begin{adjustbox}{valign=m}
    \begin{tabular}{c}
    \includegraphics*[width=55mm,keepaspectratio]{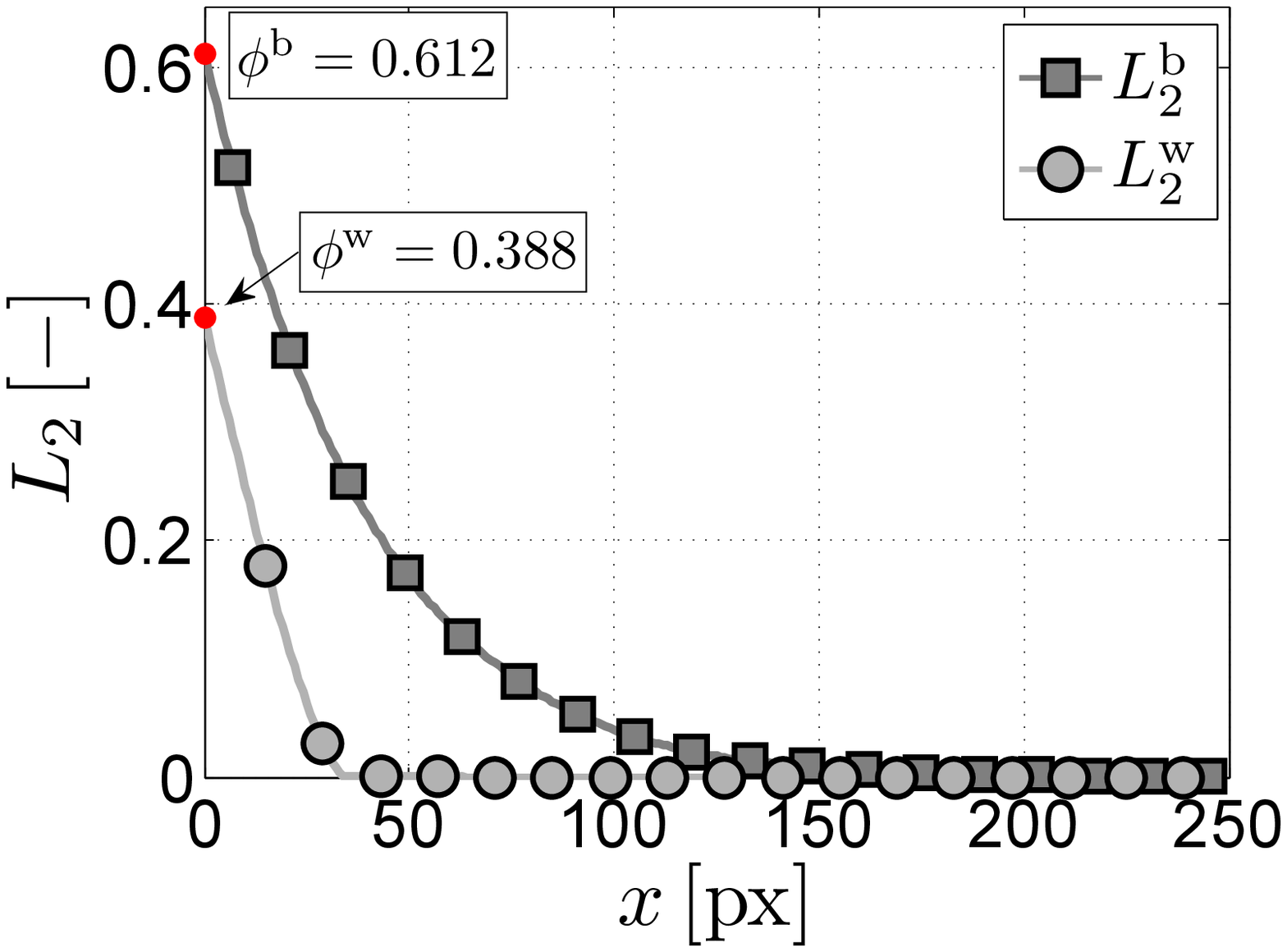} \\
    (d)\\
    \end{tabular}
\end{adjustbox}
\end{tabular}
\caption{Illustration of the two-point probability function: (a)
  Example of a homogeneous system, size $500\times500$ [px]; (b) $L_{2}^{\mathrm{w}}$-function; (c) $L_{2}^{\mathrm{b}}$-function;
  (d) Comparison of $L_2$-functions in cut 1-1.}
\label{fig:l2ilustr}
\end{figure}

Another frequently used statistical descriptor for the
microstructural morphology quantification is the lineal path
function $L_{2}^{i}({\vek{x}_1}, {\vek{x}_2})$, originally
introduced in~\cite{Lu:PR:1992} and further elaborated
in~\cite{Yeong:1998:PRE,Zeman:2003}. It is defined as a low-order
descriptor based on a more complex fundamental function
$\lambda^{i}$ able to describe certain information about the phase
connectedness and putting more emphasis on the short-range
correlations, since its value quickly vanishes to zero with
increasing $|\vek{x}|$.  The fundamental function $\lambda^{i}$ is
defined as
\begin{equation}
\lambda^{i}(\vek{x}_{1},\vek{x}_{2},\alpha)=\left\{\begin{array}{l}1,\,
\mathrm{if}\,\vek{x}_{1}\vek{x}_{2}\subset
D^{i}(\alpha),\\
0,\, \mathrm{otherwise},
\end{array}\right.
\end{equation}
i.e., a function which equals to $1$ when the segment
$\vek{x}_{1}\vek{x}_{2}$ is contained in the phase $i$ for the sample
$\alpha$ and $0$ otherwise. The lineal path function is defined as the
probability that the line segment $\vek{x}_{1}\vek{x}_{2}$ lies
entirely in the phase $i$ and it can be written as the ensemble
averaging fundamental function given as
\begin{equation}
L_{2}^{i}(\vek{x}_{1},\vek{x}_{2})=\langle\lambda^{i}(\vek{x}_{1},\vek{x}_{2},\alpha)\rangle.
\end{equation}
As mentioned above, under the assumption of statistical
homogeneity~\cite{Zeman:2003}, the function again simplifies to
$L_{2}^{i}(\vek{x}_{1},\vek{x}_{2}) = L_{2}^{i}(\vek{x})$ with
$\vek{x} = \vek{x}_2 - \vek{x}_1$ and yields
\begin{eqnarray}
  L_{2}^{i}(|\vek{x}| = 0) & = & \phi^{i} \label{eq:l2zero}\\
  \lim_{|\vek{x}|\rightarrow\infty}L_{2}^{i}(\vek{x}) & = & 0. \label{eq:l2lim}
\end{eqnarray}
Here again, the Eq.~\ref{eq:l2lim} assumes no long-range correlations
and thus the probability that the line segment
$\vek{x}_{1}\vek{x}_{2}$ lies entirely in the phase $i$ vanishes to
zero with its increasing length, see Fig.~\ref{fig:l2ilustr} for an
illustration of such a homogeneous system.

\begin{figure} [t!]
\centering
\begin{tabular}{@{}c@{}@{}c@{}@{}c@{}}
  \includegraphics*[width=35mm,keepaspectratio]{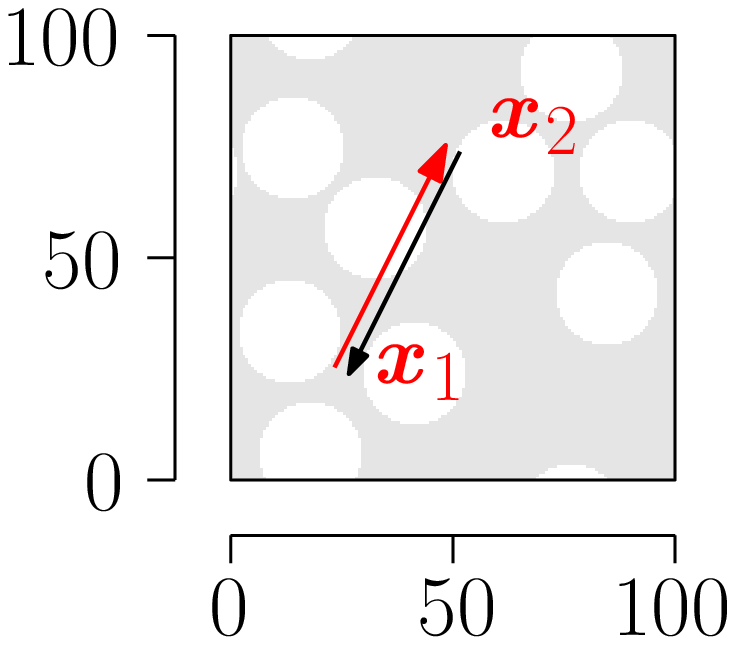} &
  \includegraphics*[width=55mm,keepaspectratio]{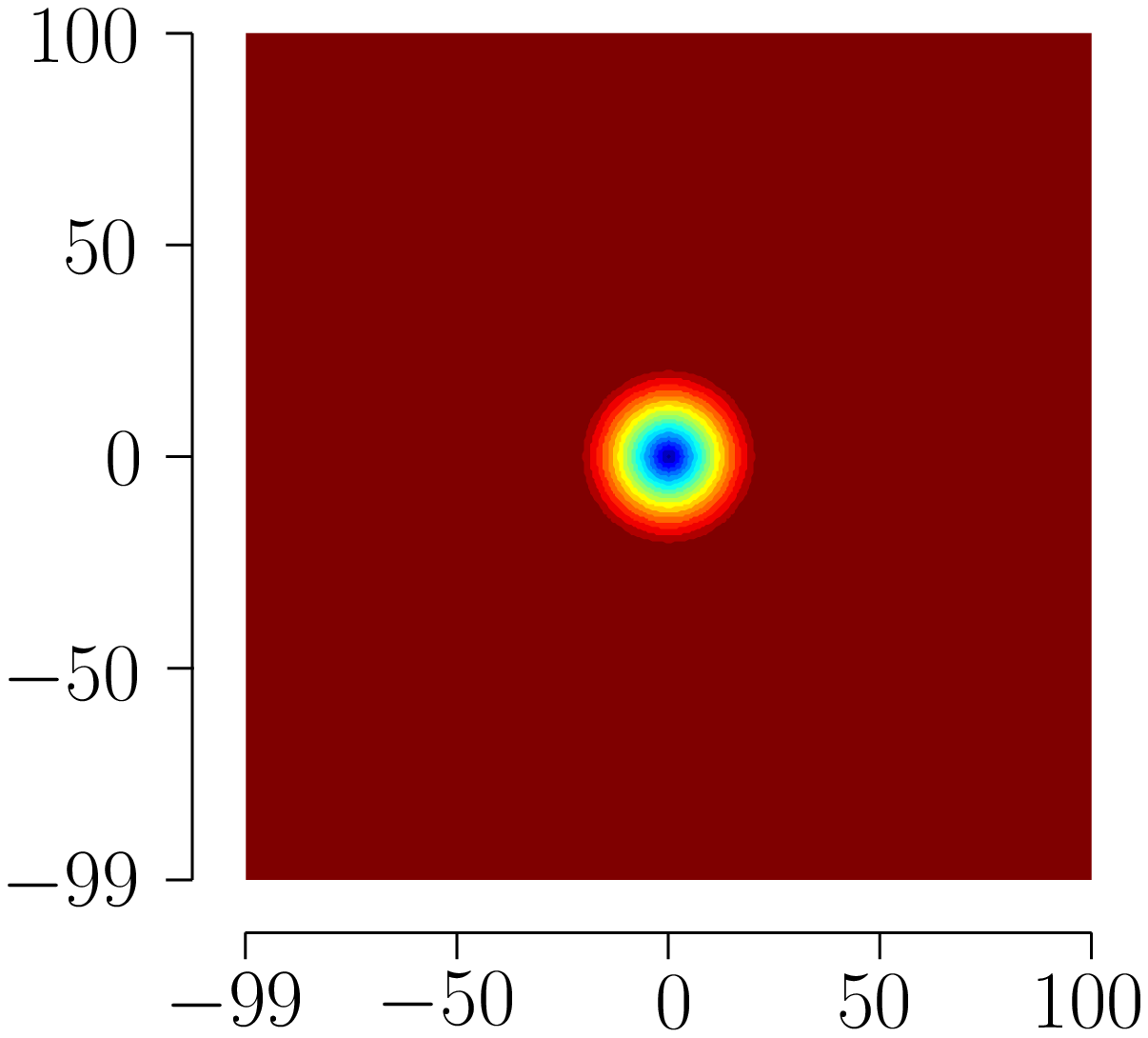} &
  \includegraphics*[width=55mm,keepaspectratio]{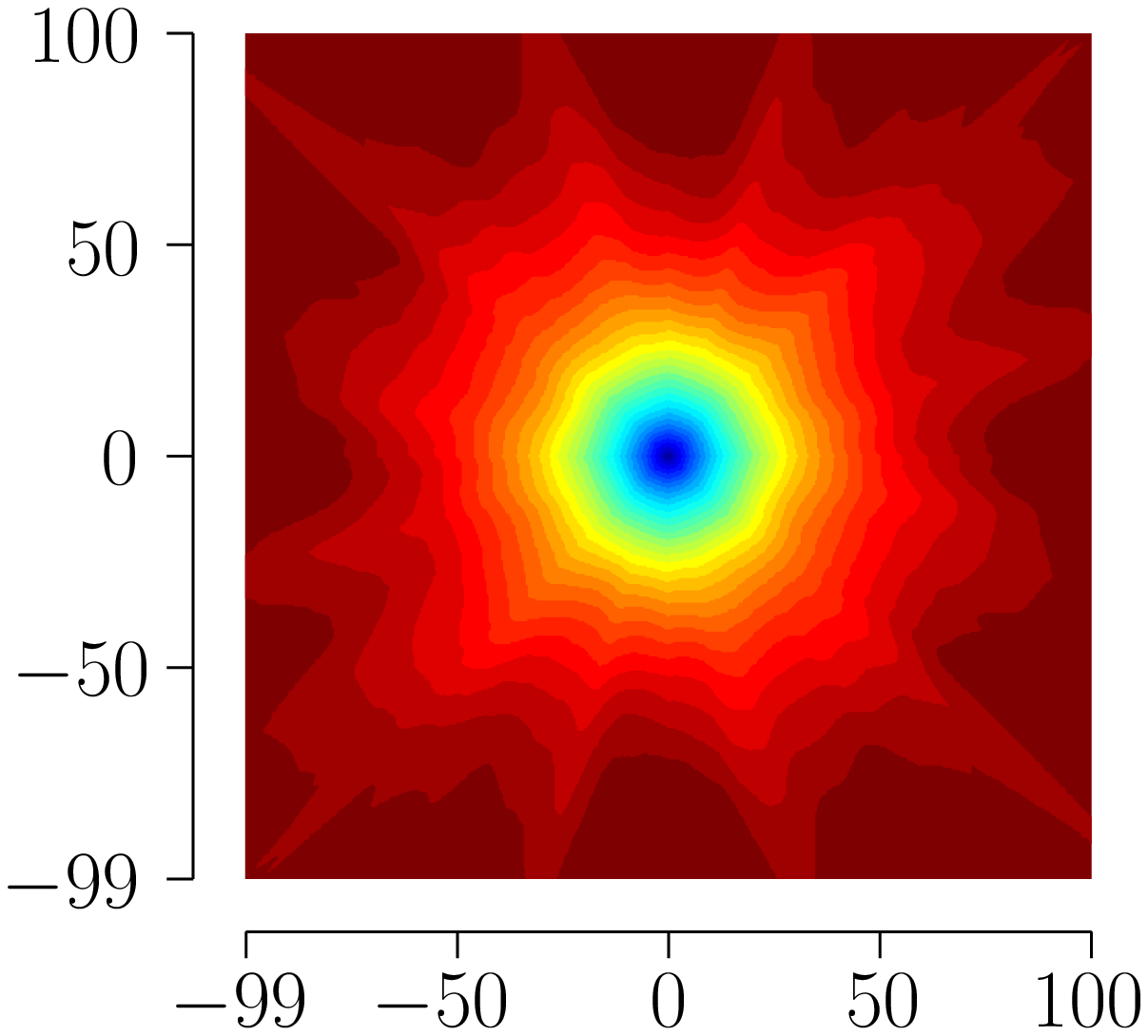}\\
  (a) & (b) & (c)\\
\end{tabular}
\caption{Vectors corresponding to identical segments (a), point
  symmetry of lineal path function of the black (b) and white (c)
  phases}
\label{fig:symmetry}
\end{figure}

For the sake of consistency with the formulation and computation of
the two-point probability function, we introduce again the assumption of
the periodicity in our numerical implementation. However, there arise
no computational benefits, since all the vectors in
Fig.~\ref{fig:periodicity}a are connecting the same points via
a different path. Nevertheless, we need to keep in mind that the line
segment $\vek{x}_{1}\vek{x}_{2}$ is identical to the line segment
$\vek{x}_{2}\vek{x}_{1}$ and thus
\begin{equation}
  L_{2}^{i}(\vek{x}) = L_{2}^{i}(-\vek{x}),
\label{eq:symmetry}
\end{equation}
which means that the lineal path posses point symmetry, see
Fig.~\ref{fig:symmetry}. Hence, we need to compute the lineal path
function only for a half of all the possible orientations of the vector
$\vek{x}$ and the rest is obtained by symmetry.

In contrast to the evaluation of $S_{2}^{i}$, see Eq.~(\ref{eq:S2add}),
the lineal path function computed for one phase does not include the
whole information about the lineal path function of the other phase,
which thus needs to be computed separately, see
Fig.~\ref{fig:symmetry}. This brings additional information about the
structural morphology, but it also means higher computational demands.

\begin{figure} [t!]
\centering
\includegraphics*[width=12cm,keepaspectratio]{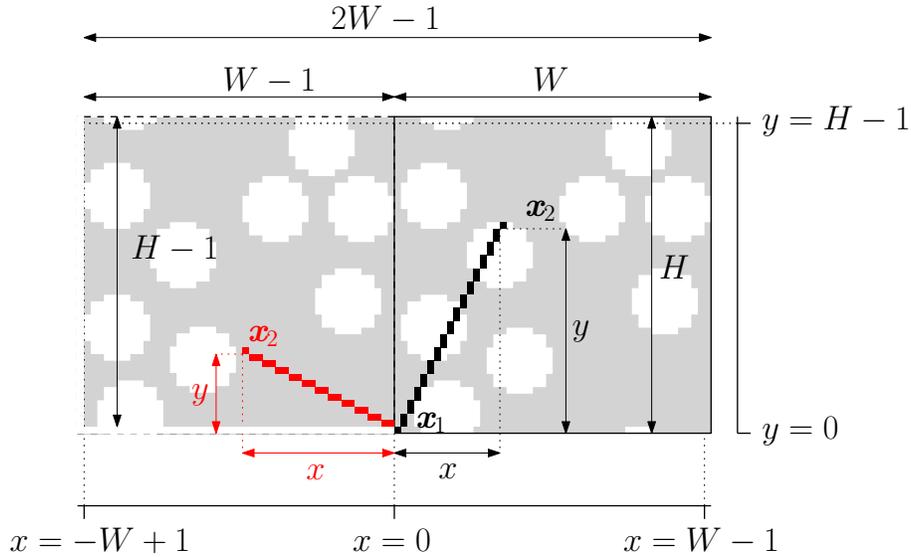}
\caption{Illustration of line segments}
\label{fig:l2px}
\end{figure}

With this in mind, the standard numerical implementation of
a sequential version of the entire $L_{2}^{i}$ starts from the
definition of line segments connecting two pixels $\vek{x}_{1}$
and $\vek{x}_{2}$ within the image with the dimensions $W\times
H$. The set of pixels representing a segment starting in
$\vek{x}_{1}=(0,0)$ and ending in $\vek{x}_{2}=(x,y)$, is
specified by an algorithm originally proposed by
Bresenham~\cite{Bresenham:SJ:1965} defining a unique solution for
any positions of boundary pixels $\vek{x}_{1},\vek{x}_{2}$. Due to
the point symmetry of the lineal path function, all the
orientations of a line segment necessary for its computation are
produced by moving the point $\vek{x}_2$ within the domain
$\mathbb{D}$ given by two rectangles specifying the left bottom corner
of a pixel, i.e.
\begin{equation}
\mathbb{D} := [-W+1; -1] \times [1; H-1] + [0; W-1] \times [0; H-1] \,
\label{eq:segments}
\end{equation}
see Fig.~\ref{fig:l2px}.  The number of segments defining the
lineal path function is thus given as a cardinality of the domain
$\mathbb{D}$, which is $|\mathbb{D}| = 2HW-H-W+1$. Once having the
defined segments, the computation of the lineal path function involves
simple translations of each segment throughout the image and the
comparison whether all pixels of the segment at a given position
correspond to image pixels with the value of the investigated
phase.  Such an intuitive description represents, however,
a computationally exhaustive procedure leading to
$\mathcal{O}(H^3W^2)$ operations for periodic media with $W\leq
H$. For our purpose, it simplifies to $\mathcal{O}(W^5)$ for
$W=H$, i.e. a square shape of SEPUC/SSRVE. In order to reduce the
computational cost, several algorithmic and hardware acceleration
steps are introduced and described in the following section.


\section{Numerical implementation of $L_{2}$}
\label{sec:imple}

\begin{figure} [h!]
\begin{center}
\begin{tabular}{@{}c@{}|@{\hspace{0.1cm}}c@{}}
    \begin{adjustbox}{valign=b}
        \begin{tabular}{c}
            A.~\includegraphics*[width=15mm,keepaspectratio]{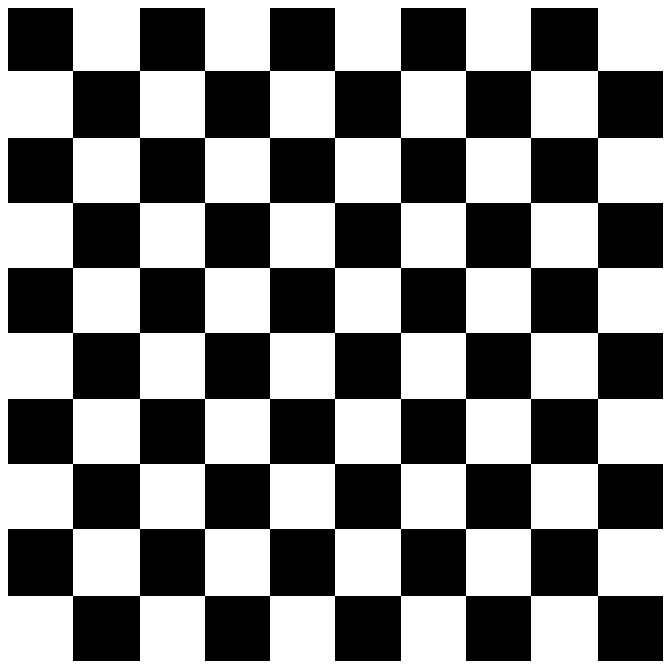}\\[20mm]
            B.~\includegraphics*[width=15mm,keepaspectratio]{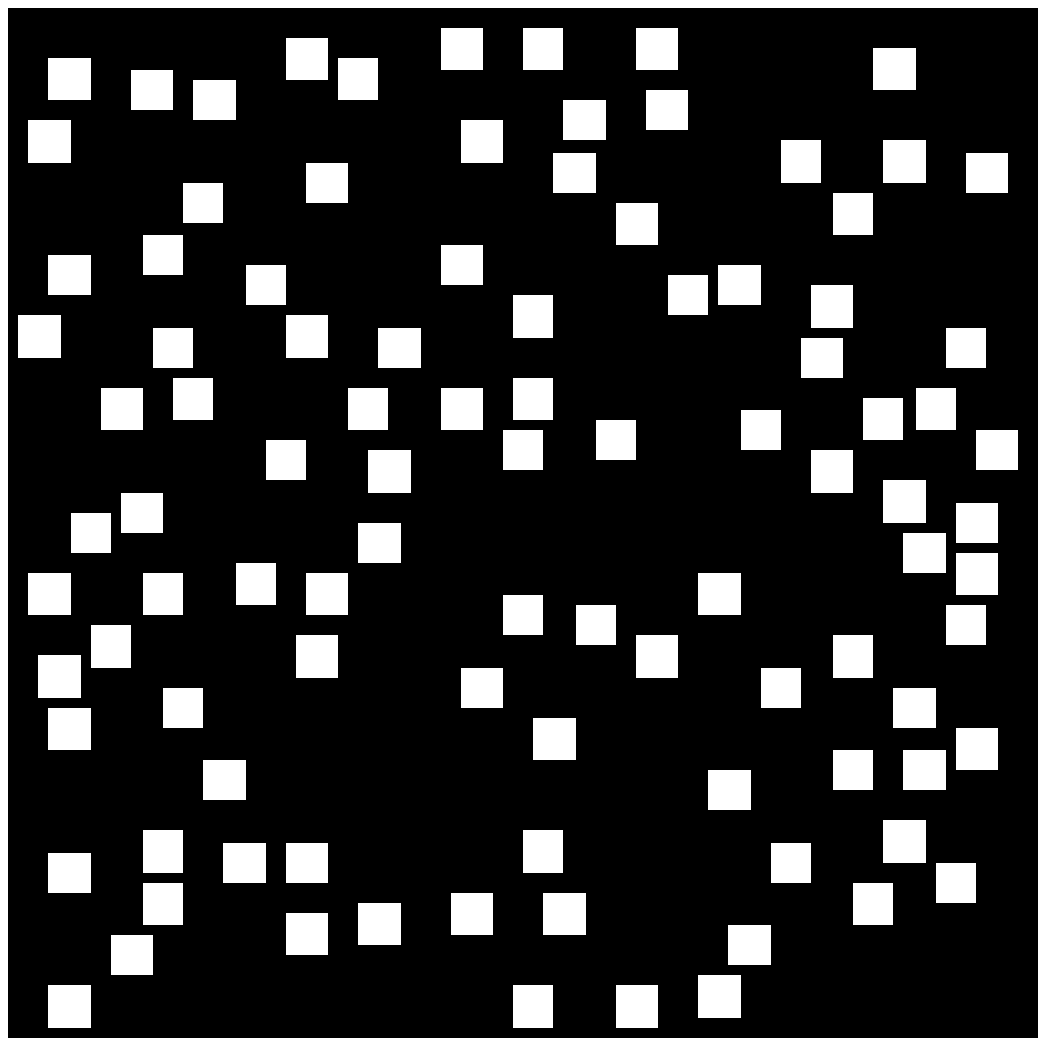}\\[20mm]
            C.~\includegraphics*[width=15mm,keepaspectratio]{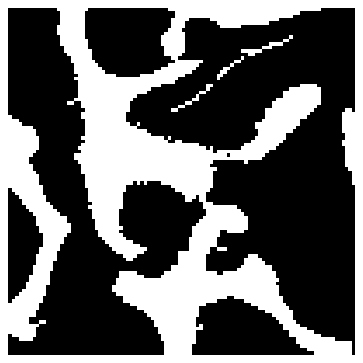}\\[10mm]
        \end{tabular}
    \end{adjustbox}
    &
    \begin{adjustbox}{valign=b}
        \begin{tabular}{c}
            \includegraphics*[width=80mm,keepaspectratio]{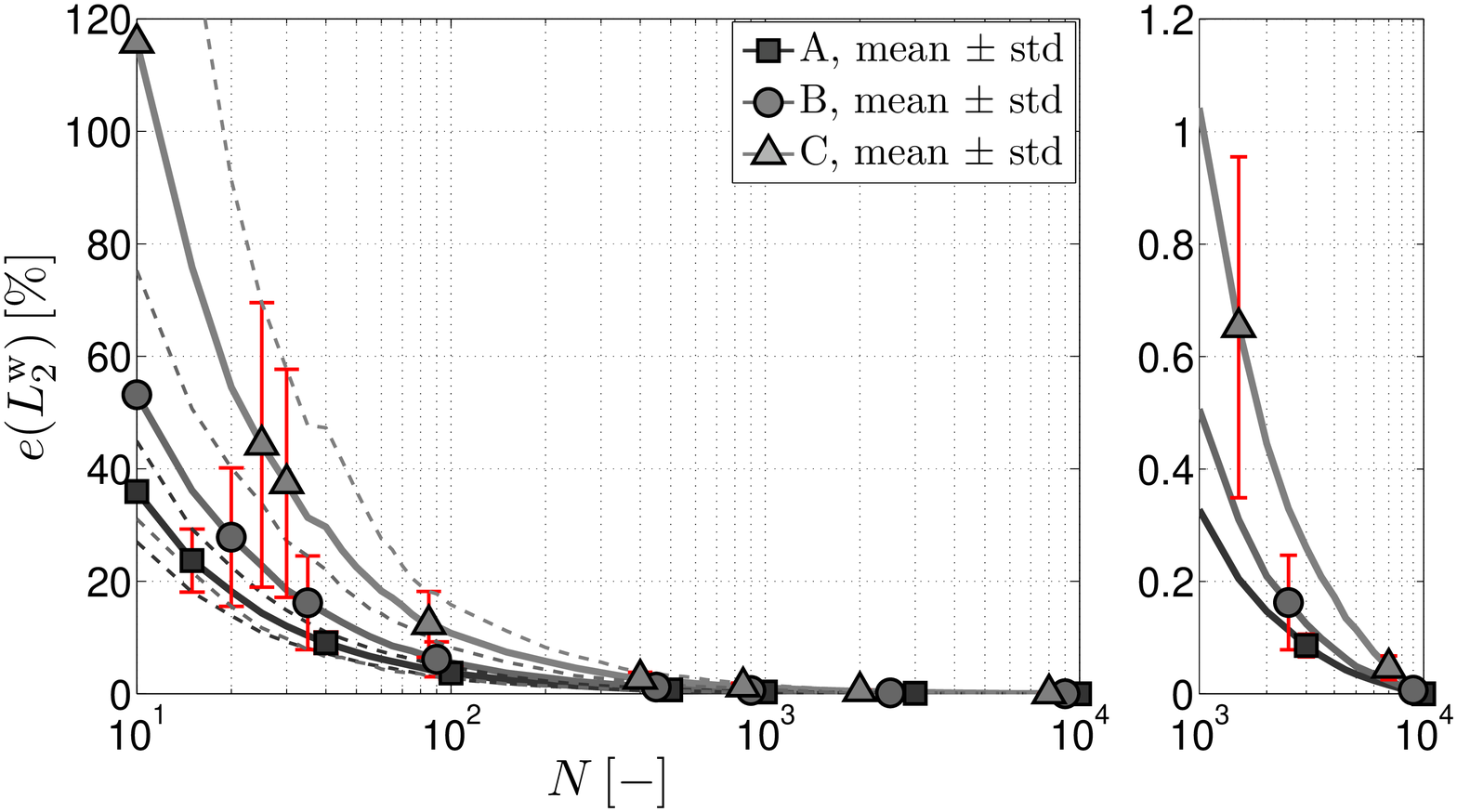}\\
            (a) \\
            \includegraphics*[width=80mm,keepaspectratio]{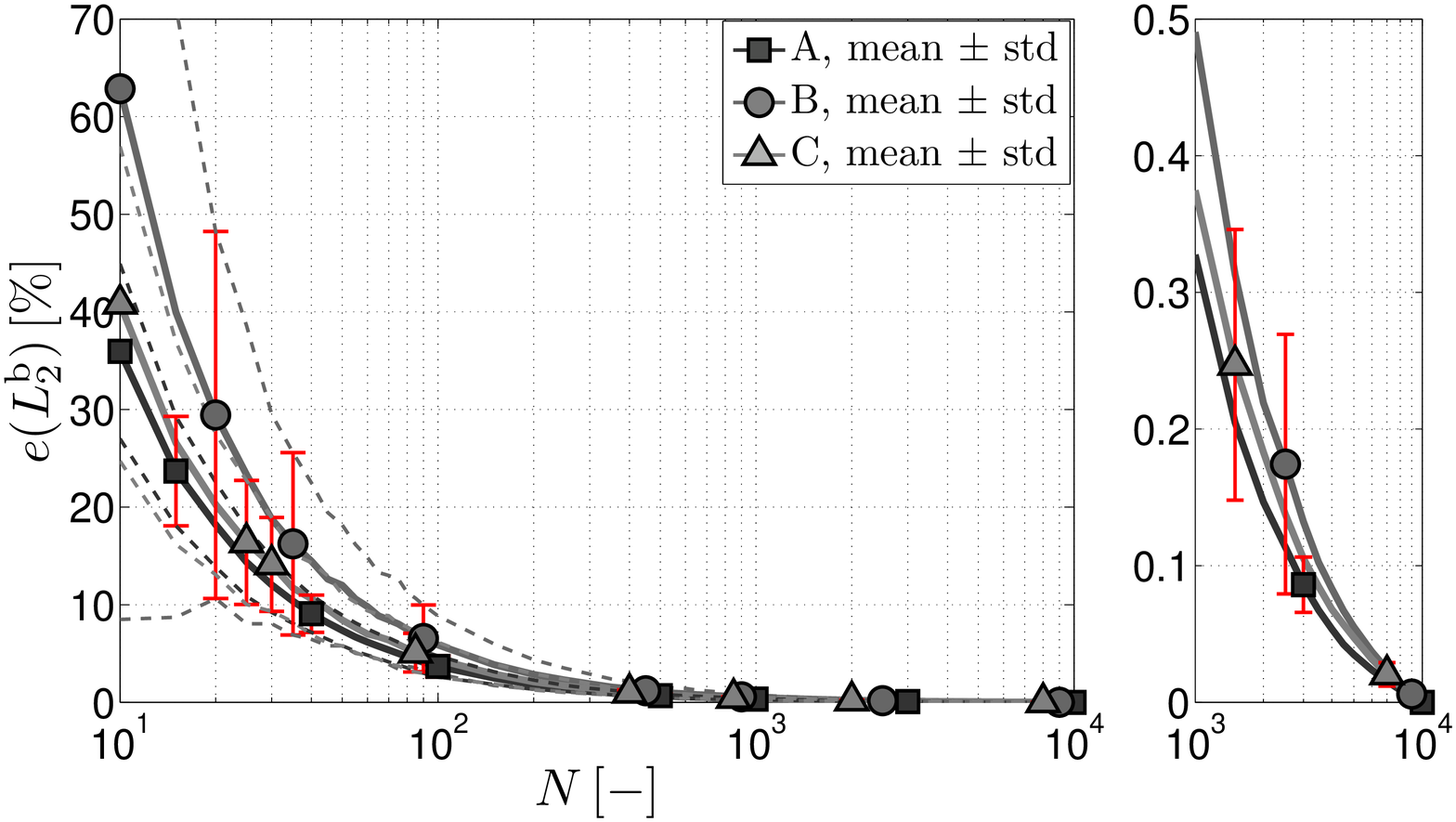}\\
            (b)
        \end{tabular}
    \end{adjustbox}
\end{tabular}
\end{center}
\caption{Convergence analysis of MC-based approximation of lineal
path functions computed for (a) white phase and (b) black phase}
\label{fig:MC_LP}
\end{figure}

In order to avoid huge computational requirements of the entire $L_2$
evaluation, some authors (see e.g. \cite{Zeman:2003}) compute
only its approximation using a Monte Carlo-based procedure. In such
a case, the line segments are not compared with the image at all
available positions equal to a number of all pixels in the image,
but only at a limited number of randomly selected positions. The
error produced by such an approximation is illustrated in Figure
\ref{fig:MC_LP} for three different types of microstructures as a
function of the number $N$ of selected positions.
The values on the vertical axis are the least square errors between the exact
lineal path function $L_2$ and its approximation $\widetilde{L}_2$ given
as
\begin{equation}
e(L_2^i) = \sum_{p\in\mathbb{D}}(L_2^i(\vek{x}_{p})-\widetilde{L}_2^i(\vek{x}_{p}))^2,
\label{eq:LSE1}
\end{equation}
where the superscript $i$ denotes the phase and the subscript $p$ covers
all the oriented segments defining the lineal path function given
by \eqref{eq:segments}. Dimensions of all three microstructural
images are $100 \times 100$ pixels. Since the lineal path function
is evaluated at $N$ positions obtained as a random $N$-combination
of a set $|\mathbb{D}|$ without repetition, the error converges to
zero for $N = |\mathbb{D}|$.


\subsection{Parallelisation on GPU using CUDA environment}

The key idea described here is a porting part of the code to GPU
device. Parallel computations on GPUs have become very popular
within the last decade thanks to the GPU's high performance at
a relatively low financial cost. Moreover, the programming
environment called CUDA (Compute Unified Device Architecture)
simplifies the GPU-based software development by using the
standard C/C++ language, see.~\cite{Nvidia}. In order to clearly
describe the GPU parallelism, we start with the algorithmic
structure of the $L_{2}$ evaluation consisting of several
computational steps:
\begin{enumerate}
\item generating line segments for given input dimensions,

\item allocating the inputs (e.g. the input representing a binary
image),

\item calculating the lineal path function based on translations
of each segment and its comparison with the image.
\end{enumerate}

Regarding computational requirements of particular steps, one
needs to keep in mind that the $L_2$ is supposed to be called
repeatedly within an optimisation process for new feasible
solutions (i.e. new binary images) of the same dimensions $W
\times H$. It means that the definition of line segments remains
the same during the whole optimisation process thus allowing to
run the step 1 only once at the beginning of the optimisation,
while the steps 2 -- 3 need to be called repeatedly. It also
further implies that step 1 is critical namely from the memory
usage point of view, while steps 2 -- 3 needs to be optimised with
respect to the computational time.

\begin{figure} [t!]
\centering
\includegraphics*[width=8cm,keepaspectratio]{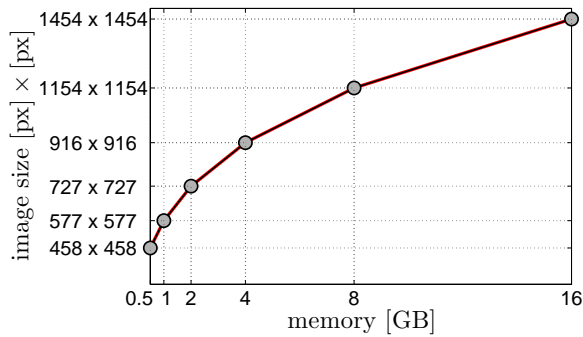}
\caption{Illustration of memory requirements for line segments storage.}
\label{fig:memory}
\end{figure}

Before starting an implementation of Bresenham's algorithm for definition of line
segments, one needs to decide about the line segments
coding. While the definition of a particular pixel by its $(x,y)$
coordinates is very intuitive, it is excessively memory demanding. It
is much more efficient to index all pixels in the image by only one
integer value from $0$ to $WH-1$. Then the number of integer values
required for definition of all line segments is given as
\begin{equation}
M = \sum_{i=1}^W \sum_{j=1}^{H} \max(i,j) + \sum_{i=2}^W \sum_{j=2}^{H} \max(i,j),
\end{equation}
which leads to
\begin{equation}
M = \frac{W(3H^2 + 3H + W^2 -
1)}{3}-\frac{W(W+1)}{2}-\frac{H(H+1)}{2}+1
\end{equation}
for $W \leq H$ and to
\begin{equation}
M = \frac{4W^3 - 4W + 3}{3}
\end{equation}
for $H=W$. Figure \ref{fig:memory} shows the dimensions of square
images which can be handled by cards with a given memory size
assuming that one integer takes 4 bytes.

As mentioned, the consecutive steps 2 and 3 are supposed to be called
repeatedly within the optimisation process and thus represent the
principal requirements on computational time, see
Alg.~\ref{alg:L2_eval} for a more detail algorithmic structure of the
step 3. The image enters the algorithm as the matrix $A$ twice wider and
twice higher than the original one, because it is periodically copied
on a grid $2 \times 2$ to allow easily translate the segments starting
within the image and ensure the segments to never end outside the
image. The translation is thus defined by moving the starting point of
a segment within one quadrant of the entering image. To facilitate
the repeatedly called computations, the indices of moves within one
quadrant are precomputed and stored in the separate matrix $C$.
\begin{algorithm}[hbt!]
 \KwData{ \\
   $A \dots$ a binary image  defined as an integer vector of size $2W \cdot 2H$ \\
   $B \dots$ an irregular 2D integer matrix defining pixels of Bresenham's line segments of size $S\times$segment size, where $S = 2WH-W-H+1$ \\
   $C \dots$ an integer vector of size $WH$ defining a translation within an image $W \times H$ mapped onto a periodically copied image of size $2W \times 2H$ \\
   $D \dots$ an integer vector of size $S$ defining a size of particular segments
   $phase \dots$ an integer defining phase, for which the $L_2$ is evaluated
}
 \KwResult{ $L \dots$ an integer vector of size $S$ defining the $L_2$}
 \For{seg=0 \KwTo $S-1$}{
 \For{transl=0 \KwTo $WH-1$}{
 \For{pix=0 \KwTo D[seg]}{
   \If{A[B[seg][pix]+C[transl]] $\neq$ phase}{
     break;
   }
 }
 \If{pix = D[seg]}{$L[seg] = L[seg]+1$\;}
}}
\caption{Algorithmic structure of implementation
designed for CPU device; $seg$, $transl$ and $pix$ represent
  integer variables used to govern the corresponding
  for loops.}
\label{alg:L2_eval}
\end{algorithm}

The structure of Alg.~\ref{alg:L2_eval} suggests several ways of
possible parallelisation. One way is a parallelisation over
particular segments (line 1), which would, however, lead to a very
asynchronous computation due to large differences among lengths of
the segments. Parallelisation over translations (line 2) is not
completely synchronous, because its inner cycle over pixels of the
segment (line 3) is stopped when proceeds to a pixel which is not
lying in a given phase, which depends on a particular image
morphology. Nevertheless, the computation have at least a chance
to be more synchronous than the surely highly asynchronous
parallelisation over segments.

Fortunately, the algorithm clearly consists of a huge number of very
simple logic and arithmetic operations and is thus well-suited for
parallelisation on GPU because of following reasons:
\begin{itemize}
\item [(i)] It allows for a nearly synchronous parallelisation scheme
  thus respecting the basic GPU programming rule -- memory
  coalescence;
\item [(ii)] It corresponds to the SIMD (single instruction, multiple
  data) architecture: a single instruction is an index of segment to be
  compared with the image at all possible positions, which thus
  represents the multiple data;
\item [(iii)] The most of the memory transfer corresponding to copying of
  image, translations, line segments and their sizes is done only once
  and in large chunks thus reducing related system overhead.
\end{itemize}

The parallel algorithmic structure proposed to increase the numerical
efficiency of the $L_2$ computation is given in Alg.~\ref{alg:GPU}. The
crucial step for the implementation efficiency concerns line 9, where
all $n_{\mathrm{t}} = WH$ translations are distributed into available
multiprocessors (MP).  Since particular GPU architectures
significantly differ among each other, here we concentrate on Fermi
compute architecture \cite{Nvidia}, where each MP has $32$
single-precision CUDA cores.  It means that each MP can simultaneously
solve up to $32$ tasks -- so-called threads -- defining one
warp. Besides currently computing threads, the MP can already load and
prepare other threads up to maximally $1536$ threads = $48$ warps. The
tasks are sent to the MP in blocks, where the particular translation is
assigned to the particular thread automatically according to its position
within the block.  Storing the translations in a 1D vector instead of
a 2D matrix thus allows for more even distribution of translations among
the MPs. Each MP can handle at the same time maximally $8$ blocks.
Particular size of a block can be chosen by a programmer, but finding
an optimum is not so straightforward. So as to maximise the occupancy
of the MPs, it is convenient to define the size of block B as
\begin{eqnarray}
\mathrm{B} = \left\{
\begin{array}{l}
1\times6 \, \mathrm{warps} = 1\times192  \, \mathrm{threads}, \quad \mbox{if } \left\lceil\frac{n_{\mathrm{t}}}{n_{\mathrm{mp}}}\right\rceil \geq 1536, \\
1\times\left\lceil\frac{n_{\mathrm{t}}}{8n_{\mathrm{mp}}}\right\rceil \mathrm{threads}, \quad \mbox{otherwise,}
\end{array}
\right.
\label{eq:block}
\end{eqnarray}
where $\left\lceil \cdot \right\rceil$ denotes the round-up
operation to the nearest integer and $n_{\mathrm{mp}}$ is a number
of available MPs. Nevertheless, other aspects related to shared
memory and registers \cite{Nvidia} may move the preferences
towards bigger blocks. More detailed study on the optimal block
size is beyond the scope of this paper. In our computations, we
focused on occupancy maximisation only and the block size is set
according to Eq.~\eqref{eq:block}.
\begin{algorithm}[hbt!]
 CPU: \hspace{12mm} calculating line segments: indices in $B$ and sizes in $D$\;
 CPU$\rightarrow$GPU: copying $B$ and $D$ into GPU\;
 CPU: \hspace{12mm} loading and copying binary image onto grid $2 \times 2$ saved \\
      \hspace{25mm}into $A$, defining translations $C$ \;
 CPU$\rightarrow$GPU: copying $C$ and $phase$ into GPU\;
 CPU$\rightarrow$GPU: copying $A$ into GPU\;
 \For{$seg=0$ \KwTo $S-1$}{
    CPU$\rightarrow$GPU: copying $seg$ into GPU\;
    GPU calls threads: \For{$transl=0$ \KwTo $WH-1$}{
        $L[seg] = 0$\;
        \For{pix=0 \KwTo D[seg]}{
          \If{A[B[seg][pix]+C[transl]] $\neq$ phase}{
            break;
          }
        }
        \If{pix = D[seg]}{$L[seg] = L[seg]+1$\;}
    }
 }
 CPU$\leftarrow$GPU: copying $L$ to CPU\;
\caption{Simplified algorithmic structure of implementation designed
  for single GPU device; All variables are defined in
  Alg.~\ref{alg:L2_eval}.} \label{alg:GPU}
\end{algorithm}

\subsection{Algorithmic acceleration of $L_2$ evaluation}

\begin{figure} [t!]
\centering
\includegraphics*[width=5.5cm,keepaspectratio]{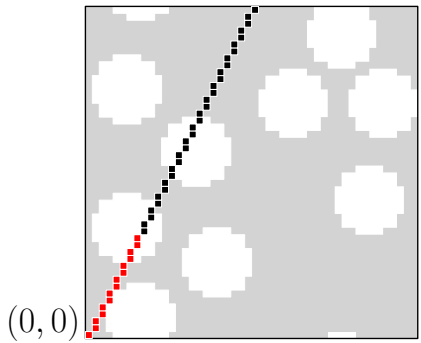}
\caption{Illustration of overlapping line segments}
\label{fig:overlap}
\end{figure}

Besides the parallelisation, we also propose one simple
algorithmic acceleration of the lineal path computation. The idea
comes from the discrete nature of segments and a fact that some
shorter segments are overlapped by some longer segments. See Fig.
\ref{fig:overlap}, where all segments start at $\vek{x}_1 = (0,0)$
and those ending in red pixels are overlapped by segments ending
in black pixels.
If a segment never falls entirely in a given phase and its $L_2$
value is zero, it is obvious that all longer overlapping segments
will have the zero value as well. This simple logic brings
additional significant time savings in the $L_2$ evaluation. It
only needs to precalculate a vector containing the indices of the
longest shorter overlapping segments (LSOS). Such a precalculation
is computational expensive, but is done only once at the beginning
of the algorithm. Having in mind that steps 6 to 21 in
Alg.~\ref{alg:GPU} are supposed to be called repeatedly within an
optimisation process, this precalculation should take place before
step 3.  Then one simple if-condition is added before translating
and comparing the segments with the image. If the LSOS
corresponding to the current segment has zero value of $L_2$, then
the $L_2$ value of the current segment is automatically assigned
to zero value too and the translating and comparing phase is
skipped. To be more specific, there are two possibilities where
this crucial if-condition can be solved. It would be an intuitive
solution to solve this if-condition on CPU, i.e. before line 8 of
the Alg.~\ref{alg:GPU} so as to skip the whole calling of GPU.
However, in such a case, the CPU needs to have knowledge about
previously computed segments, which means that the value of the
lineal path function has to be sent to CPU for every segment
separately inside the for-loop before line 18. Our computations,
however, revealed that repeated sending of one integer from GPU to
CPU is time-consuming and it is faster to call repeatedly the GPU,
solve the if-condition there (i.e. before line 10) and store all
the computed values of the lineal path function only on GPU until
the last segment is computed. Then sending of the whole vector of
the lineal path values brings significant time savings. This
latter variant was implemented and is further called as {\it
enhanced}, while the original version of the algorithm without any
algorithmic acceleration is called {\it standard}.

The performance of GPU parallelism is demonstrated on evaluation
of the $L_2$-function on three different microstructures: (i)
chess-type morphology with dimensions of squares $10\times 10$
[px], (ii) particulate suspension consisting of equal-sized
squares with dimensions $4\times 4$ [px] and (iii) metal foam
taken from ~\cite{Jirousek:JI:2013}. Tab.~\ref{tab:comptime}
compares the amount of time necessary averaged over five
evaluations of the lineal path function for both phases on single
CPU or GPU devices depending on the image size and chosen variant
of the algorithm. In particular, the computational times
correspond to a part of the lineal path function computation,
which is called repeatedly within the optimisation process, i.e.
evaluation of lines 1 to 5 in Alg.~\ref{alg:GPU} is excluded. It
is shown that for very small images the use of CPU outperforms the
GPU because of additional time spent by communicating with the
GPU. Nevertheless, for images of $50\times 50\,\mathrm{[px]}$ the
GPU achieves an evident speed-up which mostly further increases
with the increasing dimensions of the image. The exception is the
chess-type microstructure where a specific phase distribution
limits the length of the most of the line segments to $10$ [px].
This significantly elevates the acceleration obtained for the
enhanced variant of the algorithm and even the CPU version is so
fast that communication with GPU leads again to deceleration which
increases with the image dimensions.

\cleardoublepage
\renewcommand{\arraystretch}{0.92}
\begin{table}[h!]
\begin{center}
\begin{tabular}{p{2cm}|p{1.4cm}p{1.4cm}l|p{1.4cm}p{1.4cm}l|l}
\hline
\includegraphics*[width=15mm,keepaspectratio]{MC_figA.eps} & \multicolumn{3}{l}{\textit{Standard}} &  \multicolumn{3}{|l|}{\textit{Enhanced}} & \\
\hline dim. & GPU & CPU & $S$  & GPU
& CPU  & $S$ & $S^{*}$\\
$\mathrm{[px]}$ & $\mathrm{[s]}$ & $\mathrm{[s]}$ &
$\mathrm{[-]}$ & $\mathrm{[s]}$ & $\mathrm{[s]}$ & $\mathrm{[-]}$ & $\mathrm{[-]}$\\
\hline
$10\times 10$   & 2.4$\cdot 10^{-3}$  & 0.59$\cdot 10^{-3}$    & \textbf{0.2}$\times$   & 2.4$\cdot 10^{-3}$ & 0.39$\cdot 10^{-3}$    & \textbf{0.2}$\times$ & \textbf{0.2}$\times$\\
$20\times 20$ & 11.9$\cdot 10^{-3}$   & 11.9$\cdot 10^{-3}$    & \textbf{1.0}$\times$  & 11.1$\cdot 10^{-3}$  & 7.8$\cdot 10^{-3}$    & \textbf{0.7}$\times$ & \textbf{1.1}$\times$\\
$50\times 50$   & 0.15  & 0.55    & \textbf{3.7}$\times$   & 0.10  & 0.24    & \textbf{2.5}$\times$ & \textbf{5.8}$\times$\\
$100\times 100$ & 2.1   & 7.0    & \textbf{3.3}$\times$  & 0.26  & 0.27    & \textbf{1.0}$\times$ & \textbf{27.0}$\times$\\
$200\times 200$ & 32.6  & 110.9   & \textbf{3.4}$\times$  & 2.0  & 1.0   & \textbf{0.5}$\times$ & \textbf{53.9}$\times$\\
$500\times 500$ & 318.4 & 1071.1 & \textbf{3.4}$\times$  & 16.3 & 6.4  & \textbf{0.4}$\times$ & \textbf{65.8}$\times$\\
\hline\hline
\includegraphics*[width=15mm,keepaspectratio]{MC_figB.eps} & \multicolumn{3}{l}{\textit{Standard}} &  \multicolumn{3}{|l|}{\textit{Enhanced}} & \\
\hline dim. & GPU & CPU & $S$ & GPU
& CPU  & $S$ & $S^{*}$ \\
$\mathrm{[px]}$ & $\mathrm{[s]}$ & $\mathrm{[s]}$ &
$\mathrm{[-]}$ & $\mathrm{[s]}$ & $\mathrm{[s]}$ & $\mathrm{[-]}$ & $\mathrm{[-]}$ \\
\hline
$10\times 10$   & 2.5$\cdot 10^{-3}$  & 0.71$\cdot 10^{-3}$    & \textbf{0.3}$\times$   & 2.5$\cdot 10^{-3}$  & 0.38$\cdot 10^{-3}$    & \textbf{0.2}$\times$ & \textbf{0.3}$\times$\\
$20\times 20$ & 12.9$\cdot 10^{-3}$   & 16.0$\cdot 10^{-3}$    & \textbf{1.2}$\times$  & 12.5$\cdot 10^{-3}$  & 8.2$\cdot 10^{-3}$    & \textbf{0.7}$\times$ & \textbf{1.3}$\times$\\
$50\times 50$   & 0.16  & 0.81    & \textbf{5.0}$\times$   & 0.13  & 0.36    & \textbf{2.7}$\times$ & \textbf{6.0}$\times$\\
$100\times 100$ & 2.5   & 16.1    & \textbf{6.4}$\times$  & 1.7  & 7.2    & \textbf{4.1}$\times$ & \textbf{9.3}$\times$\\
$200\times 200$ & 41.9  & 256.3   & \textbf{6.1}$\times$  & 20.8  & 78.0   & \textbf{3.7}$\times$ & \textbf{12.3}$\times$\\
$500\times 500$ & 411.6 & 2544.0 & \textbf{6.2}$\times$  & 205.1 & 789.1  & \textbf{3.9}$\times$ & \textbf{12.4}$\times$\\
\hline\hline
\includegraphics*[width=15mm,keepaspectratio]{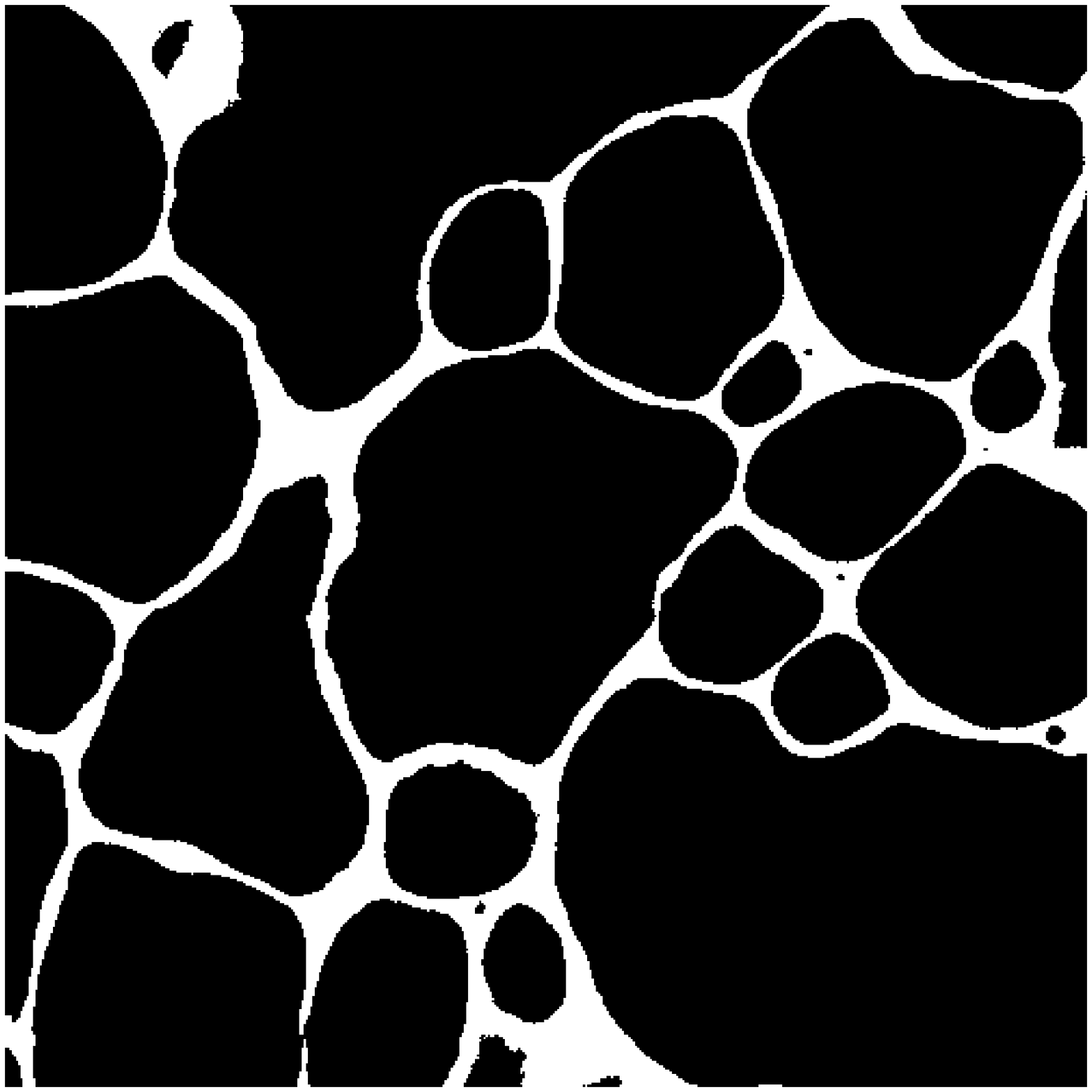} & \multicolumn{3}{l}{\textit{Standard}} &  \multicolumn{3}{|l|}{\textit{Enhanced}} & \\
\hline dim. & GPU & CPU & $S$  & GPU
& CPU  & $S$ & $S^{*}$\\
$\mathrm{[px]}$ & $\mathrm{[s]}$ & $\mathrm{[s]}$ &
$\mathrm{[-]}$ & $\mathrm{[s]}$ & $\mathrm{[s]}$ & $\mathrm{[-]}$ & $\mathrm{[-]}$ \\
\hline
$10\times 10$   & 2.4$\cdot 10^{-3}$  & 0.73$\cdot 10^{-3}$    & \textbf{0.3}$\times$   & 2.4$\cdot 10^{-3}$  & 0.65$\cdot 10^{-3}$    & \textbf{0.3}$\times$ & \textbf{0.3}$\times$\\
$20\times 20$ & 12.0$\cdot 10^{-3}$   & 13.8$\cdot 10^{-3}$    & \textbf{1.1}$\times$  & 11.9$\cdot 10^{-3}$  & 9.7$\cdot 10^{-3}$    & \textbf{0.8}$\times$ & \textbf{1.2}$\times$\\
$50\times 50$   & 0.17  & 1.31    & \textbf{7.6}$\times$   & 0.15  & 0.67    & \textbf{4.4}$\times$ & \textbf{8.7}$\times$\\
$100\times 100$ & 2.8   & 31.9    & \textbf{11.5}$\times$  & 2.22  & 17.2    & \textbf{7.7}$\times$ & \textbf{14.3}$\times$\\
$200\times 200$ & 48.3  & 577.9   & \textbf{12.0}$\times$  & 32.9  & 241.7   & \textbf{7.3}$\times$ & \textbf{17.5}$\times$\\
$500\times 500$ & 542.1 & 7911.7 & \textbf{14.6}$\times$  & 445.2.1 & 3884.2  & \textbf{8.7}$\times$ & \textbf{17.7}$\times$\\
\hline
\end{tabular}
\caption{Comparison of CPU and GPU performance averaged over five
evaluations ($S$ stands for speedup and $S^{*}$ represents overall
speedup obtained by hardware and software accelaration) }
\label{tab:comptime}
\end{center}
\end{table}

The particular computations presented in Tab.~\ref{tab:comptime}
were performed on $2\times$ INTEL Xeon E$5-2620$ @ $2.0$ GHz, $96$
GB RAM, $2\times$ GPU - NVIDIA QUADRO $4000$ with Micrsosoft
Windows~$7$ $64$-bit operating system and the CUDA v. $6.5$.
Furthermore, the algorithm is also designed for dual GPUs,
unfortunately scalability towards the multiple GPU devices is not
considered here. The logical step for the dual GPU algorithm is to
uniformly distribute the generated segments, so that each device
holds only a certain amount of them. This improvement thus results
in lower memory requirements.

\section{Optimisation procedure} \label{sec:optim}
Before proceeding to the comparative study of the lineal path and
two-point probability function, we briefly describe the
optimisation procedure employed in our computations. Here, we used
the framework firstly introduced by Yeong and
Torquato~\cite{Yeong:1998:PRE} for digitised media. The algorithm
is based on simulated annealing method independently developed by
Kirkpatrick et al.~\cite{Kirkpatrick:S:1983} and
\v{C}ern\'{y}~\cite{Cerny:JOTA:1985}. It starts with some randomly
generated microstructure and quantification of its quality by a
chosen statistical descriptor. The microstructure is then modified
by a chosen operator and its new quality is evaluated. The
acceptance of the proposed modification is governed by the
Metropolis rule, which allows with a certain probability to accept
a worse solution and thus to escape from a local extreme. Such a
generic optimisation scheme opens the possibility to define
modification operator suitable for a given microstructure. For
instance, a particulate suspension consisting of equal-sized discs
can be modified by moving a centre of an arbitrarily chosen disc,
see e.g. \cite{Novak:PR:2012,Novak:MSMSE:2013}. Such a move
affects the whole set of pixels and allows preserving the known
shape of particles, thus accelerating the optimisation procedure.
Most of the microstructures are, however, not consisting of
particles having a specific known shape. Then the simplest
modification operator is based on interchanging two randomly
chosen pixels from different phases, which at least allows to
preserve their volume fraction~\cite{Yeong:1998:PRE}. Very simple
acceleration employed in our implementation consists in a random
selection of interfacial pixels which leads to a significant
increase of accepted modifications, as presented in
\cite{Rozman:2001:PRE}.

\begin{algorithm}[hbt!]
 \KwData{binary image with dimensions $W\times H$}
 \KwResult{optimised SEPUC corresponding to given image} 
 $create\_random\_image(P)$\;
 $SDP = evaluate(P)$\;
 $T = T_{\max}$\;
 $T_{\mathrm{mult}} = (T_{\mathrm{min}}/T_{\mathrm{max}})^{(succ_{\mathrm{max}}/N_{\mathrm{step}})}$\;
 \While{$c<N_{\mathrm{step}}$}{
   $c = s = 0$\;
   \While{$c<c_{\max} \quad \& \quad s<s_{\max}$}{
     $c = c + 1$\;
     $Q = modify(P)$ \;
     $SDQ = evaluate(Q)$\;
     \If{random\_number U[0,1] $< \exp((SDQ-SDP)/T)$}{
       $s = s + 1$\;
       $P = Q$\;
       $SDP = SDQ$\;
     }
   }
   $T = T \cdot T_{\mathrm{mult}}$\;
 }
\caption{Algorithmic structure of simulated annealing} \label{alg:annealing}
\end{algorithm}
Since the proposed way of porting the lineal path evaluation onto
GPU counts with copying the whole image from CPU to GPU for any
new proposed modification, the modification operator can be
designed in any convenient way. Nevertheless, our further
computations use solely the interchanging of two pixels.
The particular structure of the employed optimisation algorithm is given
in Alg.~\ref{alg:annealing}. First of all, a random digitised
image $P$ is created with the same volume fractions of phases as
the original morphology. Its statistical similarity to the original
image is then evaluated using the chosen statistical descriptor SD
as the least square error:
\begin{equation}
e(\mathrm{SD}^i) = \sum_{p\in \mathbb{D}}
(\mathrm{SD}_{\mathrm{original}}^{i}(\vek{x}_{p})-\mathrm{SD}^{i}(\vek{x}_{p}))^2,
\label{eq:LSE}
\end{equation}
where the superscript $i$ denotes the phase for which the SD is evaluated
and the subscript $p$ corresponds to the component of a discretised
descriptor. If the superscript $i$ is missing in the following text,
the SD is evaluated for both phases. Note that the least square error
in Eq.~\eqref{eq:LSE} also consists of a large number of simple
arithmetic operations, which are again efficiently evaluated in
parallel on GPU.

Control parameters of the algorithm were set to following values:
\begin{table}[htb]
\centering
\begin{tabular}{ll}
$N_{\mathrm{step}} = 4\cdot 10^6$ & $c_{\max} = 0.1 N_{\mathrm{step}}$ \\
$T_{\min} = 0.01 T_{\max}$ & $s_{\max} = 0.01 N_{\mathrm{step}}$ \\
\end{tabular}
\caption{Control parameters of simulated annealing method}
\label{tab:annealing}
\end{table}
The value of $T_{\max}$ was manually changed for every particular
computation so as to achieve approximately the ration $s/c = 0.5$
within the first few steps of the algorithm. Some other
recommendations for setting these parameters can be found e.g. in
\cite{Leps:2003:CS}.

\section{Microstructure reconstruction}
\label{sec:recon}

Reconstruction of a microstructure from its statistical
description is an inverse problem addressed by several authors in
different ways, see
\cite{Rozman:2001:PRE,Capek:2009:TPM,Davis:2011:PRE} and the
references therein. Here we follow the concept proposed in
\cite{Yeong:1998:PRE}, where the discretised randomly generated
microstructure is optimised with respect to the prescribed
statistical descriptor.
\begin{figure} [b!]
\begin{center}
\begin{tabular}{c|c|ccc}
\begin{adjustbox}{valign=m}
    \includegraphics*[width=45mm,keepaspectratio]{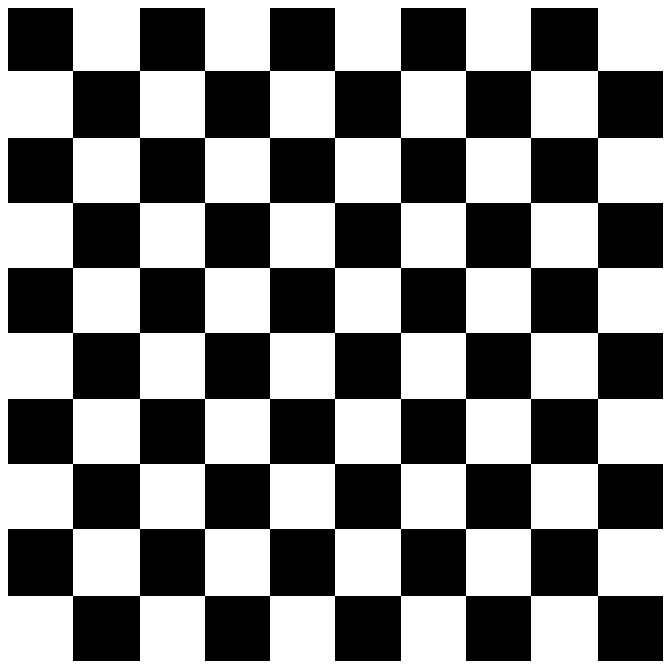}
\end{adjustbox}
&
\begin{adjustbox}{valign=m}
    \includegraphics*[width=22.5mm,keepaspectratio]{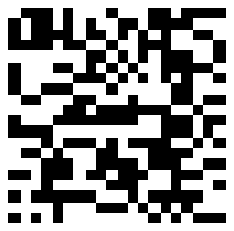}
\end{adjustbox}
&
\begin{adjustbox}{valign=m}
    \begin{tabular}{c}
    \includegraphics*[width=22.5mm,keepaspectratio]{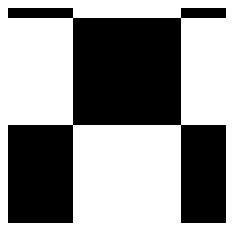} \\
    (c) \\
    \includegraphics*[width=22.5mm,keepaspectratio]{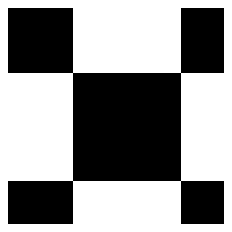} \\
    \end{tabular}
\end{adjustbox}
&
\begin{adjustbox}{valign=m}
    \begin{tabular}{c}
    \vspace{18.5mm}\\
    \\
    \includegraphics*[width=22.5mm,keepaspectratio]{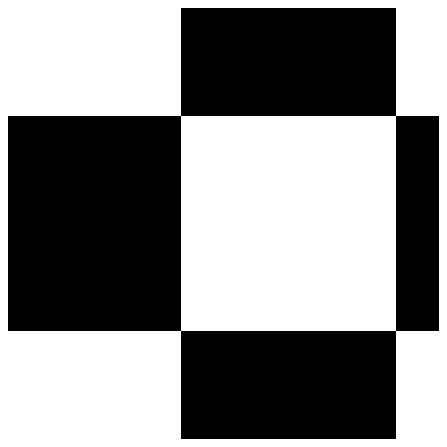}
    \end{tabular}
\end{adjustbox}
&
\begin{adjustbox}{valign=m}
    \begin{tabular}{c}
    \vspace{18.5mm}\\
    \\
    \includegraphics*[width=22.5mm,keepaspectratio]{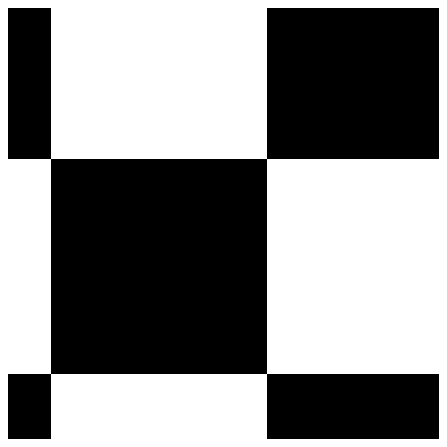} \\
    \end{tabular}
\end{adjustbox}
\\
(a) & (b) & (d) & (e) & (f) \\
\end{tabular}
\end{center}
\caption{Chess microstructure: (a) Original medium with size
  $100\times 100$ [px] and characteristic lengths $20 \times 20$ [px];
  (b) Random initial structure, size $20\times 20$ [px]; (c)
  $S_{2}$-based reconstructed image; (d) $L_{2}$-based reconstructed
  image; (e) $L_{2}^{\mathrm{b}}$-based reconstructed image; (f)
  $L_{2}^{\mathrm{w}}$-based reconstructed image; obtained within less
  than $5\cdot 10^5$ iterations.}
\label{fig:chess}
\end{figure}
The authors in \cite{Rozman:2002:PRL} presented numerical evidence
that a periodic medium discretised into pixels is completely
specified by its two-point correlation function, up to a
translation and, in some cases, inversion.
\renewcommand{\arraystretch}{1.3}
\begin{figure} [b!]
\begin{center}
\begin{tabular}{@{\hspace{0.5mm}}c@{\hspace{0.5mm}}@{\hspace{0.5mm}}c@{\hspace{0.5mm}}|cc}
\begin{adjustbox}{valign=m}
    \begin{tabular}{c}
    \includegraphics*[width=40mm,keepaspectratio]{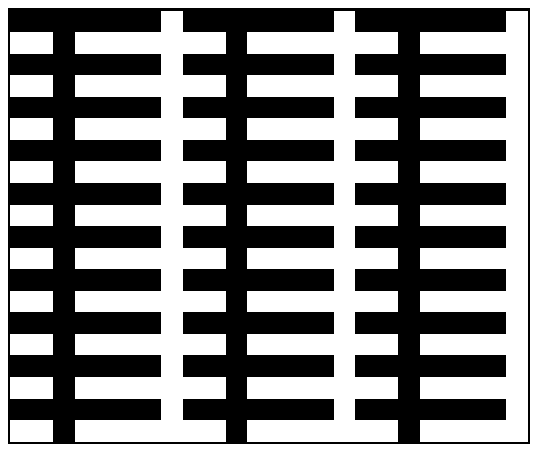}\\
    (a) \\
    \includegraphics*[width=40mm,keepaspectratio]{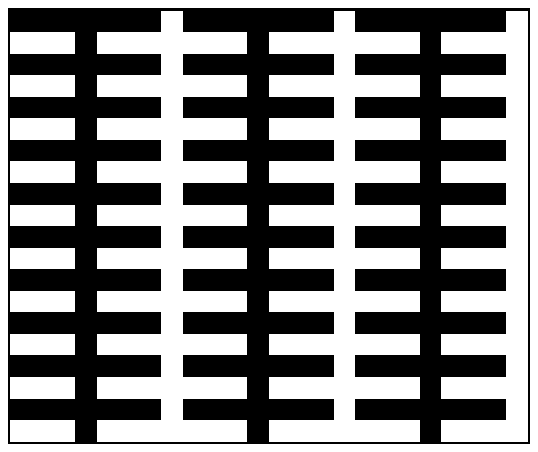}\\
    (g) \\
    \end{tabular}
\end{adjustbox}
&
\begin{adjustbox}{valign=m}
    \begin{tabular}{c}
    \includegraphics*[width=20mm,keepaspectratio]{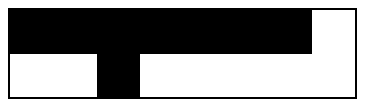}\\
    \vspace{30mm}
    (b) \\
    \includegraphics*[width=20mm,keepaspectratio]{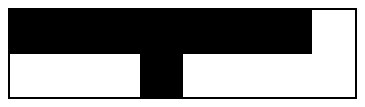}\\
    (h) \\
    \end{tabular}
\end{adjustbox}
&
\begin{adjustbox}{valign=m}
    \begin{tabular}{c}
    \includegraphics*[width=40mm,keepaspectratio]{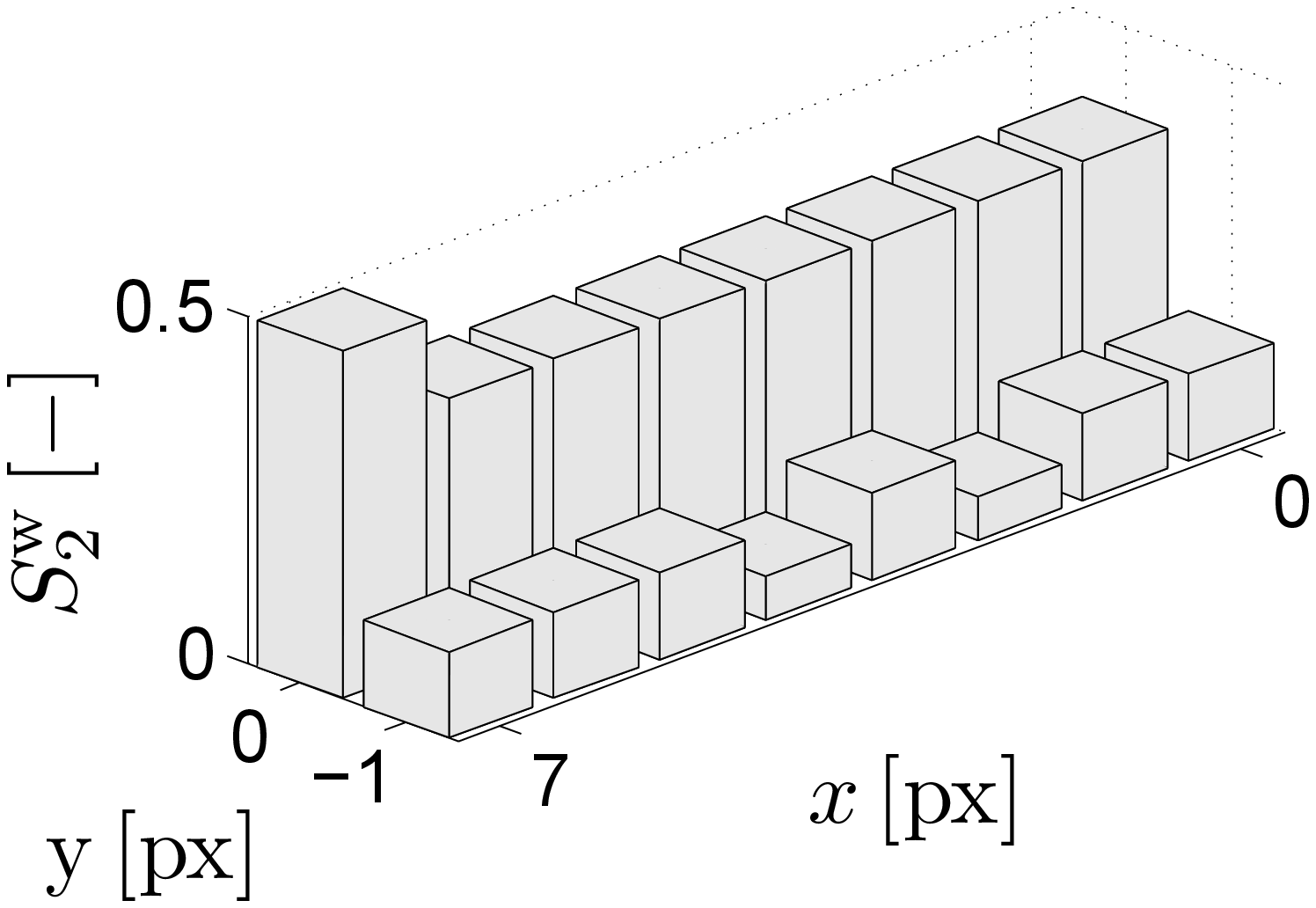}\\
    (c)\\
    \includegraphics*[width=40mm,keepaspectratio]{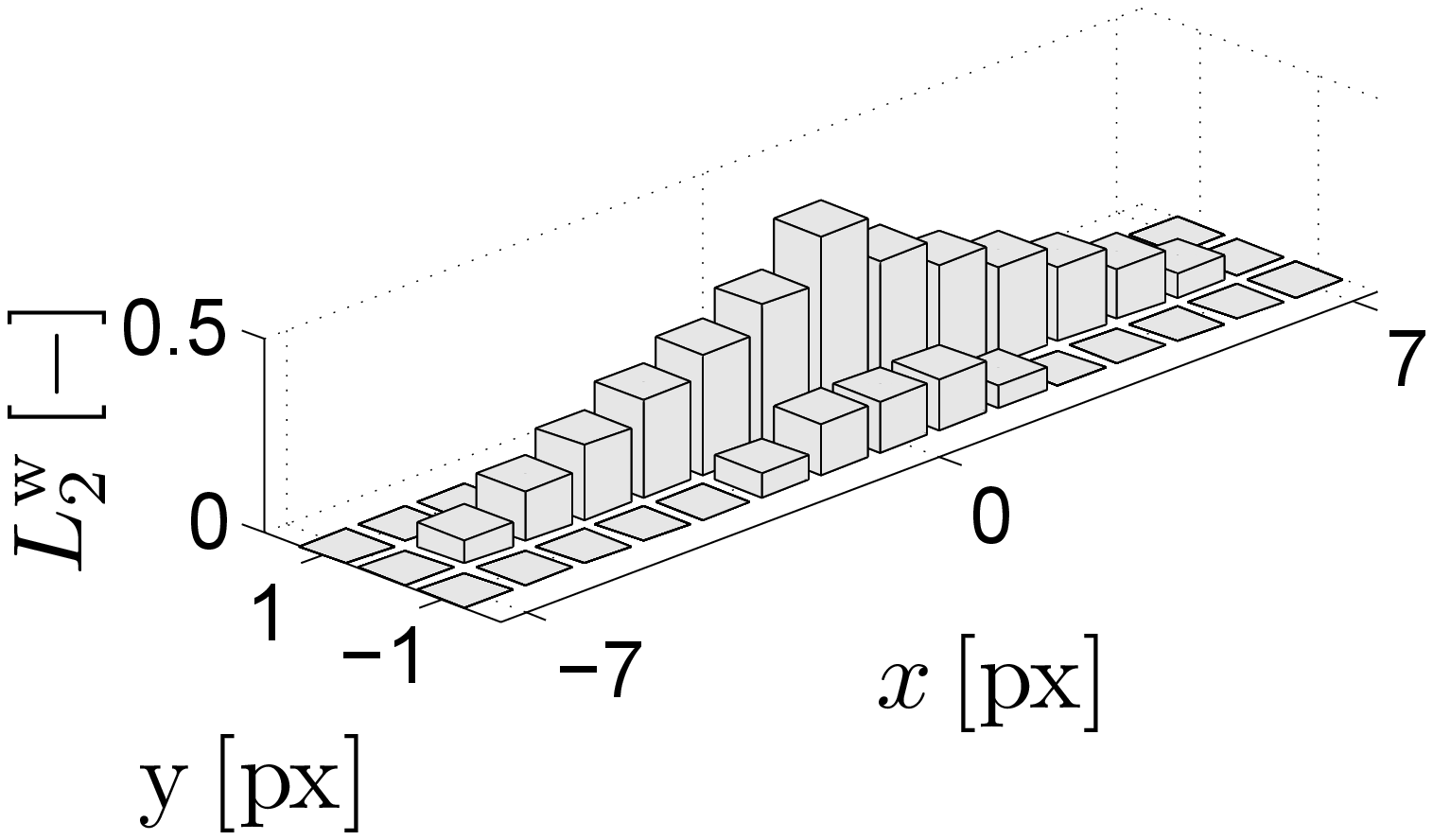}\\
    (e)\\
    \hline
    \\
    \includegraphics*[width=40mm,keepaspectratio]{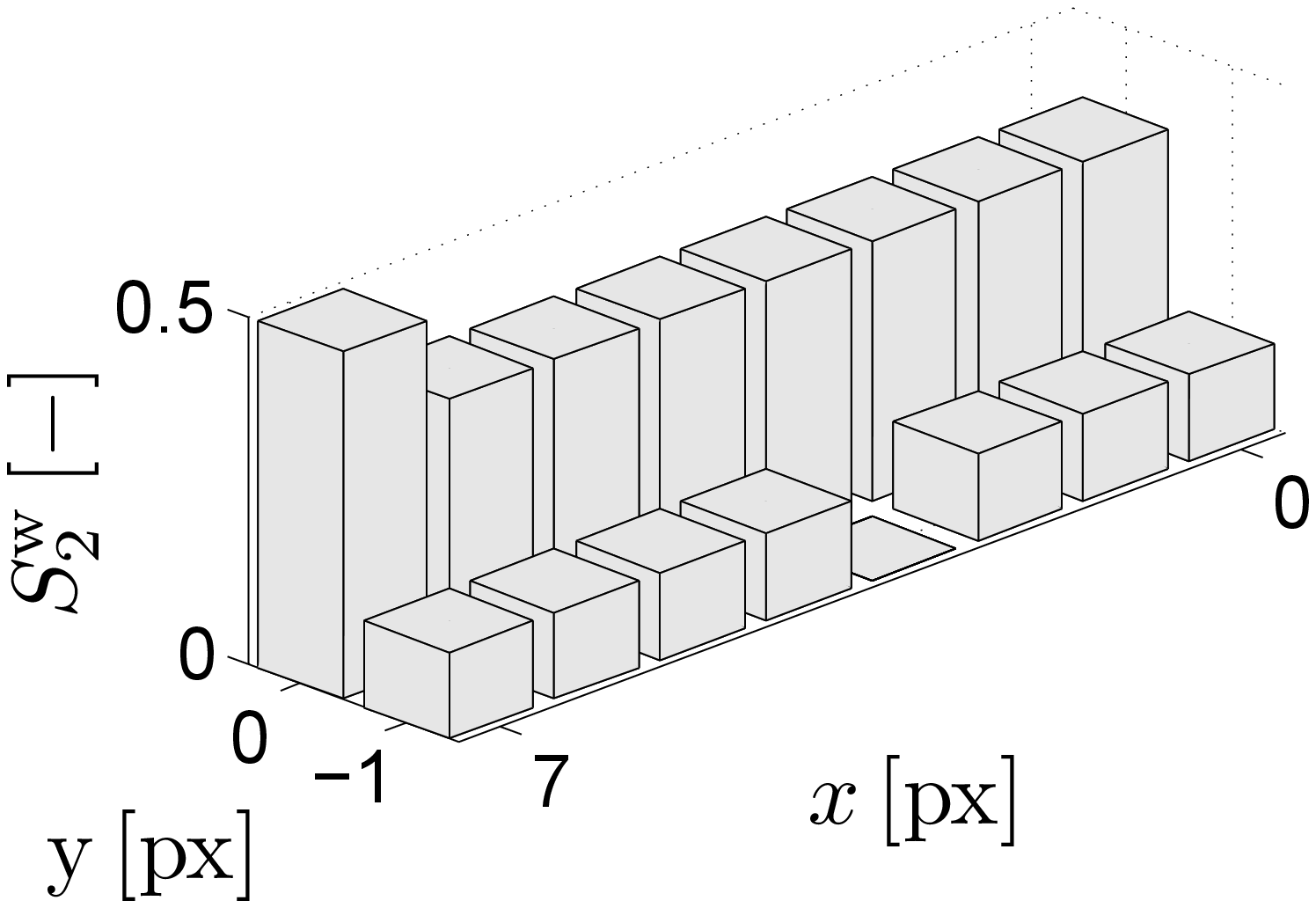}\\
    (i)\\
    \includegraphics*[width=40mm,keepaspectratio]{L2w_test_n.eps}\\
    (k)\\
    \end{tabular}
\end{adjustbox}
&
\begin{adjustbox}{valign=m}
    \begin{tabular}{c}
    \includegraphics*[width=40mm,keepaspectratio]{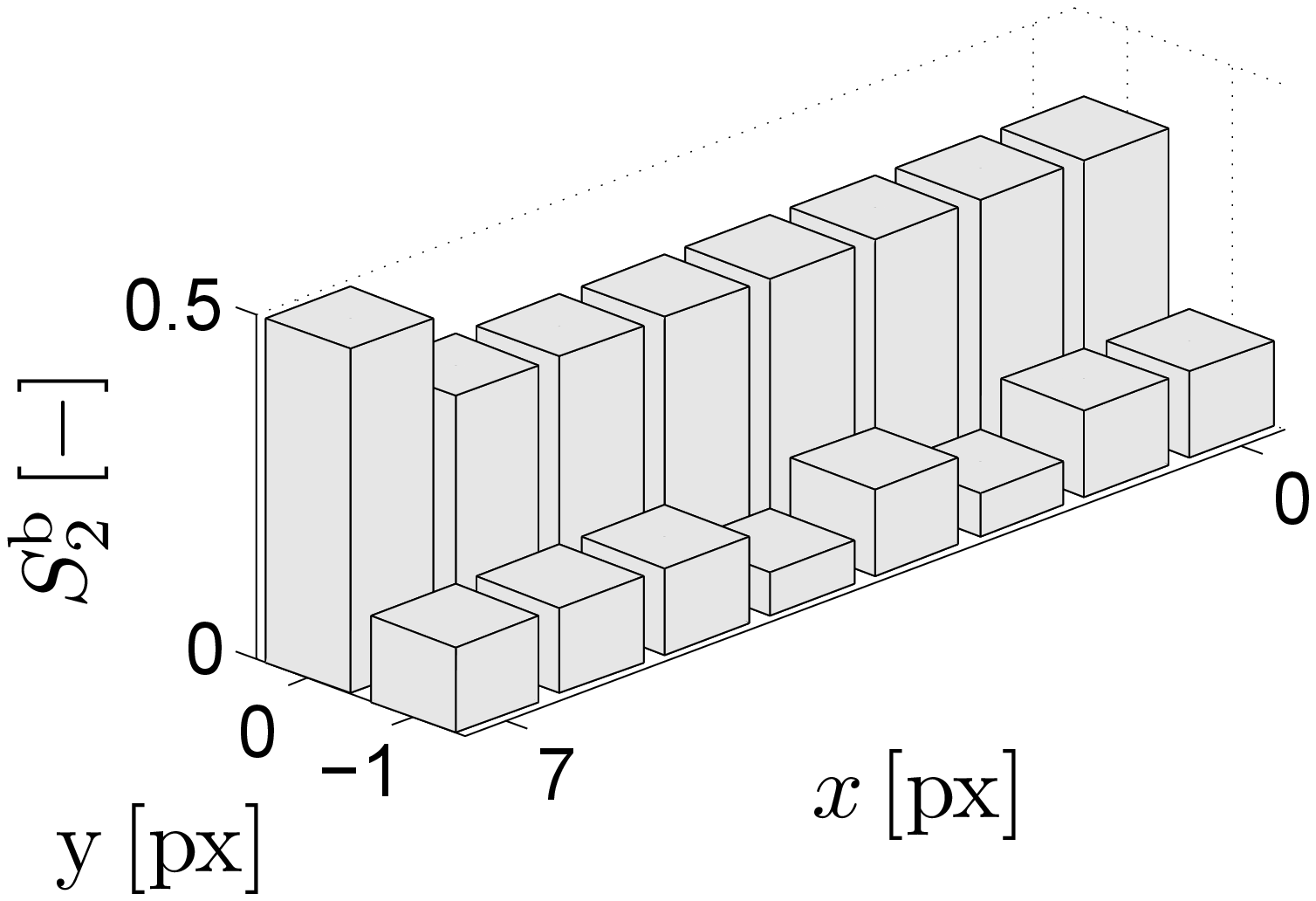}\\
    (d)\\
    \includegraphics*[width=40mm,keepaspectratio]{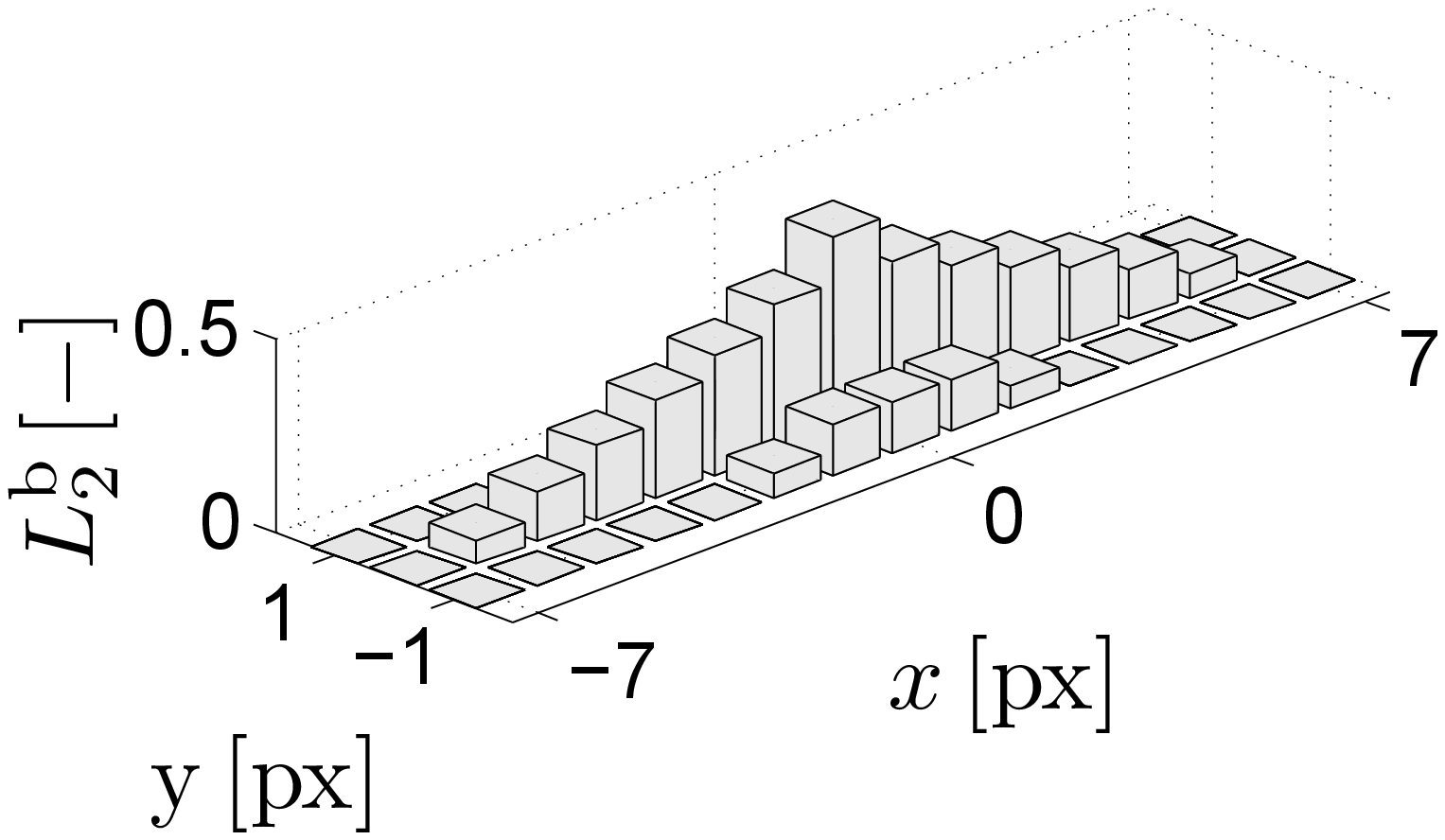}\\
    (f)\\
    \hline
    \\
    \includegraphics*[width=40mm,keepaspectratio]{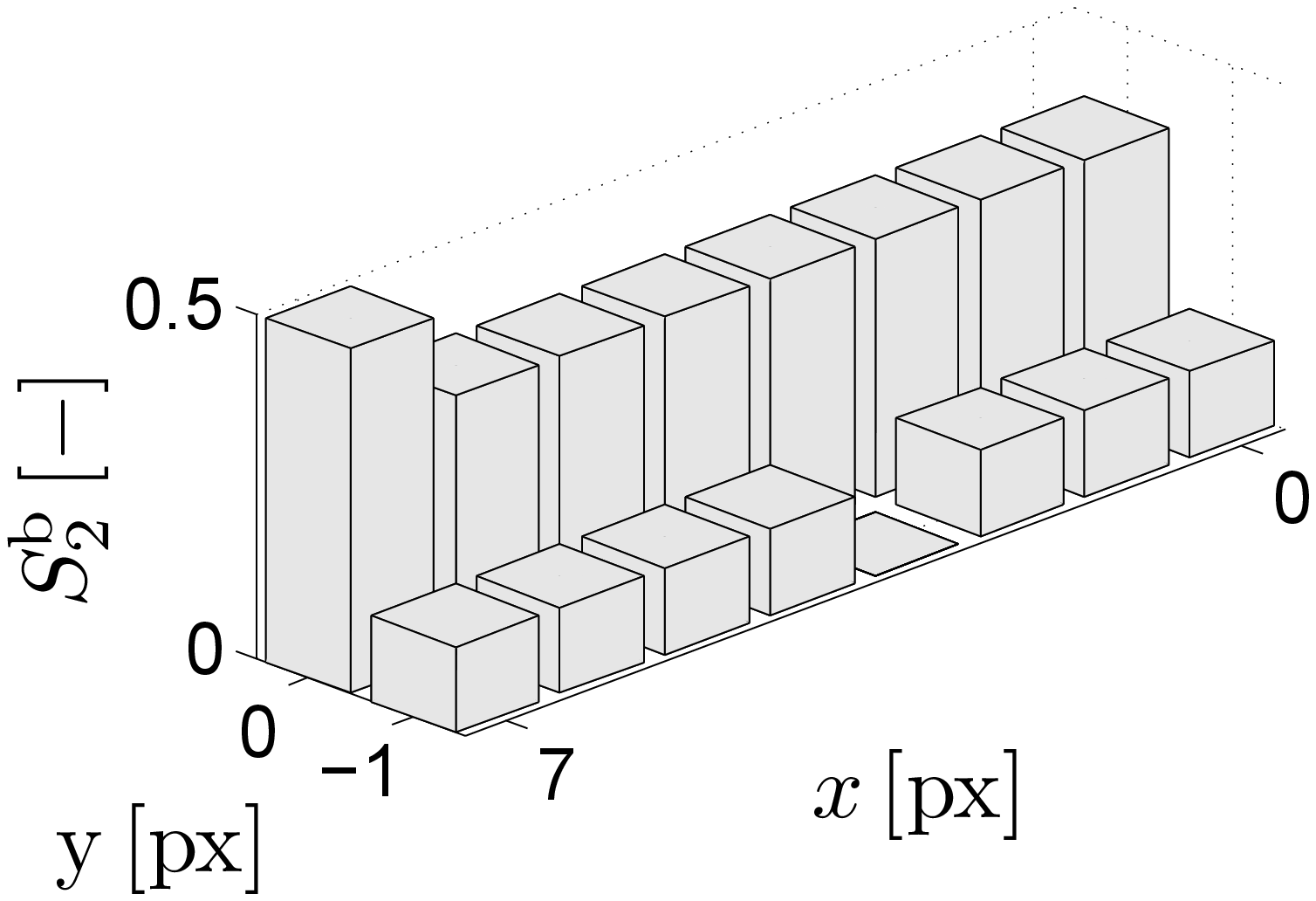}\\
    (j)\\
    \includegraphics*[width=40mm,keepaspectratio]{L2b_test_n.eps}\\
    (l)\\
    \end{tabular}
\end{adjustbox}
\end{tabular}
\end{center}
\caption{(a) Original medium, size $20\times 24$ [px]; (b)
Periodic unit cell (PUC), size $2\times 8$ [px];
(c)~$S_{2}^{\mathrm{w}}$-function of PUC;
(d)~$S_{2}^{\mathrm{b}}$-function of PUC;
(e)~$L_{2}^{\mathrm{w}}$-function of PUC;
(f)~$L_{2}^{\mathrm{b}}$-function of PUC; (g) Original medium,
size $20\times 24$ [px]; (h) Periodic unit cell, size $2\times 8$
[px]; (i)~$S_{2}^{\mathrm{w}}$-function of PUC;
(j)~$S_{2}^{\mathrm{b}}$-function of PUC;
(k)~$L_{2}^{\mathrm{w}}$-function of PUC;
(l)~$L_{2}^{\mathrm{b}}$-function of PUC;} \label{fig:unicity}
\end{figure}
This conclusion implies that the reconstruction process based on
the discretised two-point probability function has a unique
solution. For many microstructural morphologies, the same holds
also for reconstruction from the lineal path function. For
instance, the chess-type morphology is fully defined not only by
lineal path function computed for both phases, but only one of the
phase is fully sufficient to completely define the morphology, see
Figure~\ref{fig:chess}. Nevertheless, generic evidence for a
unique solution of the lineal path function-based reconstruction
is missing and is suggested just by findings concerning
orientation-dependent chord length distributions in continuous
domains \cite{Nagel:1993}. On the contrary, we can demonstrate
that employing Bresenham's algorithm for the line segments'
definition, the lineal path function does not define a unique
solution for a reconstruction process based on a discretised
medium.

Fig.~\ref{fig:unicity} shows an example of two different periodic
cells of dimensions $2 \times 8$ pixels. Due to the same volume
fraction of both phases, the two-point probability functions
obtained for both phases in Fig.~\ref{fig:unicity}c-d are
identical, which is in agreement with the Eq.~\ref{eq:S2add}, but
both functions differ from the corresponding ones obtained for the
other cell in Fig.~\ref{fig:unicity}i-j. The lineal path functions
are, on the other hand, identical for both phases in
Fig.~\ref{fig:unicity}e-f as well as for both phases obtained for
the second cell in Fig.~\ref{fig:unicity}k-l. This proves a
non-unique solution of a reconstruction process for the chosen
highly rough discretisation. Of course, for a higher resolution,
the difference between the lineal path function obtained for both
cells can be again revealed. As a conclusion, the reconstruction
process based on the discretised medium has always a unique
solution in case of the two-point probability function and also
mostly in case of the lineal path function where the differences
among a potential set of solutions are decreasing with increasing
resolution.



Based on this conclusion, we can proceed to the comparison of the
entire lineal path function $L_2$ with its Monte Carlo-based
approximation $\tilde{L}_2$ withing the reconstruction
process. Considering the microstructures depicted in
Fig.~\ref{fig:MC_LP}, we may assume that their reconstruction based on
the entire lineal path function will lead to almost the same
microstructures as the original ones in case of the microstructures B
and C and to exactly same one in case of A. To decrease computational
demands of the comparison, we reduced the dimensions of the
microstructures to $50\times 50$ pixels. The reconstruction process is
driven again as the minimisation of the least square error given in
Eq.~\ref{eq:LSE1}. The Monte Carlo-based approximation -- described in
Sec.~\ref{sec:imple} -- is applied here to evaluate both the lineal
path functions of the original as well as of the reconstructed image,
respectively. In order to investigate the influence of the
approximation quality, we have considered three levels corresponding
to the Monte Carlo evaluation based on $N = 10, 100$ and $1000$
samples. The results are compared with the reconstruction based on the
entire lineal path function, where the number of samples is identical
with the number of pixels in the images, i.e. $N = 2500$.

\begin{figure} [t!]
\begin{center}
\begin{tabular}{@{}c@{}|@{\hspace{0.1cm}}c@{}}
    \begin{adjustbox}{valign=b}
        \begin{tabular}{c}
            A.~\includegraphics*[width=15mm,keepaspectratio]{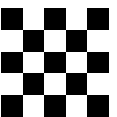}\\[2mm]
            B.~\includegraphics*[width=15mm,keepaspectratio]{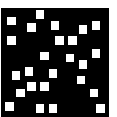}\\[2mm]
            C.~\includegraphics*[width=15mm,keepaspectratio]{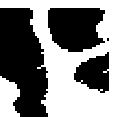}\\[1mm]
        \end{tabular}
    \end{adjustbox}
    &
    \begin{adjustbox}{valign=b}
        \begin{tabular}{c}
            \includegraphics*[width=90mm,keepaspectratio]{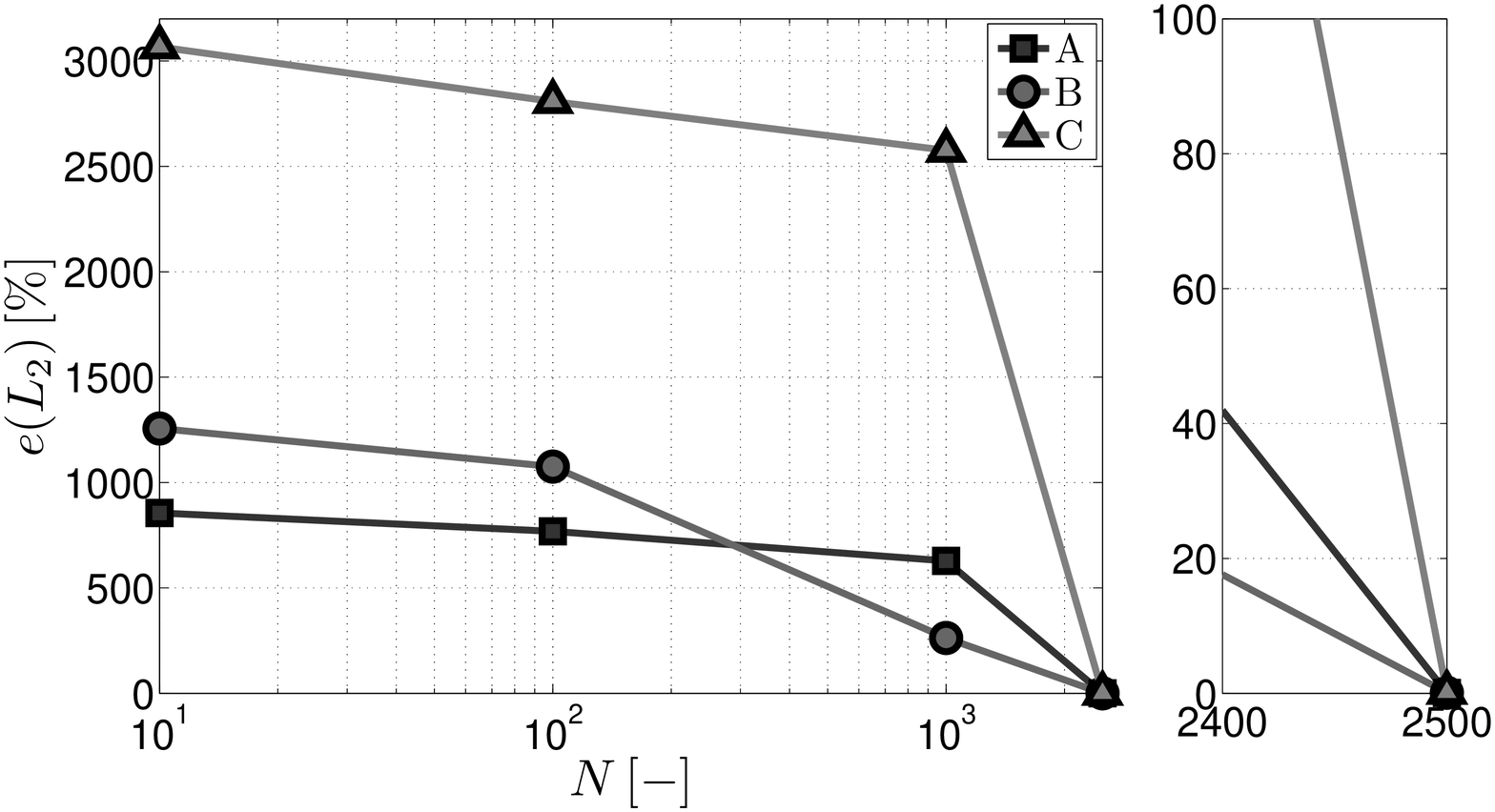}\\
        \end{tabular}
    \end{adjustbox}
\end{tabular}
\end{center}
\caption{Comparison of the entire $L_2$ and its approximation
$\tilde{L}_2$ used in microstructure reconstruction for different
values of $N$} \label{fig:ComMicRec}
\end{figure}

The relative errors of the final reconstructed $L_2$-based images
related to the entire lineal path function $L_2$ of original images
are displayed in Fig.~\ref{fig:ComMicRec}. The displayed values reveal
that the reconstruction procedure based on the lineal path
approximation converges very slowly with the number of evaluated
samples and the reconstruction process thus leads to images with the
lineal path function, which is highly different from the prescribed
one.

%

\section{Microstructure compression} \label{sec:compress}

While the reconstruction process aims at rediscovering of
a microstructure with dimensions and spatial statistics defined by
the given descriptor, the compression process tries to reduce the
information content of the given morphology and searches for its
compressed representation by a small statistically similar
periodic cell \cite{Zeman:2007:MSMSE,Lee:2009:PRE} or a set of
compatible cells \cite{Novak:PR:2012}. After evaluating a chosen
statistical descriptor over the whole available domain of the original
medium, one needs to decide about the cells' dimensions and
accordingly cut the dimensions of the descriptor.  Then the
compression process proceeds in an exactly same manner as the
microstructure reconstruction.

As our numerical implementations of the two-point probability function and
the lineal path function are both based on the assumption of periodicity, they
will not provide precise results when applied to original random and
non-periodic microstructure. Nevertheless, as already mentioned
previously, in \cite{Gajdosik:2006:PEM} it was shown that the
assumption of periodicity does not introduce a systematic bias in the
values of the descriptors.

\subsection{Particulate suspension}

\begin{figure} [h!]
\begin{center}
\begin{tabular}{c|c|@{}c@{}@{}c@{}@{}c@{}}
\begin{adjustbox}{valign=m}
    \includegraphics*[width=45mm,keepaspectratio]{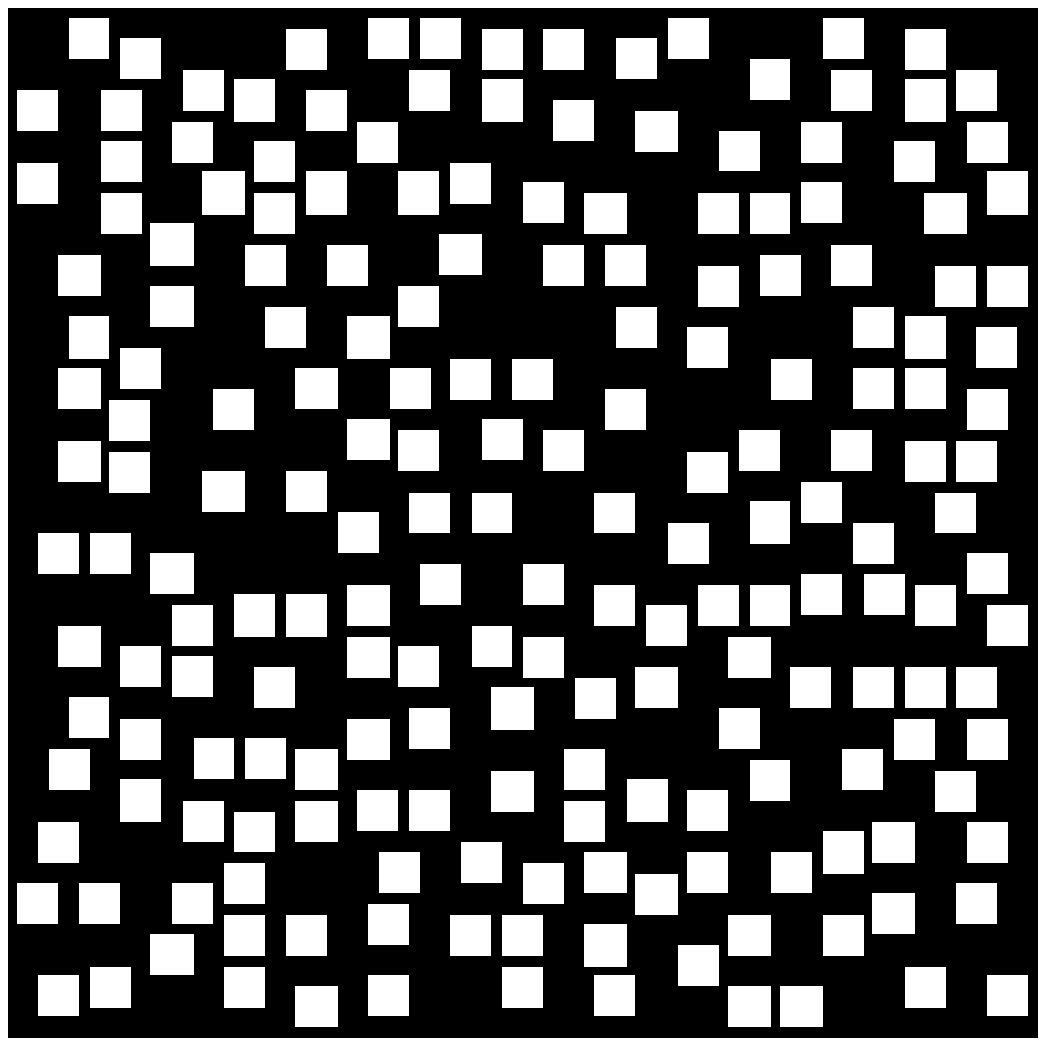}
\end{adjustbox}
&
\begin{adjustbox}{valign=m}
    \includegraphics*[width=22.5mm,keepaspectratio]{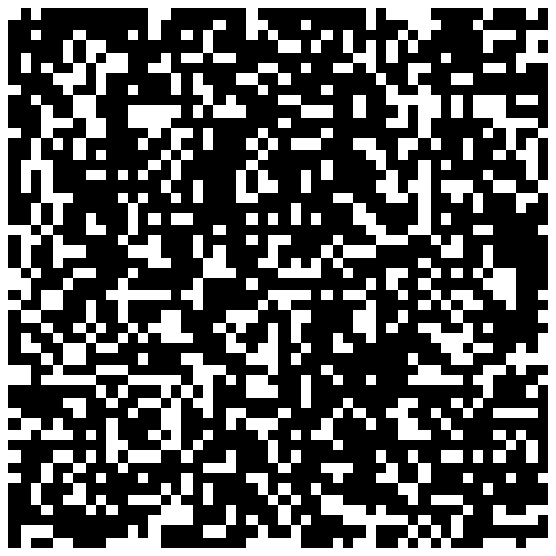}
\end{adjustbox}
&
\begin{adjustbox}{valign=m}
    \begin{tabular}{c}
    \includegraphics*[width=22.5mm,keepaspectratio]{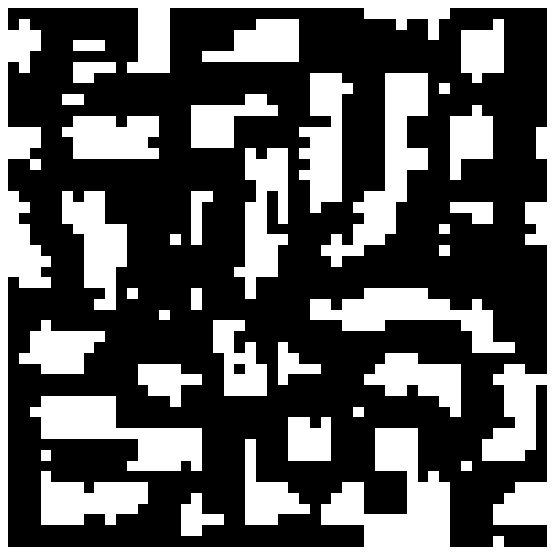} \\
    (c) \\[5mm]
    \includegraphics*[width=22.5mm,keepaspectratio]{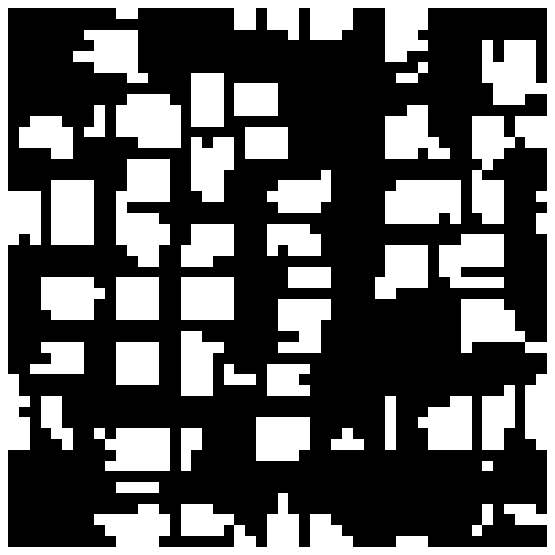} \\
    \end{tabular}
\end{adjustbox}
&
\begin{adjustbox}{valign=m}
    \begin{tabular}{c}
    \vspace{23mm}\\
    \\
    \includegraphics*[width=22.5mm,keepaspectratio]{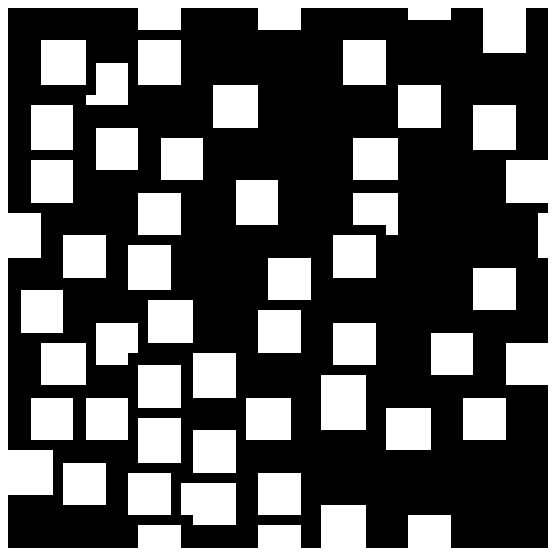} \\
    \end{tabular}
\end{adjustbox}
&
\begin{adjustbox}{valign=m}
    \begin{tabular}{c}
    \vspace{23mm}\\
    \\
    \includegraphics*[width=22.5mm,keepaspectratio]{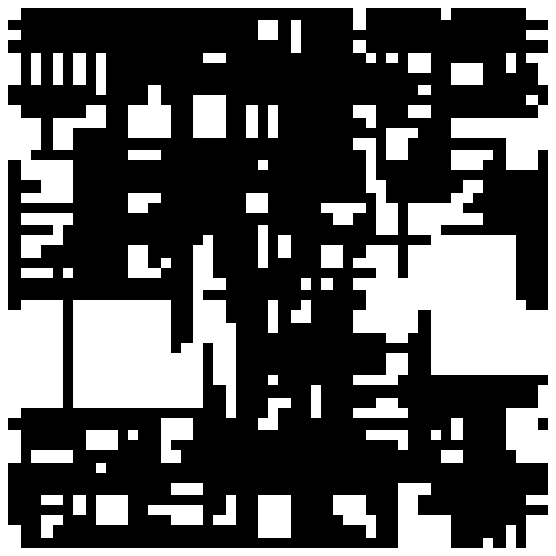} \\
    \end{tabular}
\end{adjustbox}
\\
(a) & (b) & (d) & (e) & (f)\\
\end{tabular}
\end{center}
\caption{Particulate suspension: (a) Original medium, size $100\times
  100$ [px], particles $4 \times 4$ [px]; (b) Random initial
  structure, size $50\times 50$ [px]; (c) Compressed $S_{2}$-based
  image, size $50\times 50$ [px]; (d) Compressed $L_{2}^{\mathrm{b}}$
  and $L_{2}^{\mathrm{w}}$-based image, size $50\times 50$ [px]; (e)
  Compressed $L_{2}^{\mathrm{w}}$-based image, size $50\times 50$
  [px]; (f) Compressed $L_{2}^{\mathrm{b}}$-based image, size
  $50\times 50$ [px]} \label{fig:squares}
\end{figure}
\begin{figure} [h!]
\begin{center}
\begin{tabular}{c@{}@{}c@{}@{}c@{}}
\begin{adjustbox}{valign=m}
    \includegraphics*[width=50mm,keepaspectratio]{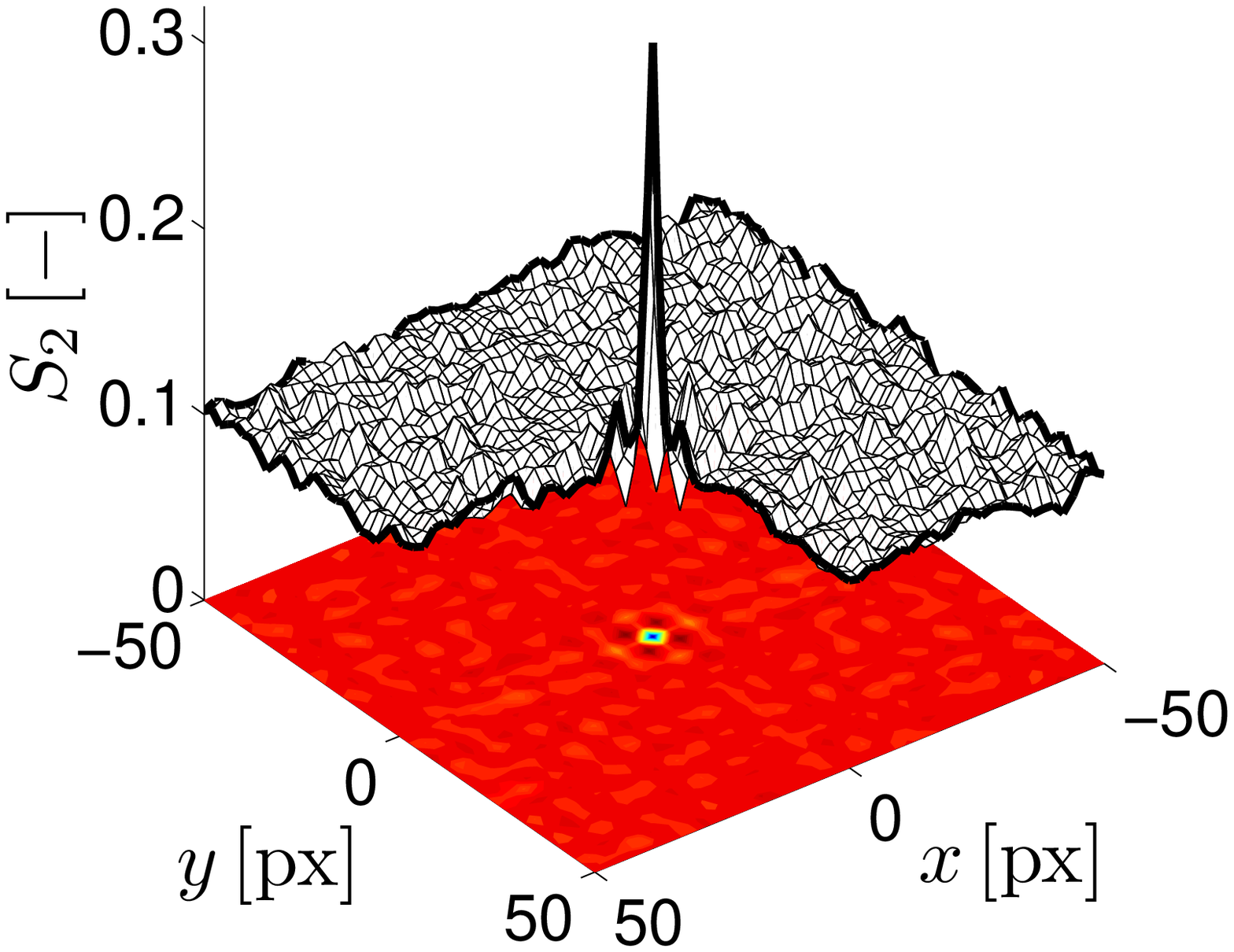}
\end{adjustbox}
&
\begin{adjustbox}{valign=m}
    \includegraphics*[width=50mm,keepaspectratio]{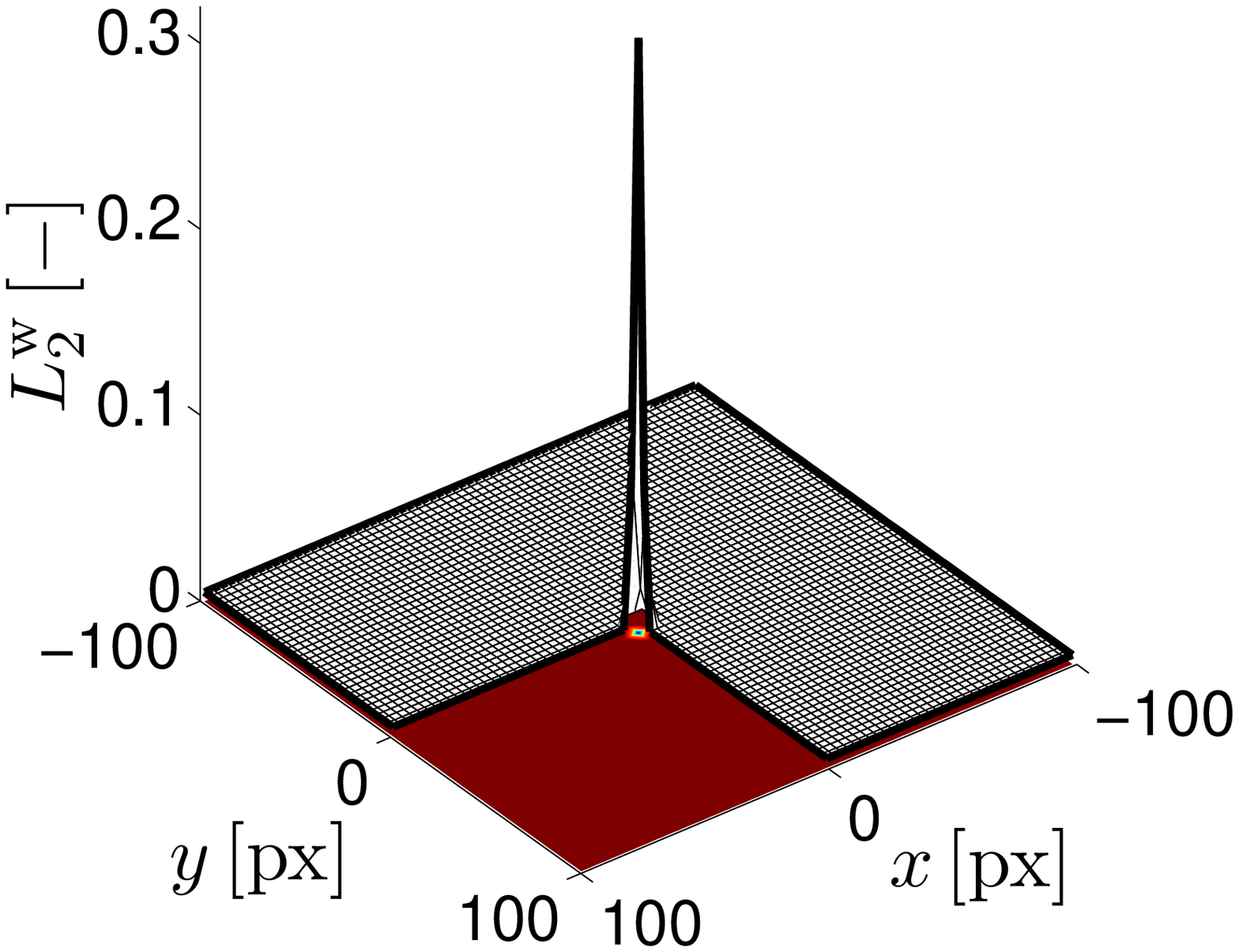}
\end{adjustbox}
&
\begin{adjustbox}{valign=m}
    \includegraphics*[width=50mm,keepaspectratio]{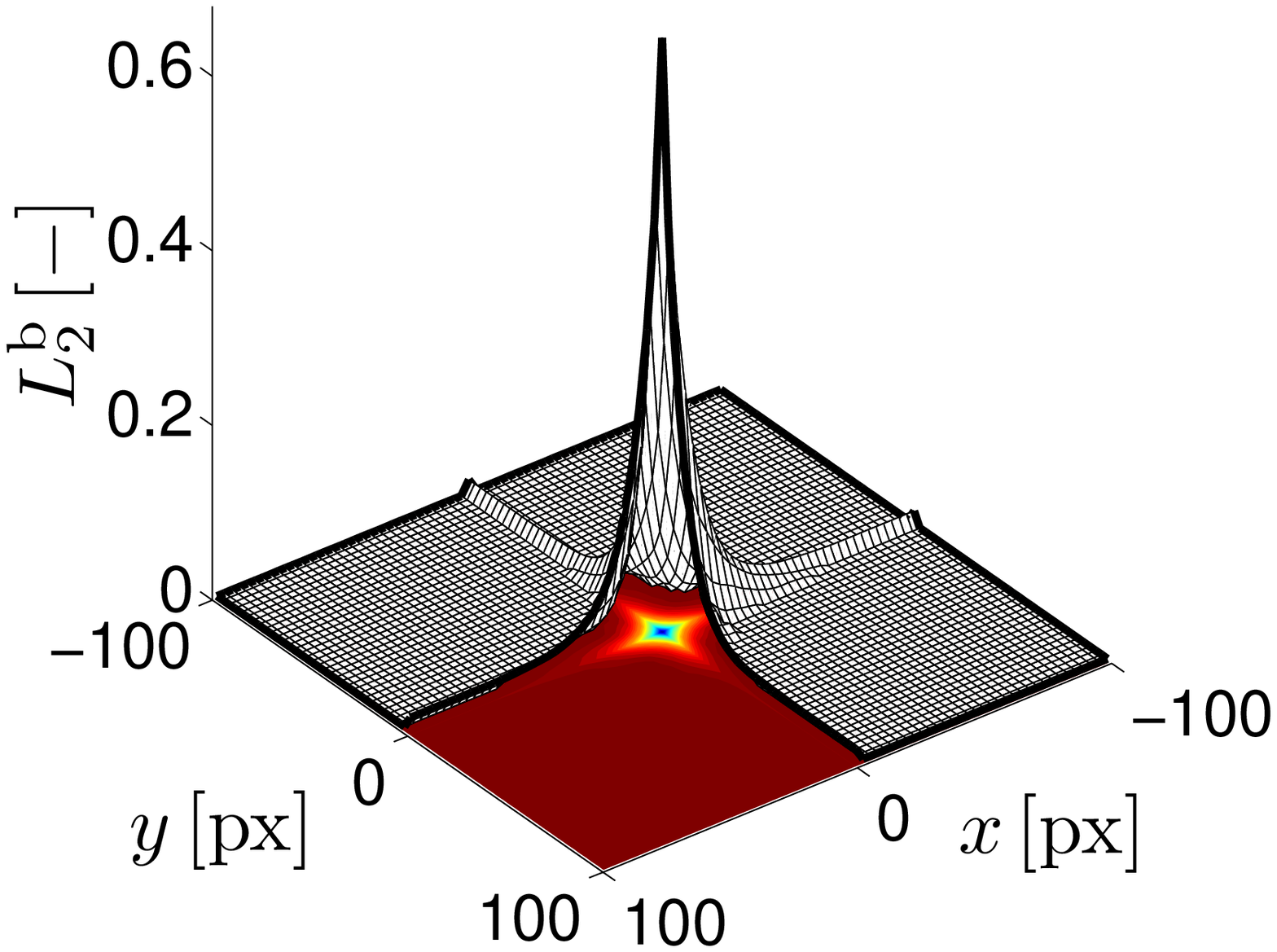}
\end{adjustbox}
\\
(a) & (b) & (c)
\end{tabular}
\end{center}
\caption{Particulate suspension: (a) Original medium, size
$100\times 100$ [px], particles $4\times 4$ [px]; (b)
$S_{2}$-function; (c) $L_{2}^{\mathrm{w}}$-function; (d)
$L_{2}^{\mathrm{b}}$-function} \label{fig:sqBst}
\end{figure}

The first example comparing properties of the $S_2$- and
$L_2$-based compression concerns an artificially created
particulate suspension consisting of equal-sized white squares
randomly distributed within a black matrix, see
Fig.~\ref{fig:squares}. The shape of particles is a very
significant property of such a microstructure, which can be easily
preserved by modifying the optimisation algorithm so as to start
with randomly distributed particles and then to move their centres
within the optimisation process. Nevertheless, here we aimed at
testing the descriptors in their ability to capture such an
important property within the compression process.

Fig.~\ref{fig:squares}c shows that the $S_2$-based compression
leads to significant deterioration of the shape of particles. This
is caused by the very small ratio of the particles over the size
of the PUC, i.e. $4 \times 4$ [px] vs. $50 \times 50$ [px]. The
information about their shape is thus saved on a small portion of
the descriptor's domain corresponding to short-range correlations
(see Fig.~\ref{fig:sqBst}a), while the most of the domain defines
long-range correlations corresponding to the mutual distances of
the particles. The lineal path function allows to separate the
information about the shape of particles and their mutual
positions, since the $L_{2}^{\mathrm{w}}$ in Fig.~\ref{fig:sqBst}b
contains only the first, while the $L_{2}^{\mathrm{b}}$ in
Fig.~\ref{fig:sqBst}c defines mostly the latter. According to
that, the $L_{2}^{\mathrm{w}}$-based compression in
Fig.~\ref{fig:squares}e leads obviously to the well compressed
shape of particles, while the $L_{2}^{\mathrm{b}}$-based
compression in Fig.~\ref{fig:squares}f does not capture the shape
of particles at all and the $L_2$-based compression in
Fig.~\ref{fig:squares}d provides a compromise solution. It is hard
to evaluate the quality of obtained structures in an objective
manner, but we can conclude that the $L_2$ function allows a user
to emphasise short-range effects as needed.

Another interesting aspect concerns the mutual comparison of the
compressed microstructures and the corresponding errors in
describing original medium according to Eq.~\eqref{eq:LSE1}, which
are listed in Tab.~\ref{tab:mutual}. While the optimisation of
$S_2$ leads to a microstructure which is relatively good also with
respect to $L_2$ and comparable with microstructures obtained for
the $L_{2}^{\mathrm{w}}$- or $L_{2}^{\mathrm{w}}$-based optimisation,
the opposite is not true. The microstructures optimised w.r.t. one
or both phases of $L_2$ manifest very bad correlations, which are
comparable or even worse than those obtained for a random image.
The lineal path function thus cannot be applied for a
correlations-based compression.
\begin{table}[htb!]
\centering
\begin{tabular}{lll}
\hline
Compressed medium & $e(S_2)$  & $e(L_2)$ \\
\hline
Random Fig.~\ref{fig:squares}b & $3.12 \cdot 10^{-1}$ & $1.68 \cdot 10^{1}$ \\
$S_{2}$-based Fig.~\ref{fig:squares}c & $9.20 \cdot 10^{-3}$ & $2.35 \cdot 10^{0}$ \\
$L_{2}$-based Fig.~\ref{fig:squares}d & $3.09 \cdot 10^{-1}$ & $2.73 \cdot 10^{-2}$ \\
$L_{2}^{\mathrm{w}}$-based Fig.~\ref{fig:squares}e & $ 3.53 \cdot 10^{-1}$ & $ 1.08 \cdot 10^{0}$ \\
$L_{2}^{\mathrm{b}}$-based Fig.~\ref{fig:squares}f & $ 6.52 \cdot 10^{-1}$ & $ 9.60 \cdot 10^{-1}$ \\
\hline
\end{tabular}
\caption{Mutual comparison of compressed microstructures.}
\label{tab:mutual}
\end{table}

\subsection{Epithelial cells}

\begin{figure} [h!]
\begin{center}
\begin{tabular}{c|c|@{}c@{}@{}c@{}@{}c@{}}
\begin{adjustbox}{valign=m}
    \begin{tabular}{c}
    \includegraphics*[width=45mm,keepaspectratio]{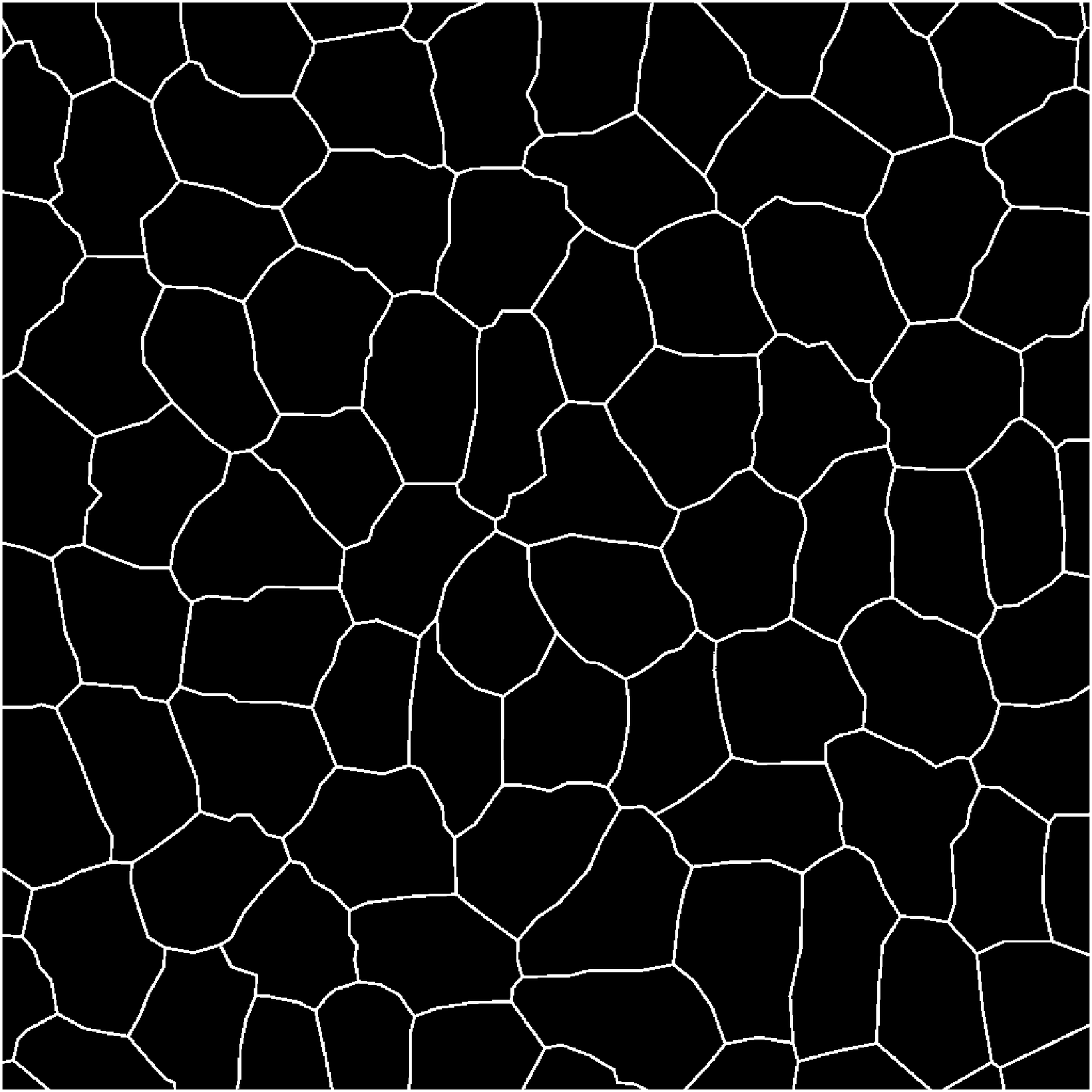} \\
    (a) \\
    \end{tabular}
\end{adjustbox}
&
\begin{adjustbox}{valign=m}
    \begin{tabular}{c}
    \vspace{15mm}\\
    \\
    \includegraphics*[width=22.5mm,keepaspectratio]{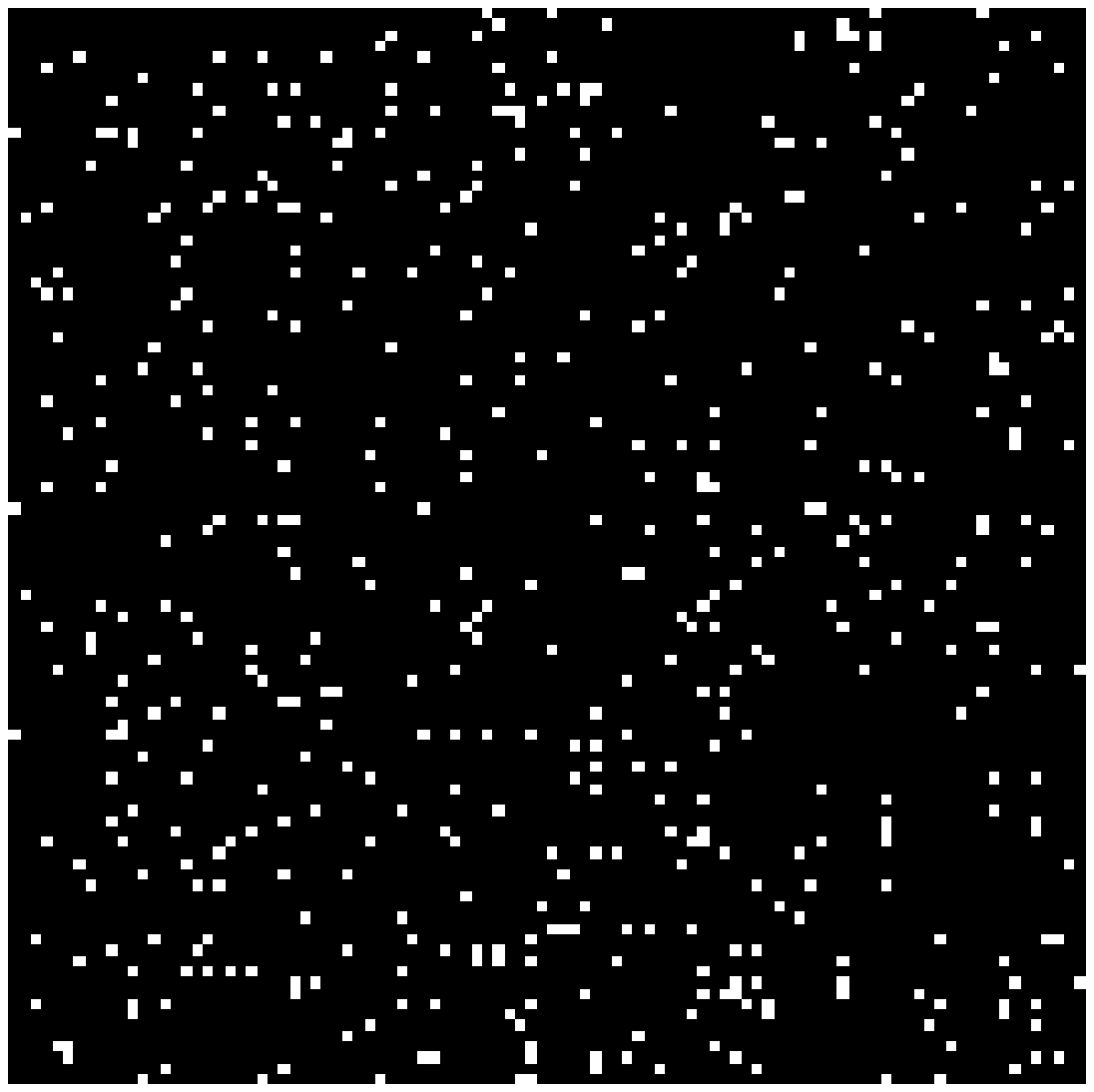} \\
    (b)
    \\
    \vspace{15mm}\\
    \end{tabular}
\end{adjustbox}
&
\begin{adjustbox}{valign=m}
    \begin{tabular}{c}
    \includegraphics*[width=22.5mm,keepaspectratio]{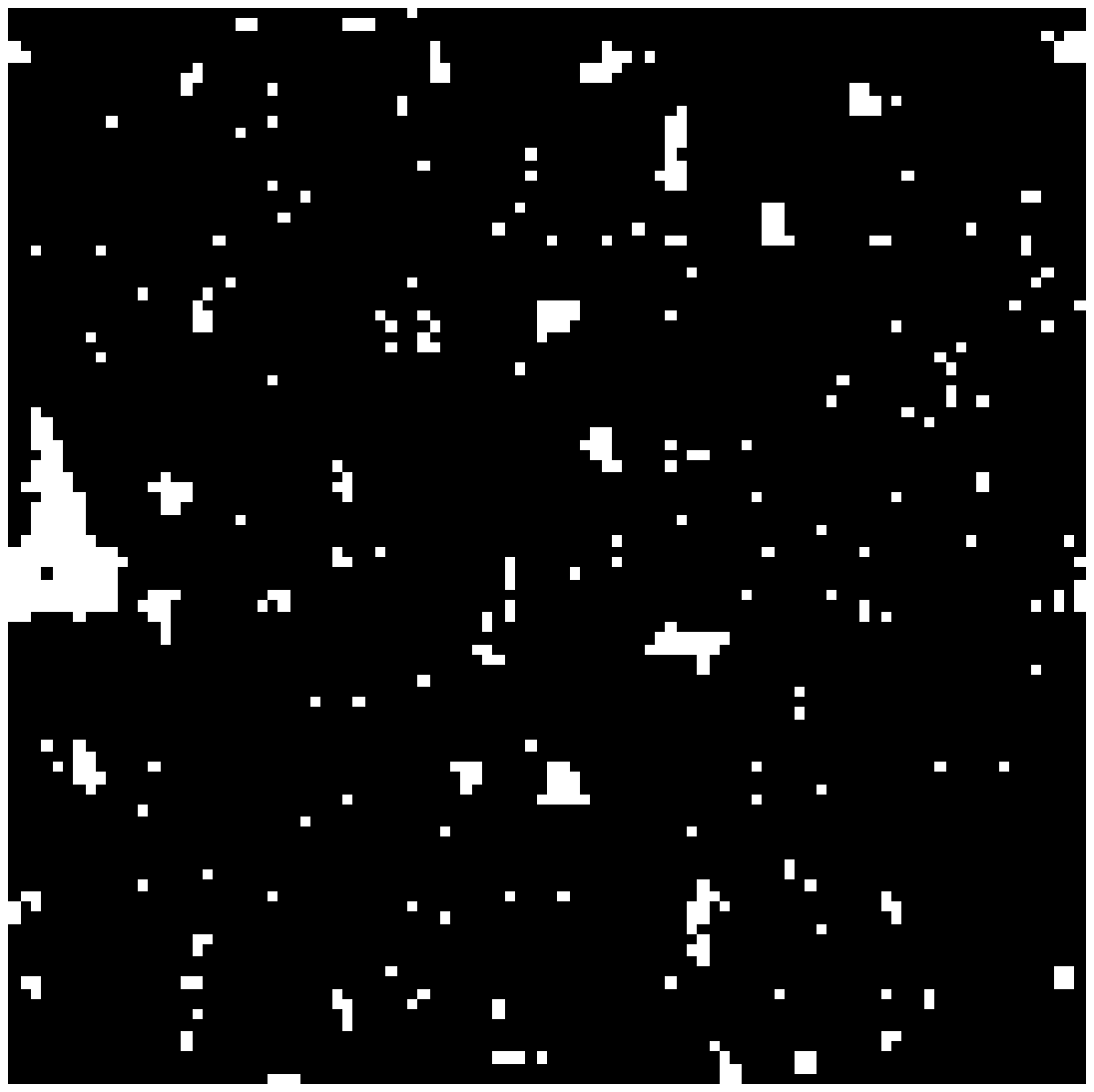} \\
    (c) \\[5mm]
    \includegraphics*[width=22.5mm,keepaspectratio]{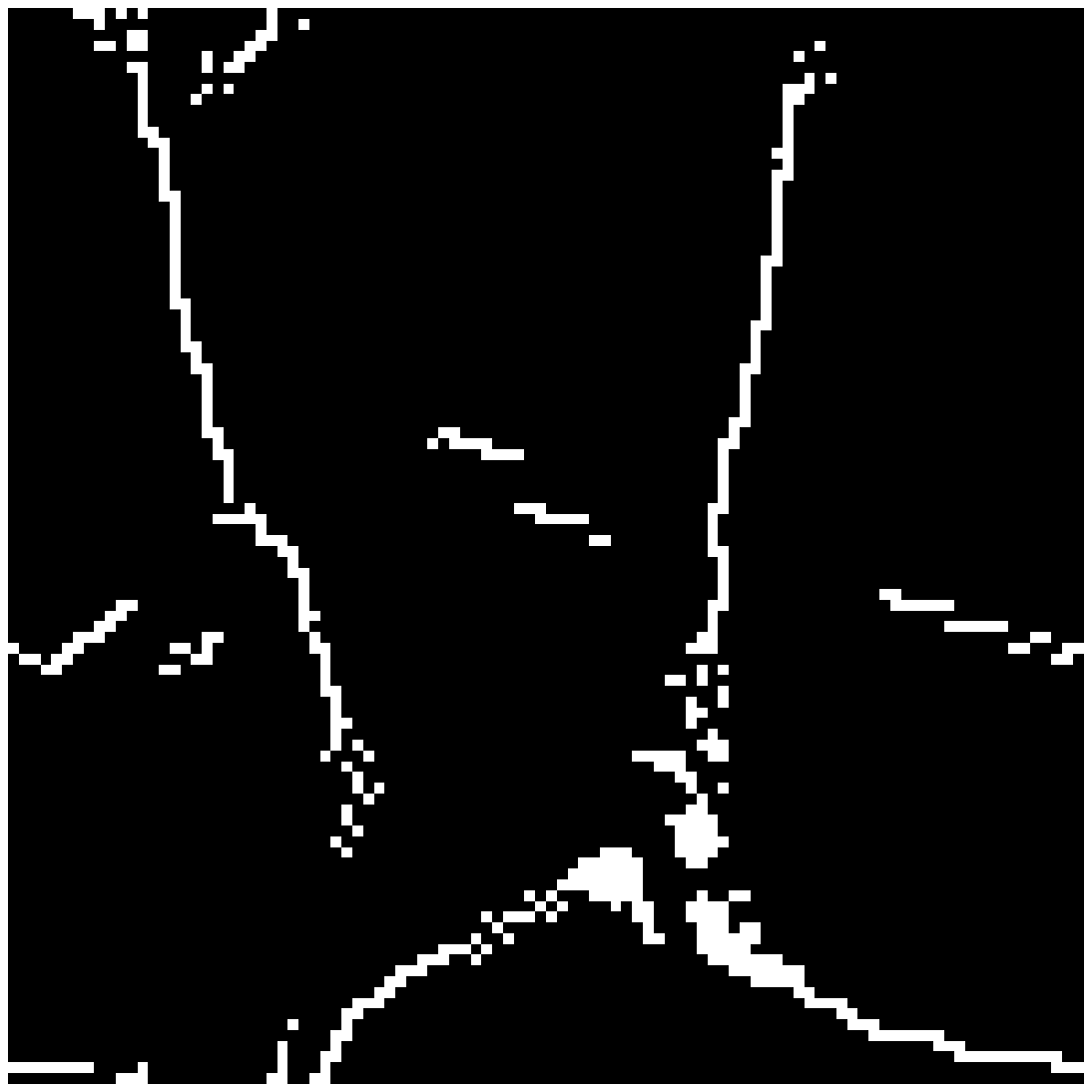} \\
    (d) \\
    \end{tabular}
\end{adjustbox}
&
\begin{adjustbox}{valign=m}
    \begin{tabular}{c}
    \vspace{23mm}\\
    \\
    \includegraphics*[width=22.5mm,keepaspectratio]{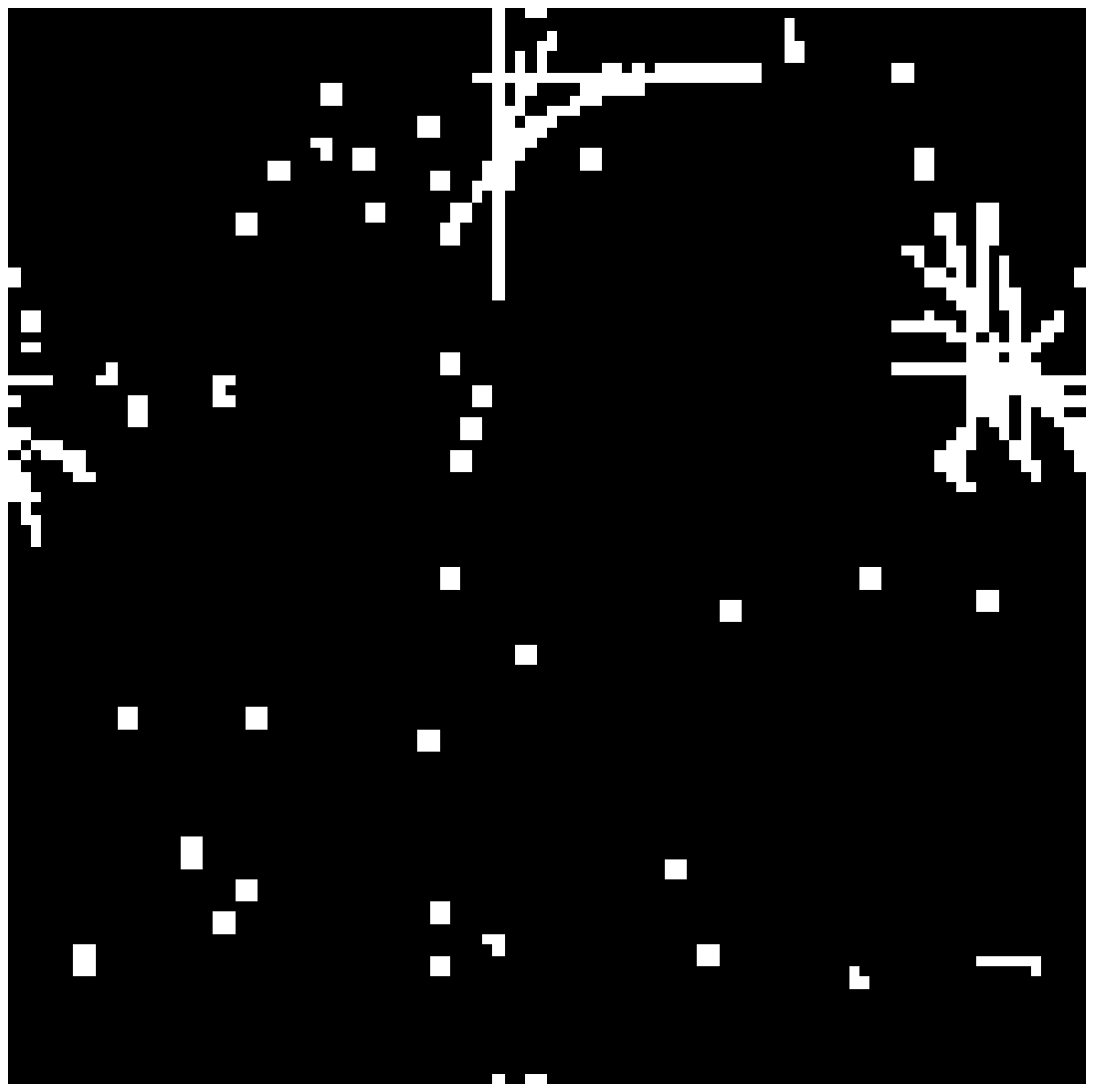} \\
    (e) \\
    \end{tabular}
\end{adjustbox}
&
\begin{adjustbox}{valign=m}
    \begin{tabular}{c}
    \vspace{23mm}\\
    \\
    \includegraphics*[width=22.5mm,keepaspectratio]{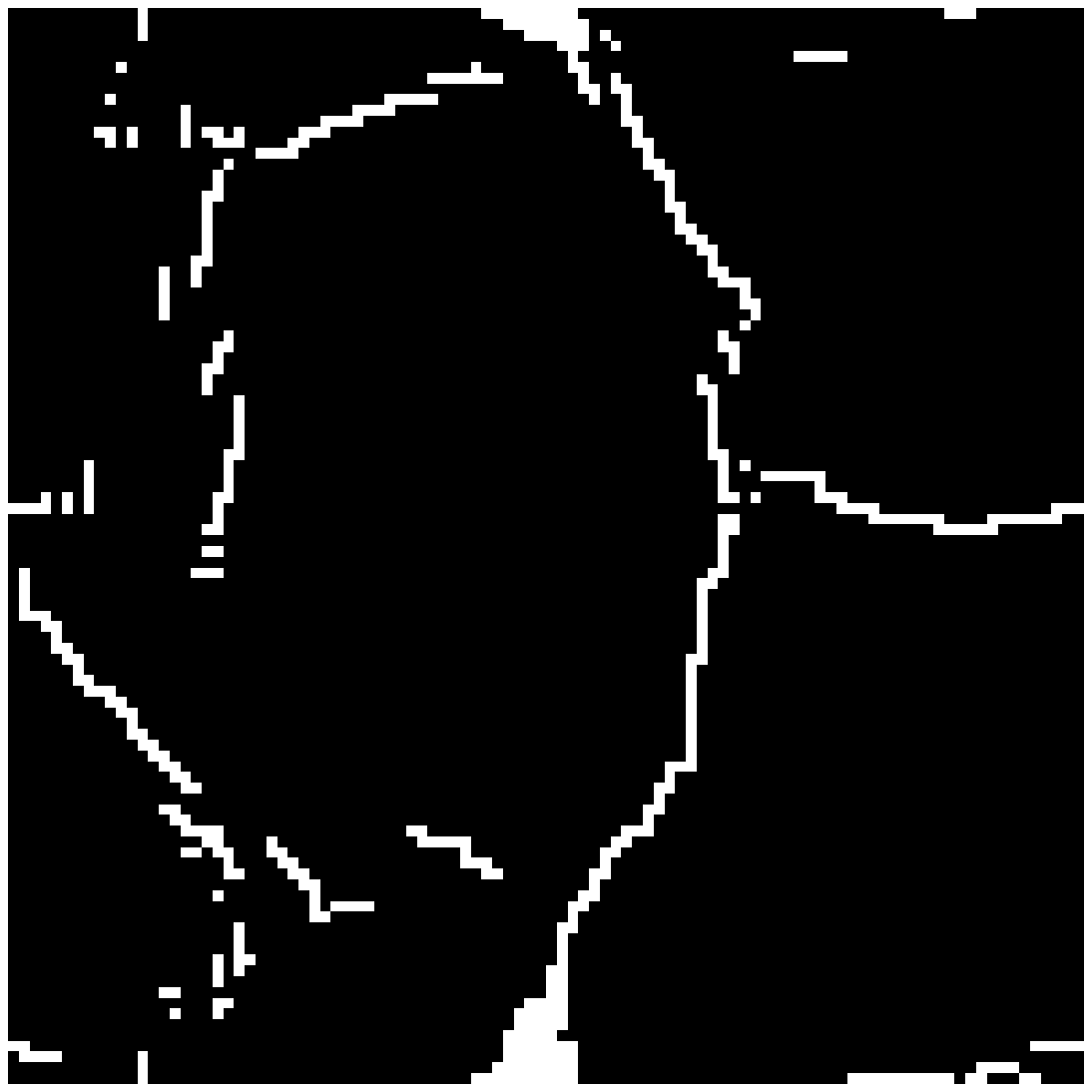} \\
    (f) \\
    \end{tabular}
\end{adjustbox}
\\
\end{tabular}
\end{center}
\caption{Epithelial cells: (a) Original medium, size $510\times
510$ [px]; (b) Random initial structure, size $100\times 100$
[px]; (c) Compressed $S_{2}$-based image, size $100\times 100$
[px]; (d) Compressed $L_{2}^{\mathrm{b}}$ and
$L_{2}^{\mathrm{w}}$-based image, size $100\times 100$ [px]; (e)
Compressed $L_{2}^{\mathrm{w}}$-based image, size $100\times 100$
[px]; (f) Compressed $L_{2}^{\mathrm{b}}$-based image, size
$100\times 100$ [px]} \label{fig:epith}
\end{figure}
\begin{figure} [h!]
\begin{center}
\begin{tabular}{c@{}@{}c@{}@{}c@{}}
\begin{adjustbox}{valign=m}
    \includegraphics*[width=50mm,keepaspectratio]{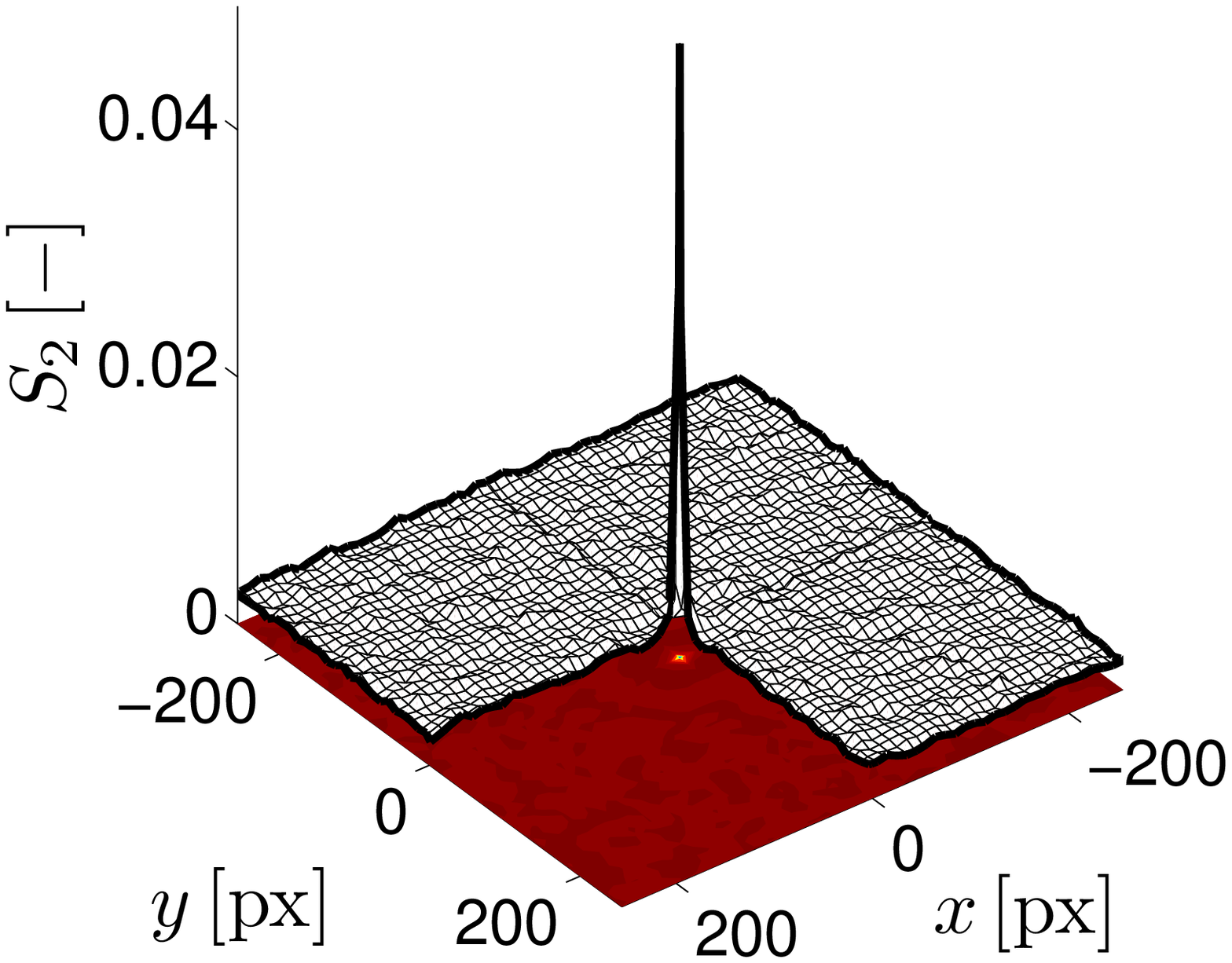}
\end{adjustbox}
&
\begin{adjustbox}{valign=m}
    \includegraphics*[width=50mm,keepaspectratio]{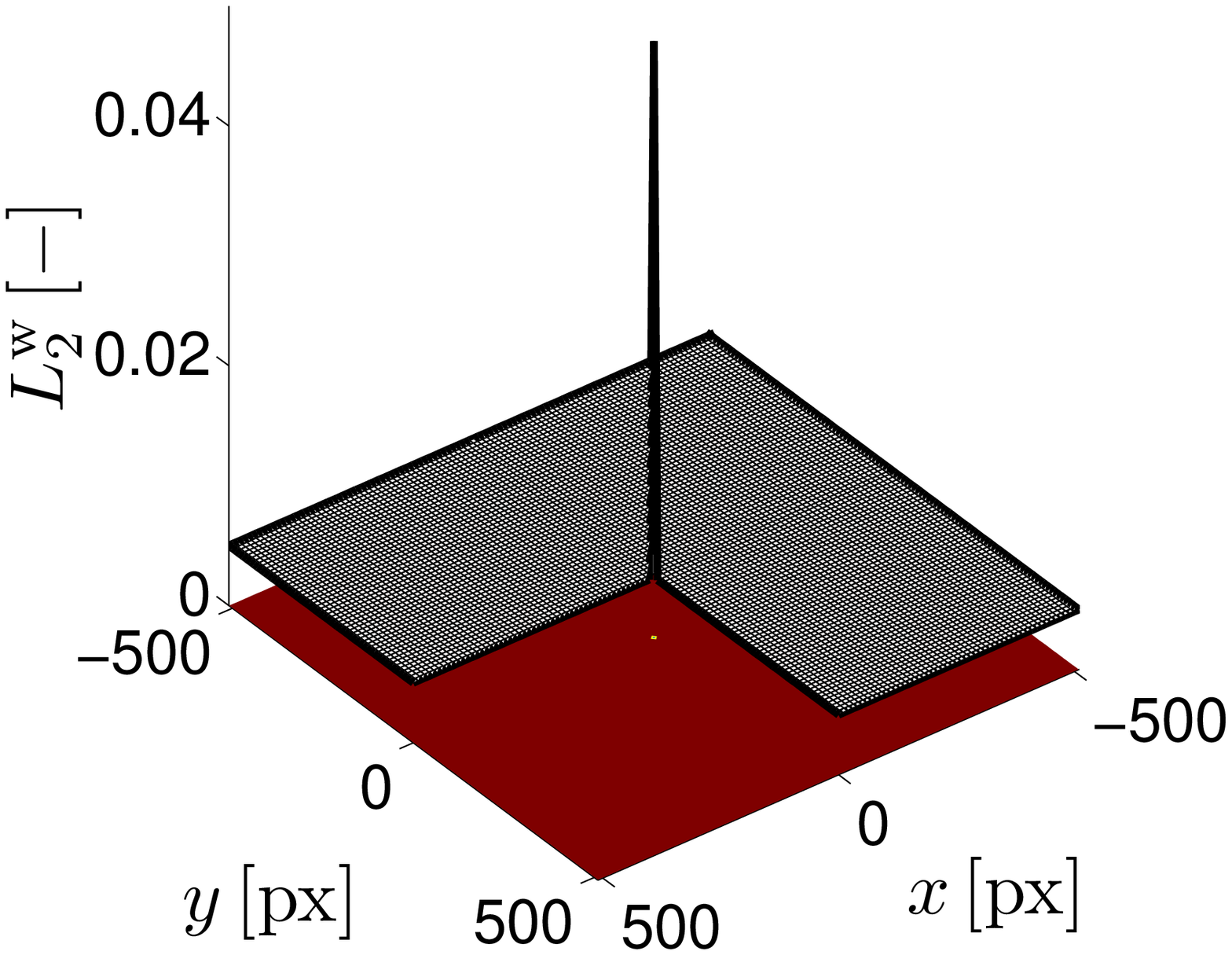}
\end{adjustbox}
&
\begin{adjustbox}{valign=m}
    \includegraphics*[width=50mm,keepaspectratio]{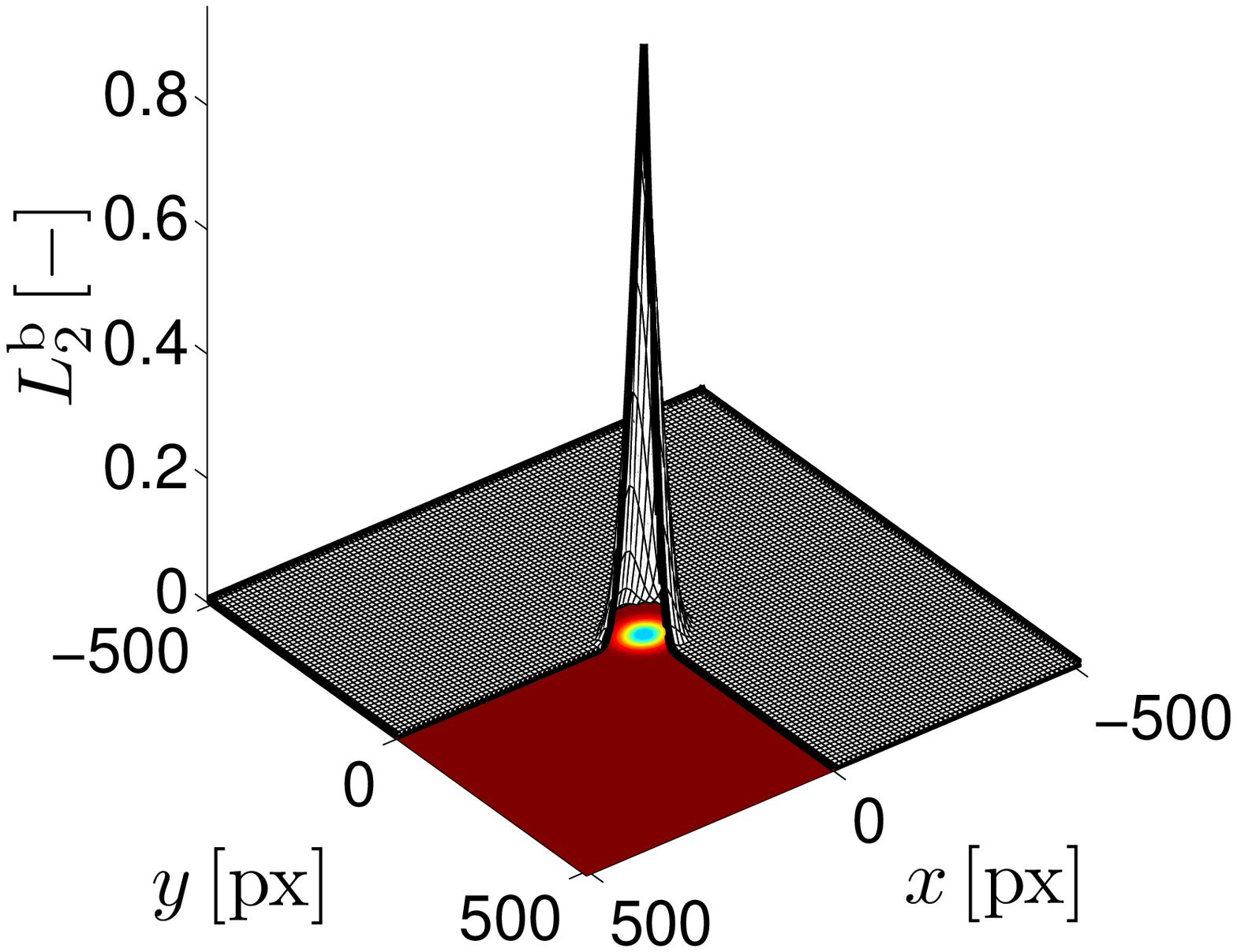}
\end{adjustbox}
\\
(a) & (b) & (c)
\end{tabular}
\end{center}
\caption{Epithalial cells: (a) Original medium, size $100\times 100$
  [px]; (b) $S_{2}$-function; (c) $L_{2}^{\mathrm{w}}$-function; (d)
  $L_{2}^{\mathrm{b}}$-function} \label{fig:epithBst}
\end{figure}

Epithalial cells are a typical example of morphology characterised
by very thin and continuous walls, see Fig.~\ref{fig:epith}a.
Their volume fraction is very small, only $4.97\,\mathrm{[\%]}$
and thickness is mostly equal to only $1$ pixel.  The assembling
of continuous walls from random initial arrangement is rather
unattainable. As can be expected, the two-point probability
function fails completely in this task, see Fig.~\ref{fig:epith}c.
Nevertheless, the assumption that the continuity of the white
walls can be captured by the lineal path computed for the white
phase is wrong. As a matter of fact, the nonlinear walls are
composed of a set of short line segments and the
$L_{2}^{\mathrm{w}}$-based compression thus leads to their random
stars-resembling arrangement as visible in Fig.~\ref{fig:epith}e.
The continuity of walls is actually closely related to cells,
whose limited size requires the continuity of the surrounding
medium. As a consequence, the information about the continuity of
walls is surprisingly hidden in $L_{2}^{\mathrm{b}}$, see results
of $L_{2}^{\mathrm{b}}$-based compression in
Fig.~\ref{fig:epith}f. Due the small volume fraction of the white
phase, its influence on the $L_2$-based compression is rather
small and the results are principally similar to the
$L_{2}^{\mathrm{b}}$-based compression, cf. Figs.~\ref{fig:epith}d
and \ref{fig:epith}f. The remaining discontinuities are very
difficult to be improved within the proposed optimisation strategy
based on random interchanging of two pixels. We can only assume
that the full connectivity of walls can be obtained using some
more sophisticated modification operator.

\subsection{Trabecular bone}

\begin{figure} [h!]
\begin{center}
\begin{tabular}{ccc}
\multicolumn{3}{c}{\includegraphics*[width=100mm,keepaspectratio]{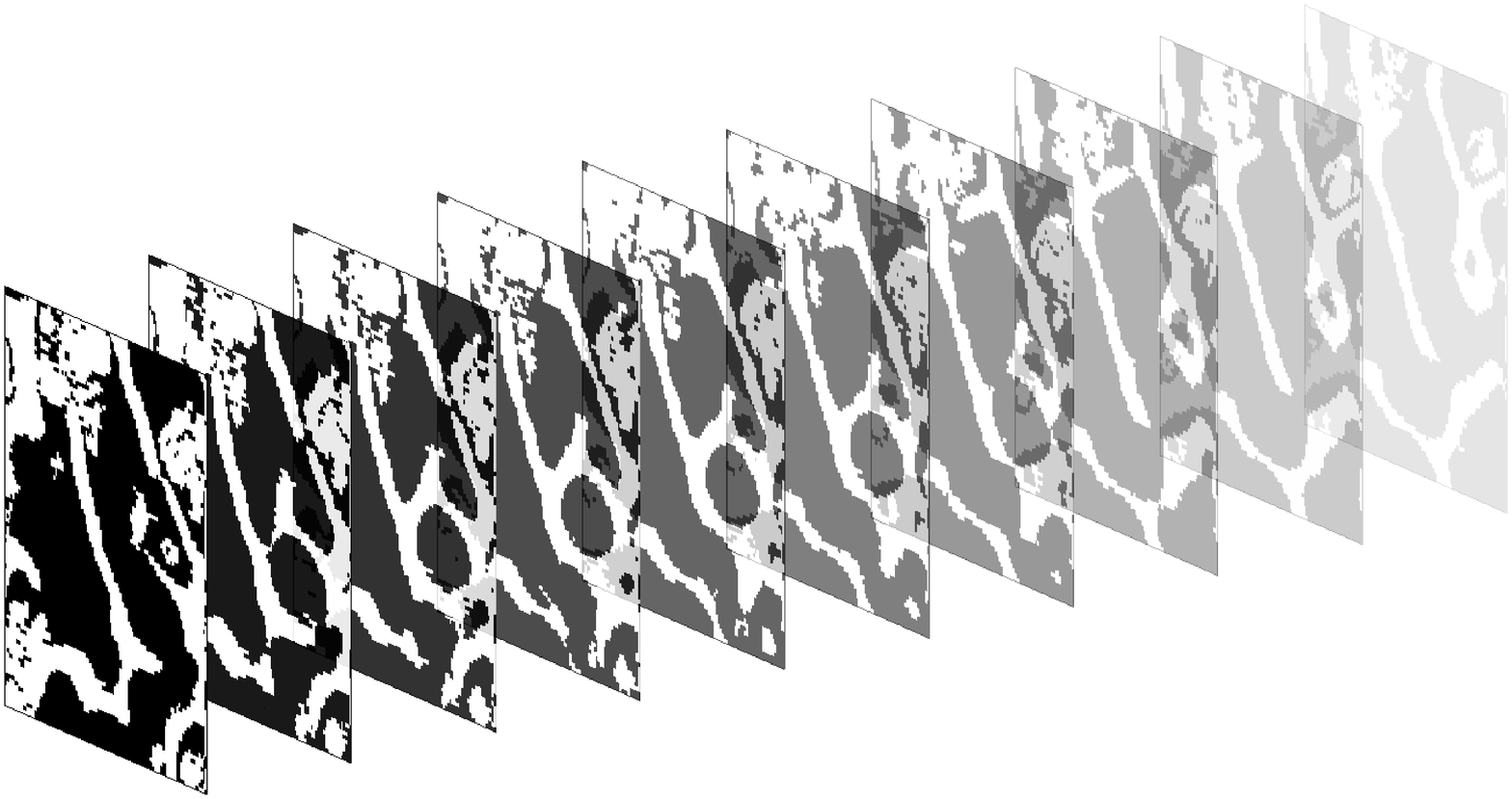}} \\
\multicolumn{3}{c}{(a)} \\
\includegraphics*[width=35mm,keepaspectratio]{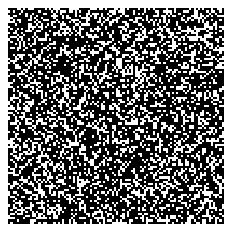} &
\includegraphics*[width=35mm,keepaspectratio]{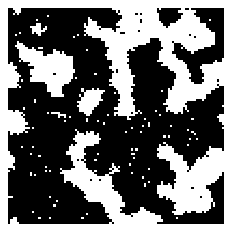} &
\includegraphics*[width=35mm,keepaspectratio]{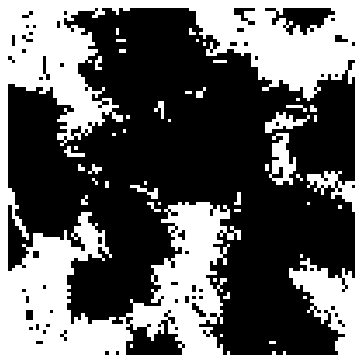} \\
(b) & (c) & (d) \\
\end{tabular}
\end{center}
\caption{Trabecular bone microstructure obtained by micro Computed
  Tomography~\cite{Jirousek:NIMPRSA:2011}: (a) $3$-D cuts of original
  structure, $100\times 100$ [px]; (b) Initial random morphology
  corresponding to volume fraction $\phi^{\mathrm{b}}$ and
  $\phi^{\mathrm{w}}$ of original medium, $100\times 100$ [px]; (c)
  Compressed $S_{2}$-based structure, $100\times 100$ [px]; (d)
  Compressed $L_{2}$-based structure, $100\times 100$ [px]}
\label{fig:bone}
\end{figure}

\renewcommand{\arraystretch}{1.1}
\begin{figure} [h!]
\begin{center}
\begin{tabular}{@{}c@{}|@{}c@{}}
    \begin{adjustbox}{valign=m}
        \begin{tabular}{c}
            \includegraphics*[width=15mm,keepaspectratio]{Fig_S2.eps}\\[40mm]
            \includegraphics*[width=15mm,keepaspectratio]{Fig_final.eps}\\ [40mm]
            \includegraphics*[width=15mm,keepaspectratio]{Fig_final.eps}
        \end{tabular}
    \end{adjustbox}
    &
    \begin{adjustbox}{valign=m}
        \begin{tabular}{cc}
            \includegraphics*[width=50mm,keepaspectratio]{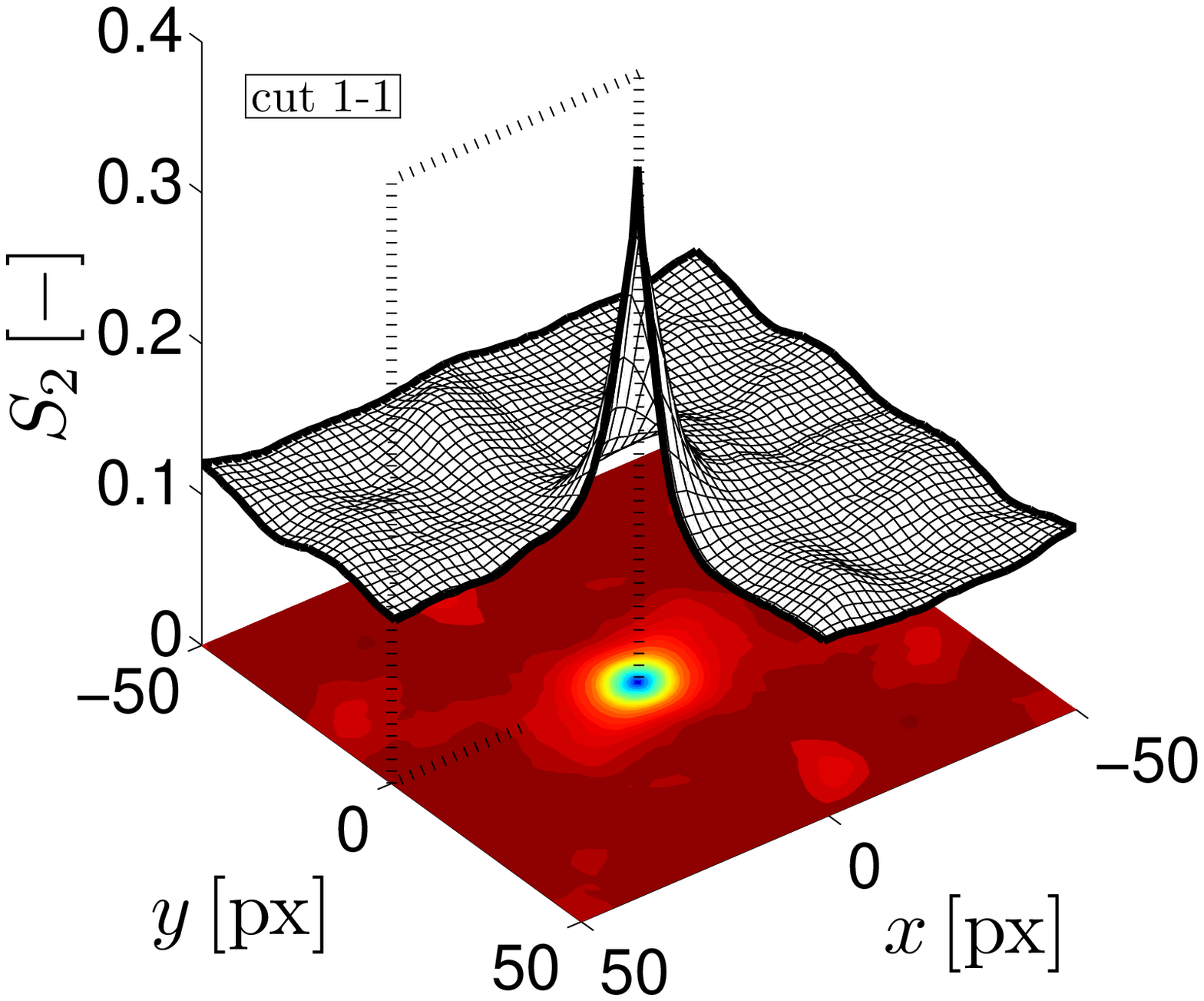} &
            \includegraphics*[width=55mm,keepaspectratio]{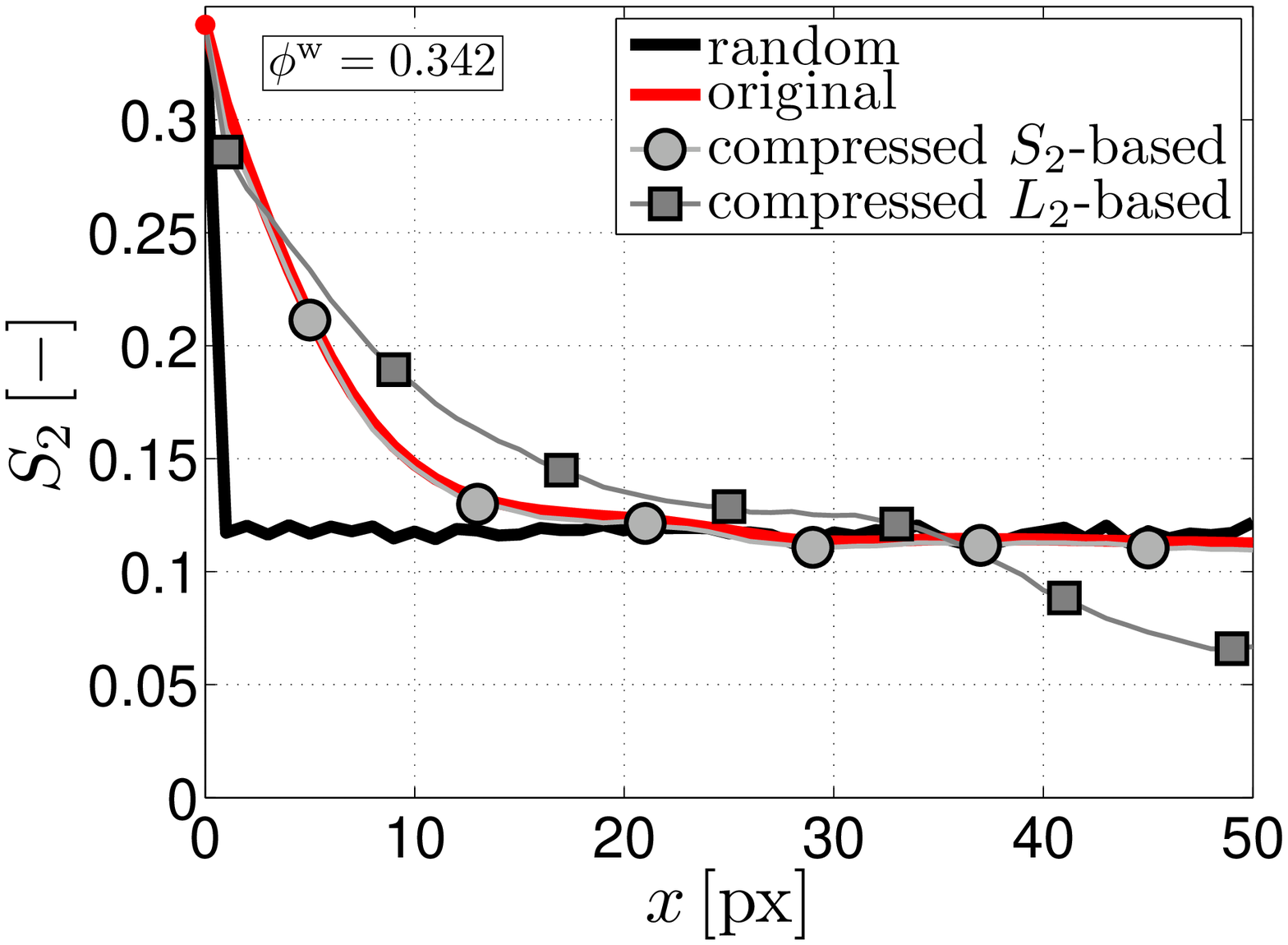} \\
            (a) & (b) \\[-0.5cm]
            \includegraphics*[width=50mm,keepaspectratio]{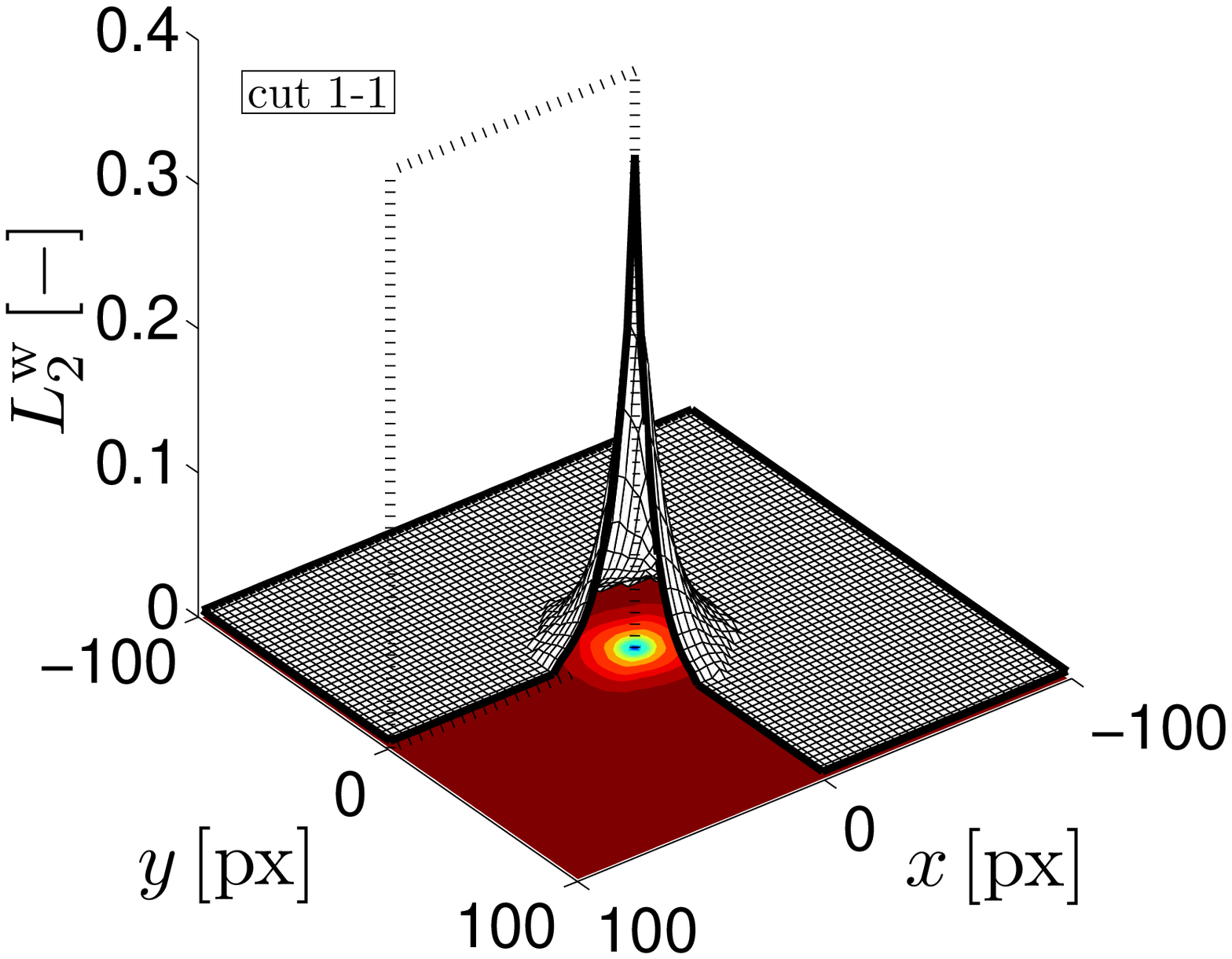}&
            \includegraphics*[width=55mm,keepaspectratio]{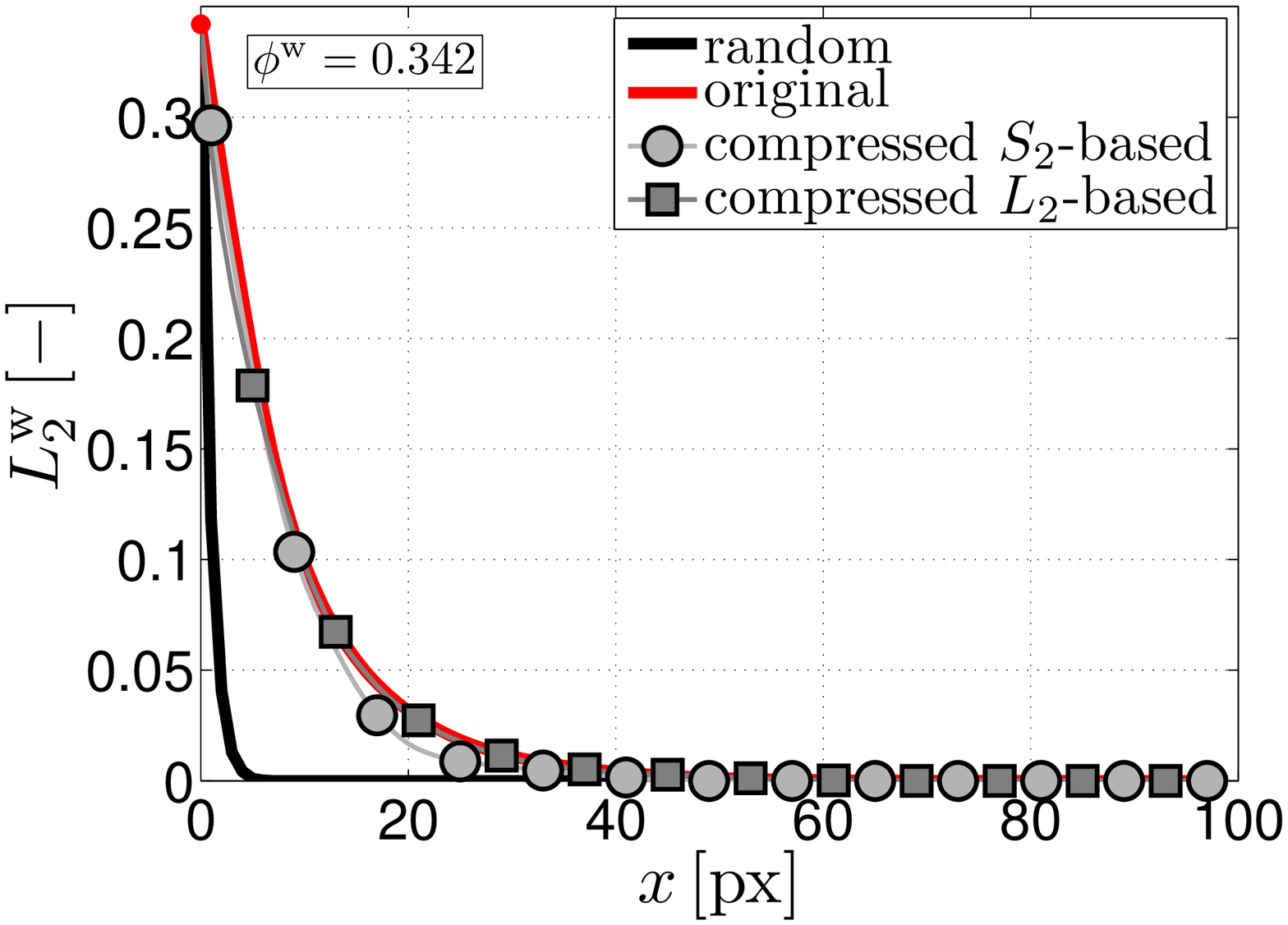} \\
            (c) & (d) \\[-0.5cm]
            \includegraphics*[width=50mm,keepaspectratio]{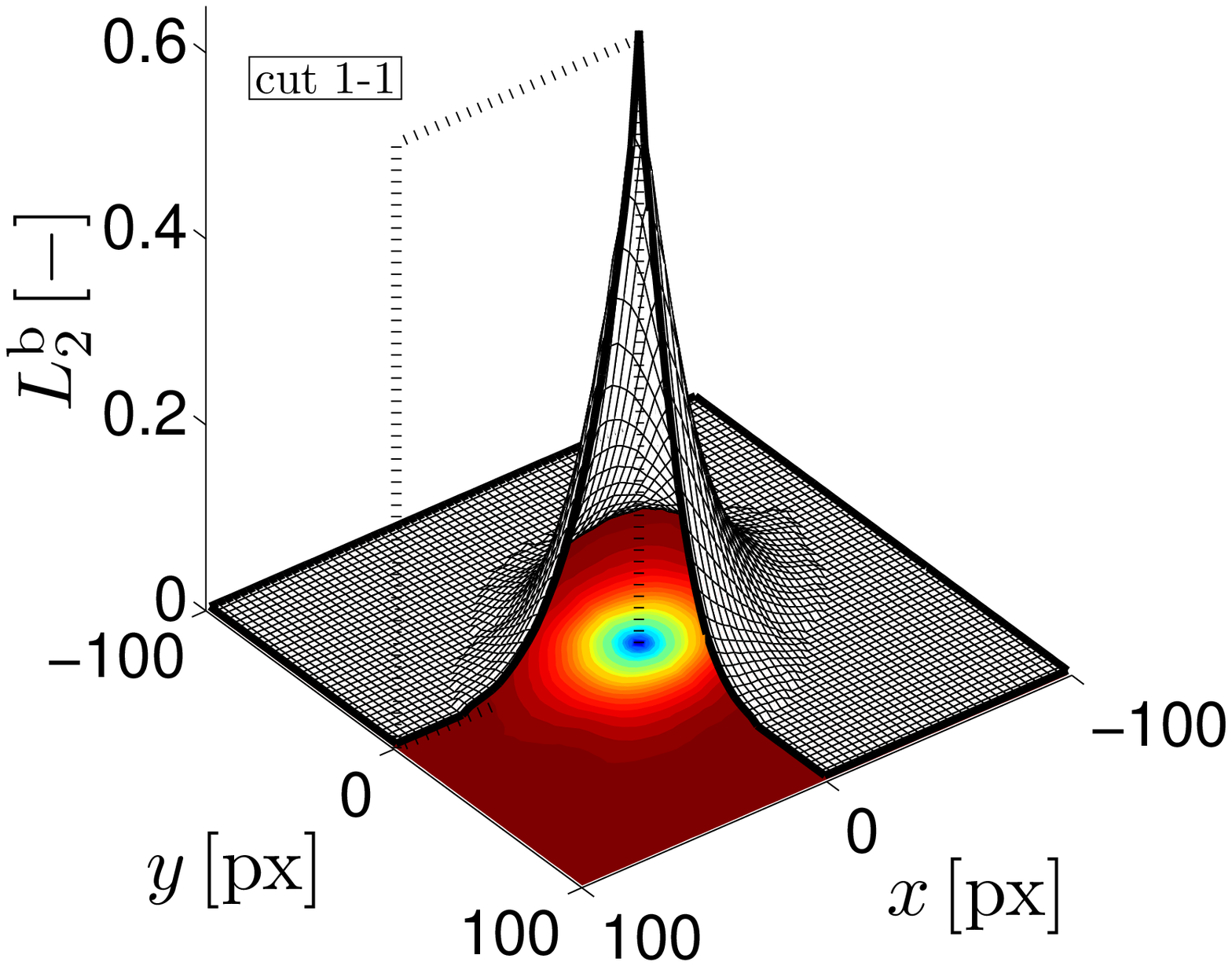}&
            \includegraphics*[width=55mm,keepaspectratio]{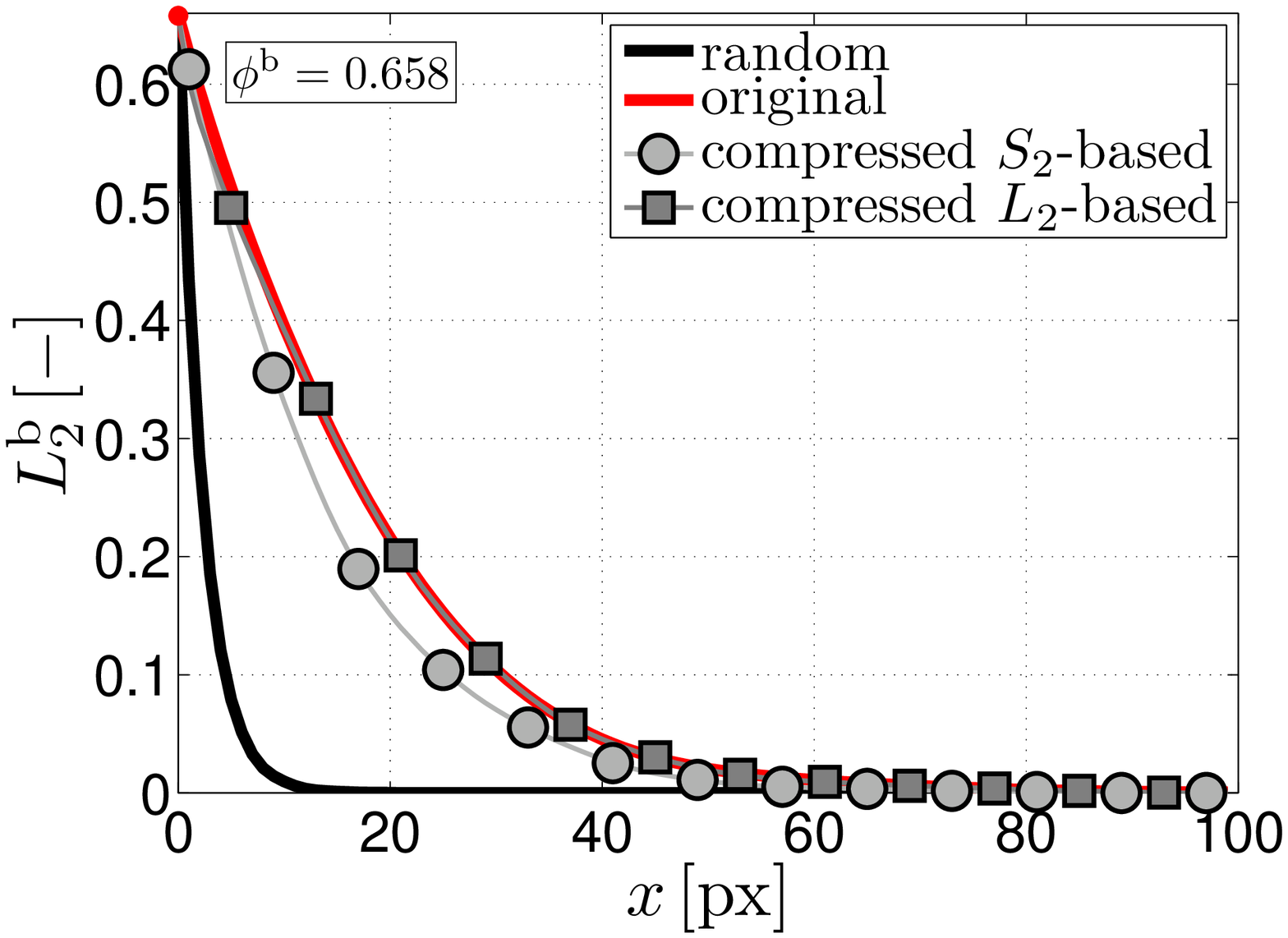} \\
            (e) & (f)
        \end{tabular}
    \end{adjustbox}
\end{tabular}
\end{center}
\caption{Results of compressed trabecular bone microstructure: (a)
  $S_2$-function of compressed $S_{2}$-based medium; (b) Comparison of
  $S_2$-functions in cut 1-1; (c) $L_{2}^{\mathrm{w}}$-function of
  compressed $L_{2}$-based system; (d) Comparison of
  $L_{2}^{\mathrm{w}}$-functions in cut 1-1; (e)
  $L_{2}^{\mathrm{b}}$-function of compressed $L_{2}$-based system;
  (f) Comparison of $L_{2}^{\mathrm{b}}$-functions in cut 1-1; }
\label{fig:boneResult}
\end{figure}

The last example concerns trabecular bone, which represents a medium with
approximately equal volume fractions of both phases creating
continuous irregular branches. Our original microstructural specimen
consists of $100 \times 100 \times 100$ [px] three-dimensional image
obtained by micro Computed Tomography~\cite{Jirousek:NIMPRSA:2011}. We
divide this data into ensemble of $100$ two-dimensional cuts $100
\times 100$ [px] and by employing the assumption of ergodicity, the
statistical descriptors are computed as an average over the ensemble.
The computational effort in case of the $L_{2}$-based reconstruction is
enormous. Although the part of the $L_{2}$ calculation was ported to GPU,
the whole compression process for image $100\times 100$ [px] lasted
days, recall Tab.~\ref{tab:comptime} for time requirements of the single
$L_{2}$ evaluation. According to relations in Sec.~\ref{sec:statdes},
the overall number of operations for single standard
calculation\footnote{The number of operations related to enhanced
  version of $L_{2}^{i}$ evaluation cannot be determined exactly
  because of missing knowledge about zero segments. However, it is
  approximately $97$ percent less operations for dimensions $100\times
  100$ [px], i.e. approximately $2.0\cdot 10^{8}$ operations.} of
the $L_{2}^{i}$-function is $6.716\cdot 10^{9}$ comparing to
$1.021\cdot 10^{5}$ operations of the $S_{2}$ evaluation.

The final compressed $S_{2}$ and $L_2$-based structures are shown
in Figs.~\ref{fig:bone}c and d, respectively.  The $S_{2}$- and
$L_{2}$-functions of original and new microstructures are then
summarised in Fig.~\ref{fig:boneResult}. It is clearly visible
that the optimised functions very well coincide with the
prescribed ones, see Fig.~\ref{fig:boneResult}b,d,f. However, the
same does not hold for a microstructure optimised w.r.t. one
descriptor, but then evaluated w.r.t. to other. The $L_2$-based
compressed microstructure manifests much stronger short-range
correlations, see Fig.~\ref{fig:boneResult}b, while the
$S_2$-based compressed microstructure underestimates the
connectivity and, especially in the black phase, consists of a smaller
number of continuous line segments, see
Fig.~\ref{fig:boneResult}f. Nevertheless, we can conclude that
both descriptors provide visually well compressed microstructures
and even the two-point probability function allows to obtain
continuous regions similar to the original medium.

\section{Conclusions}
\label{sec:concl}

The paper is devoted to comparison of the lineal path function and
the two-point probability function in reconstruction and
compression of two-phase microstructures. So as to investigate
properties of the descriptors in a sufficient detail and to avoid
some misleading conclusions based on rough discretisation of the
lineal path by an approximately evaluated sampling template, the
accelerated version of the entire lineal path function was
proposed. The acceleration involves namely reformulation of the
sequential C/C++ code for the repeatedly called part of the lineal
path function into the parallel C/C++ code with CUDA extensions
enabling the use of computational potential of the NVIDIA graphics
processing unit (GPU). Even though the algorithm requires to copy
relatively large data structures to the GPU, it was shown that the
principal limitations reside in computational time required within
the compression or reconstruction process, where the lineal path
function needs to be often called more than million times. Despite
the parallel evaluation of the lineal path function on GPU, the
evaluation of the two-point probability function remains faster
even on a single CPU thanks to its accelerated formulation based
on the Fast Fourier Transform.

The accelerated discrete versions of both descriptors were
successfully employed for microstructure reconstruction and
compression processes governed by the simulated annealing algorithm
based on interchanging of two interfacial pixels belonging to
opposite phases. It was demonstrated that unlike the two-point
probability function, the discrete version of the lineal path function
based on line segments defined by Bresenham's algorithm does not
generally ensure a unique solution of a reconstruction process.
Nevertheless, the difference among the feasible solutions is small
and decreases with the increasing resolution. On the other hand,
many different morphologies could be fully defined by the lineal path
computed for only one continuous phase.

Three particular microstructures were employed for illustration of
typical features of both descriptors. The particulate suspension
consisting of equal sized squares revealed incapability of the
$S_2$ function to capture the shape of particles, which can be
emphasised by the $L_2$ function. Example of epithelial cells
demonstrated that very thin walls also cannot be captured by the
$S_2$ function and that the computation of $L_2$ corresponding to
the phase of walls is surprisingly not needed to achieve the
mostly connected walls in the compressed cell. Trabecular bone, on
the other hand, represents an example of mutually penetrating
phases of comparable volume fractions, where both descriptors
provided visually well looking microstructures.

We may conclude that despite the proposed acceleration steps, the
lineal path function remains computationally expensive descriptor,
which can be, however, essential for compression of morphologies
consisting of specific formations such as particles or thin
walls.

\section*{Acknowledgment}
This outcome has been achieved with the financial support of the
Czech Science Foundation, projects No.~105/12/1146,
No.~105/11/P370 and No.~13-24027S. We would like to thank
Ond\v{r}ej Jirou\v{s}ek of Institute of Theoretical and Applied
Mechanics, Czech Republic for providing us the measured image data
of microstructures and Jan Zeman of Czech Technical University,
Czech Republic, for bringing our attention to coupling the spatial
statistics with GPU computations.

\bibliographystyle{elsarticle-num}
\bibliography{liter}

\end{document}